\documentclass[structabstract,a4paper]{aa}  
\usepackage{graphicx}
\usepackage{txfonts}
\usepackage{natbib}
\usepackage{longtable}
\newcommand{\reduceme}{\mbox{R\raisebox{-0.35ex}{E}D%
\hspace{-0.05em}\raisebox{0.85ex}{uc}\hspace{-0.90em}%
\raisebox{-.35ex}{{m}}\hspace{0.05em}E}}
\begin{document}
   \title{A new stellar library in the region of the CO index at 2.3~$\mu$m}

   \subtitle{New index definition and empirical fitting functions}

   \author{E. M\'{a}rmol-Queralt\'{o}\inst{1}
          \and
          N. Cardiel\inst{1}
          \and
          A. J. Cenarro\inst{2}
          \and
          A. Vazdekis\inst{2}
          \and
          J. Gorgas\inst{1}
          \and
          S. Pedraz\inst{3}
          \and
          R.F. Peletier\inst{4}
          \and
          P. S\'{a}nchez-Bl\'{a}zquez\inst{5}
          }

   \institute{Departamento de Astrof\'{\i}sica y CC. de la Atm\'{o}sfera, 
              Universidad Complutense de Madrid, E28040-Madrid, Spain \\
              \email{emq@astrax.fis.ucm.es}
         \and
             Instituto de Astrof\'{\i}sica de Canarias, V\'{\i}a L\'{a}ctea
             s/n, E38200-La Laguna, Tenerife, Spain 
         \and
             Centro Astron\'{o}mico Hispano Alem\'{a}n, Calar Alto (CSIC-MPG), 
             C/Jes\'{u}s Durb\'{a}n Rem\'{o}n 2-2, 04004-Almer\'{\i}a, Spain
         \and
             Kapteyn Astronomical Institute, University of Groningen, Postbus
             800,9700 Av Groningen, the Netherlands
         \and
             Centre for Astrophysics, University of Central Lancashire, Preston
             PR1 2HE
             }


 
  \abstract
   {The analysis of unresolved stellar populations demands evolutionary
   synthesis models with realistic physical ingredients and extended wavelength
   coverage.}
   {To obtain a quantitative description of the first CO bandhead at
   2.3~$\mu$m, to allow stellar population models to provide improved
   predictions in this wavelength range.}
   {We have observed a new stellar library with a better coverage of the
   stellar atmospheric parameter space than preceding works. We have
   performed a detailed analysis of the robustness of previous CO index
   definitions with spectral resolution, wavelength calibration,
   signal-to-noise ratio, and flux calibration.}
   {We define a new line-strength index for the first CO bandhead at
   2.3~$\mu$m, D$_{\rm CO}$, better suited for stellar population studies than
   previous index definitions. We compute empirical fitting functions for the
   CO feature as a function of the stellar parameters (T$_{\rm eff}$, $\log g$
   and [Fe/H]), showing a detailed quantitative metallicity dependence. }
   {}


   \maketitle

\section{Introduction}

One of the most important challenges in modern astrophysics is the proper
understanding of the stellar content of unresolved systems, such as
extragalactic globular clusters and galaxies in different environments.  
Since the pioneering work of \citet{1961MNRAS.122...27C} and
\citet{Tinsley72,Tinsley78,Tinsley80}, this has been accomplished through the
comparison of the photometric and spectroscopic data with so-called
evolutionary stellar population synthesis models, which make use of theoretical
isochrones and libraries of spectral energy distributions (SEDs), either
theoretical, empirical or mixed \citep[for more recent models see
e.g.][]{2003MNRAS.340.1317V,2003MNRAS.344.1000B,Maraston05}. The most powerful
approach to achieve this goal is to compare a number of observed line-strengths
indices with their model predictions, providing in this way constraints to the
relevant physical properties of the systems, namely age, metallicity, initial
mass function (IMF), and the relative abundance of different chemical species.
Since, obviously, the reliability of model predictions improves as more
realistic physical ingredients are included, an important effort has been
devoted to improve the quality of the SED libraries.  Theoretical libraries
usually exhibit systematic discrepancies among themselves and when compared
with observational data
\citep[e.g.,][]{1997A&AS..125..229L,1998A&AS..130...65L}. Although the
alternative empirical libraries constitute a coarse grained, and usually
incomplete (especially for non solar metallicities and non solar abundance
ratios) sampling of the space of stellar atmospheric parameters, the use of
empirical fitting functions
\citep[e.g.,][]{1993ApJS...86..153G,1999A&AS..139...29G,1994ApJS...94..687W,2002MNRAS.329..863C}
can help to reduce these effects
\citep[e.g.,][]{1994ApJS...95..107W,2003MNRAS.340.1317V}.

Up to date, most of the observational effort has been focused to obtain
complete libraries in the optical range. However, a full understanding of the
physical properties of integrated stellar systems cannot be achieved ignoring
other spectral windows. In this sense, the CO features in the K band have
been used by many researchers to investigate the stellar content of galaxies,
including ellipticals
\citep{1975ApJ...195L..15F,1978ApJ...220...75F,1980ApJ...240..785F,1996MNRAS.280..895M,2000MNRAS.316..507M,1999MNRAS.306..199J,2001MNRAS.326..745M,Fornax_red,Davidge08},
spirals \citep{1999A&A...350..791J,Bendo04}, compact galaxies
\citep{2007AJ....133..576D,2008ApJ...677..276M}, starbursts and active galactic
nuclei
\citep{1994ApJ...421..101D,1994ApJ...428..609R,Shier96,Puxley,Goldader97,Vanzi97,1997ApJ...482L.149M,2000ApJ...545..190I,1999AJ....117..111H,Riffel07},
among others. These strong absorptions are the bandheads formed in the first
overtone ($\Delta \nu = +2$) bands of CO \citep{KH86}. Despite the common use
of these spectral features for stellar population studies, a proper
characterization of the CO bands with stellar atmospheric parameters is still
lacking. For that reason, we present in this work an improved study of the
infrared region around 2.3~$\mu$m, where the first bandhead of the strong CO
absorptions appear. In particular, we have observed a new library of stars
which clearly surpasses preceding works (see
\S~\ref{subsec-previous-work}) in the coverage of the stellar atmospheric
parameters. After a thorough analysis of previous index definitions that have
been used to measure the first CO bandhead, we present a new index, D$_{\rm
CO}$, which is well suited for stellar population studies. This new index
depends very little on spectral resolution (or velocity dispersion), less
sensitive to uncertainties in radial velocities, and can be measured with
poorer S/N ratios.  

In Section~\ref{sec-stellar-library} we present the new stellar library,
highlighting the improvements over previous libraries, the sample selection as
well as an overview of the observations and the data reduction. A detailed
discussion of the D$_{\rm CO}$ index definition is given in
Section~\ref{sec-index-definition}. This section also includes a comparative
study of the robustness of the new index to relevant effects. The measurements
of the D$_{\rm CO}$ index for the stellar library, and their associated error
estimates appear in Section~\ref{sec-CO-measurements}.
Section~\ref{sec-atmospheric-parameters} describes the stellar atmospheric
parameters used to compute the fitting functions, which are derived in
Section~\ref{sec-fitting-functions}. Finally, Appendix~A includes the tables
with all the D$_{\rm CO}$ measurements for all the stars used for the fitting
functions, as well as their stellar atmospheric parameters.


\section{The new stellar library}
\label{sec-stellar-library}

\subsection{Previous work}
\label{subsec-previous-work}

\begin{figure*}
\centering
\includegraphics[angle=-90,width=0.80\textwidth]{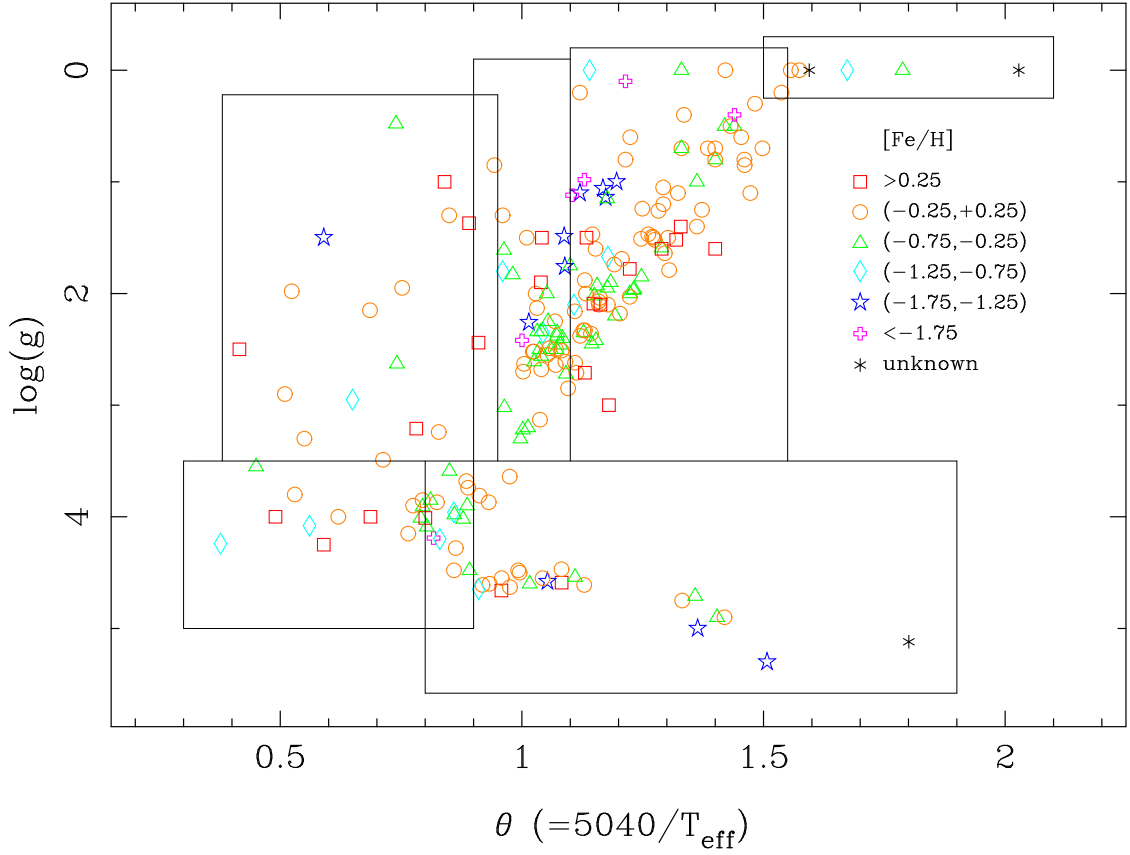}
\caption{ $\log g-\theta$ diagram for the stellar library presented in this
work, where $\theta = 5040/T_{\rm eff}$. Different symbols are used to
indicate stars of different metallicities, as shown in the key. The boxes
display the regions of the corresponding local fitting functions for the new CO
index (see \S~\ref{CO_fitting} for details).}
\label{hr_diagram}
\end{figure*}

Several authors have compiled, for different purposes, spectroscopic stellar
libraries in the K band
\citep{Joh,KH86,LR92,Ali,Han,WH96,WH97,Ram,For,Lan,Ivanov,Han2,BASI}.
Table~\ref{previous_libraries} summarizes the previous stellar libraries in the
K band, including the number of stars, spectral range, spectral resolution and
spectral types of the stars in each library. Due to the high S/N ratio of their
spectra, it is interesting to highlight the library of \citet{KH86} (hereafter
KH86), which contains 26 stars, but only with solar abundances. \citet{Ivanov}
present a library with 218 stars, which are not flux calibrated. The poor
metallicity coverage for these previous libraries (see Figs.~\ref{hr_diagram}
and \ref{comparison_libraries}) has not made it possible explore the
metallicity dependence of the spectral features in the K band. 

\begin{table*}
\caption{Main characteristics of previous spectroscopic stellar
libraries in the K band and the new stellar library presented in this work.}
\label{previous_libraries}
\centering
\begin{tabular}{lccccl}
\hline\hline
Reference     & Number of& Spectral range & Spectral resolution   & Spectral & Notes \\
              & stars    & ($\mu$m)  &($R=\lambda/\Delta\lambda$) & types    &       \\
\hline  
\citet{Joh}   &  32 & $1.2-2.5$ &        550  & A--M, I--V  & Low resolution \\
\citet{KH86}  &  26 & $2.0-2.5$ &2500 -- 3100 & F--M, I--V  & Solar abundances \\
\citet{LR92}  &  56 & $1.4-2.5$ &        550  & O--M, I--V  & Low resolution \\
\citet{Ali}   &  33 & $2.0-2.4$ &       1380  & F--M, V     & Dwarf stars\\
\citet{Han}   & 180 & $2.0-2.2$ &800 -- 3000  & O--B, I--V  & Hot stars, not CO region\\
\citet{WH96}  &  12 &$2.02-2.41$&  $\ge45000$ & G--M, I--V  & Few stars \\
\citet{Ram}   &  43 &$2.19-2.34$& 1380, 4830  & K--M, III   & Giant stars\\
\citet{WH97}  & 115 & $2.0-2.4$ &       3000  & O--M, I--V  & Solar abundances \\
\citet{For}   &  31 &$1.90-2.45$&  830, 2000  & G--M  I--III& Giant and supergiant stars\\
\citet{Lan}   &  77 & $0.5-2.5$ &       1100  & K--M, I--III& Giant and supergiant stars\\
\citet{Ivanov}& 218 &$1.48-2.45$&2000 -- 3000 & G--M, I--V  & Not flux calibrated\\
\citet{Han2}  &  37 & $2.0-2.2$ &8000--12000  & O--B, I--V  & Hot stars, not CO region\\
\citet{Cush}  &  26 & $0.6-4.1$ &       2000  & M--T, V     & Extremely cold dwarf stars\\
\citet{BASI}  & 114 & $2.05-2.19$&      2200  & O--M, I--V  & Not CO region\\
{\bf This work}&220& $2.11-2.37$&       2500  & O--M, I--V  & Improved metallicity coverage\\
\hline
\end{tabular}
\end{table*}

\subsection{Sample selection}

We have observed a new stellar library in the K band which comprises 220
stars.  The observed sample is a subset of MILES \citep[Medium-resolution
Isaac Newton Telescope Library of Empirical Spectra;][]{MILES,MILES_PARAM}, a
stellar library in the optical range with well known atmospheric parameters
for all the stars \citep[][]{MILES_PARAM}. Our final stellar sample includes
stars in the following stellar parameter ranges:
\begin{center}
$2485~{\rm K}\le T_\mathrm{eff}\le 13404~{\rm K}$,

$-0.34\;{\rm dex}\le\log g\le 5.30\;{\rm dex}$,

$-2.63\;{\rm dex}\le[{\rm Fe/H}]\le +0.98\;{\rm dex}$,
\end{center}
where $[{\rm Fe/H}]=\log Z - \log Z_{\odot}$.

\begin{figure*}
\centering
\includegraphics[angle=-90,width=0.70\textwidth]{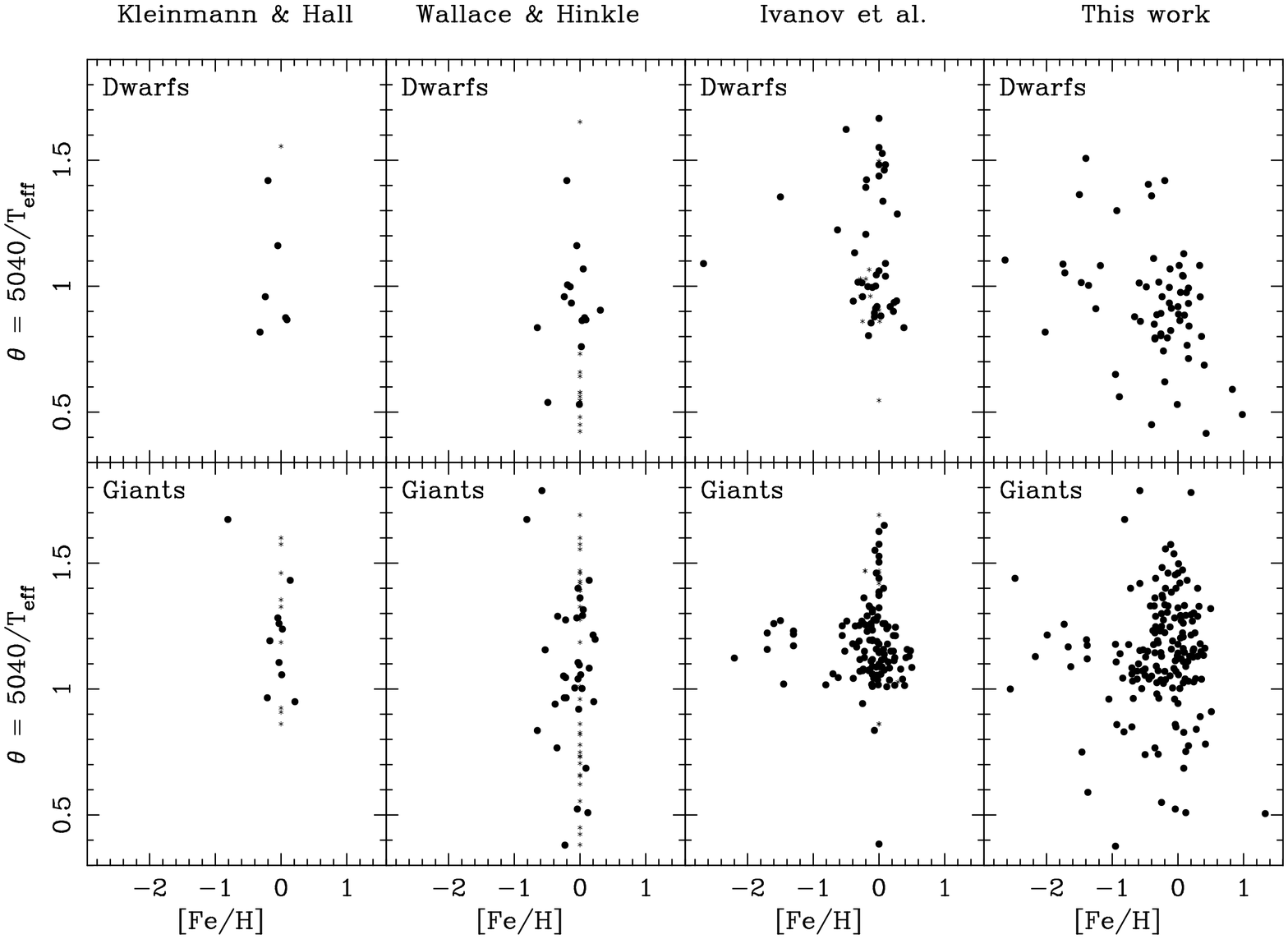}
\caption{Stellar parameter coverage in the stellar libraries of KH86, 
\citet{WH97}, \citet{Ivanov}, and this new library. We present separately dwarf
and giant (and supergiant) stars (upper and lower panels, respectively). The
stellar parameters of the stars from KH86 and \citet{WH97} were taken for 
\citet{Cayrel01} if available. Otherwise, we assigned solar metallicity
and T$_{\rm eff}$ from their spectral type following the tables of \citet{Lang} 
(small asterisks).} 
\label{comparison_libraries}
\end{figure*}

This library clearly has a larger metallicity coverage than the previous ones
(see Table~\ref{previous_libraries} and Fig.~\ref{comparison_libraries} for a
comparison between different works), and contains 8 stars in common with KH86,
23 with \citet{WH97} and 39 with \citet{Ivanov}. 

\subsection{Observations and data reduction}

\begin{table}
\caption{Observational configurations.}
\label{configuration}
\centering
\begin{tabular}{lcc}
\hline
Telescope           & CAHA 3.50 m       & TNG 3.56 m \\
\hline 
Instrument          & $\Omega$-CASS     & NICS       \\
Slit width (\arcsec)& 0.60              & 0.75       \\
Grism               & \#1               & KB         \\
Filter              & K                 & --         \\
Spectral coverage   & 2.01--2.43 $\mu$m & 1.95--2.34 $\mu$m\\
Dispersion          & 2.527 \AA /pix    &4.375 \AA /pix \\
FHWM                & 6.8 \AA           & 11.3 \AA   \\
Detector            & Hawaii-I          & Hawaii-I   \\
\hline
\end{tabular}
\end{table}

The bulk of the stellar library (217 stars) was observed during a total of 13
nights in five observing runs from 2002 to 2005 on the 3.5~m telescope at Calar
Alto Observatory (CAHA, Almer\'{\i}a, Spain) with $\Omega$--CASS. A subsample
of the stellar library (52 stars) was observed again at the Telescopio
Nazionale Galileo (TNG) at Roque de los Muchachos Observatory (La Palma, Spain)
with NICS (Near Infrared Camera Spectrometer) in February 2006 and May
2007, plus 3 new stars. The details of the instrumental configuration for both
runs are given in Table~\ref{configuration}.

Each star was observed several times at different positions of the slit
(standard procedure for infrared observations) to perform a reliable sky
subtraction. Halogen lamps (on and off) and arc lamps were observed for
flat-fielding, and C-distortion correction and wavelength calibration,
respectively. Vega type (A0) stars were observed at different airmasses during
each night in order to calibrate in relative flux and eliminate telluric lines
in the stellar spectra. 

We carried out a standard data reduction in the infrared using \reduceme\
\citep{1999PhDT........12C}, a reduction package which allows a parallel
treatment of data and error frames. The reduction process includes
flat-fielding, sky subtraction by subtracting consecutive images (A--B),
cosmetic cleaning, C-distortion correction and wavelength calibration with arc
lamps, spectrum extraction and relative flux calibration.  Atmospheric
extinction was corrected by using extinction coefficients (namely the relative
contributions of the Raleigh scattering and the aerosol extinction) derived for
CAHA Observatory by \citet{Hopp}. Those coefficients were extrapolated for La
Palma Observatory to correct the stars observed in this observatory.

Some of the reduction steps that requires more careful work are explained in
detail in the following subsections.

\subsubsection{Wavelength calibration}
Arc spectra of Argon lamps were acquired to perform the C-distortion correction
and the wavelength calibration. Due to instabilities and flexures of the
instrument and the telescope, calibration arc frames were obtained after each
star observed at CAHA. In the K band, the typical number of known lines in the
arc spectrum is rather low (just six in our instrumental configuration).
Because of that, the wavelength calibration was not accurately enough and a
second order wavelength correction was performed by identifying OH air-glow
lines in the sky spectrum of each star \citep{lineas_OH2,lineas_OH1}. For
observations at CAHA, a polynomial fit of the differences between the observed
and theoretical OH lines for the sky spectrum was necessary. The wavelength
calibration polynomial is expressed as a function of position as

\begin{equation}\label{pol_final}
W\left(x\right) = \lambda\left(x\right) + z\left(\lambda\left(x\right)\right),
\end{equation}
being $\lambda\left(x\right)$ the initial approximation to the wavelength
calibration 

\begin{equation}\label{pol_calibracion}
\lambda\left(x\right) = \sum_{i=0}^n a_i x^i,
\end{equation}
where $x$ is the position in the spectral direction, and $a_i$ are the
coefficients of the \mbox{$n^{\rm th}$-degree} calibration polynomial. The
second order correction is given by 

\begin{equation}
z\left(\lambda\left(x\right)\right) = \sum_{j=0}^m b_j \lambda^j,
\end{equation}
where $z\left(\lambda\right)$ are the computed differences between the observed
and theoretical OH lines, and $b_j$ are the coefficients of a \mbox{$m^{\rm
th}$-degree} polynomial. In our case, it is a second order polynomial. The
final wavelength calibration polynomial correction is 

\begin{equation}\label{pol_correccion}
z\left(x\right) = \sum_{j=0}^m b_j \Big( \sum_{i=1}^n a_i x^i \Big)^j
 =\sum_{k=0}^l c_k x^k,
\end{equation}
where $c_k$ are the coefficients of the new correction polynomial of order $l=m
\cdot n$ as a function of the position $x$. 

In the case of the observations at the TNG, we compared the observed sky
spectrum with the well calibrated sky spectrum from CAHA and we observed
constant wavelength differences between them. We cross-correlated both spectra
and we applied this constant shift to the wavelength calibration of TNG
spectra.

Finally, we checked the final spectra with K band spectra from KH86 and
\citet{WH97}, since they used a Fourier transform spectrometer, which implies a
very accurate wavelength calibration of the spectra. Although in the case of
observations at CAHA no differences were obtained, for the TNG observations we
had to apply a constant shift in order to achieve the correct wavelength
calibration. The origin of this discrepancy is found in the lack of OH sky
lines in the reddest wavelength region of the K band, which prevented from an
accurate cross-correlation of the sky spectra in that region.

\subsubsection{Flux calibration and telluric correction}\label{flux}

\begin{figure}
\centering
\includegraphics[angle=-90,width=0.90\columnwidth]{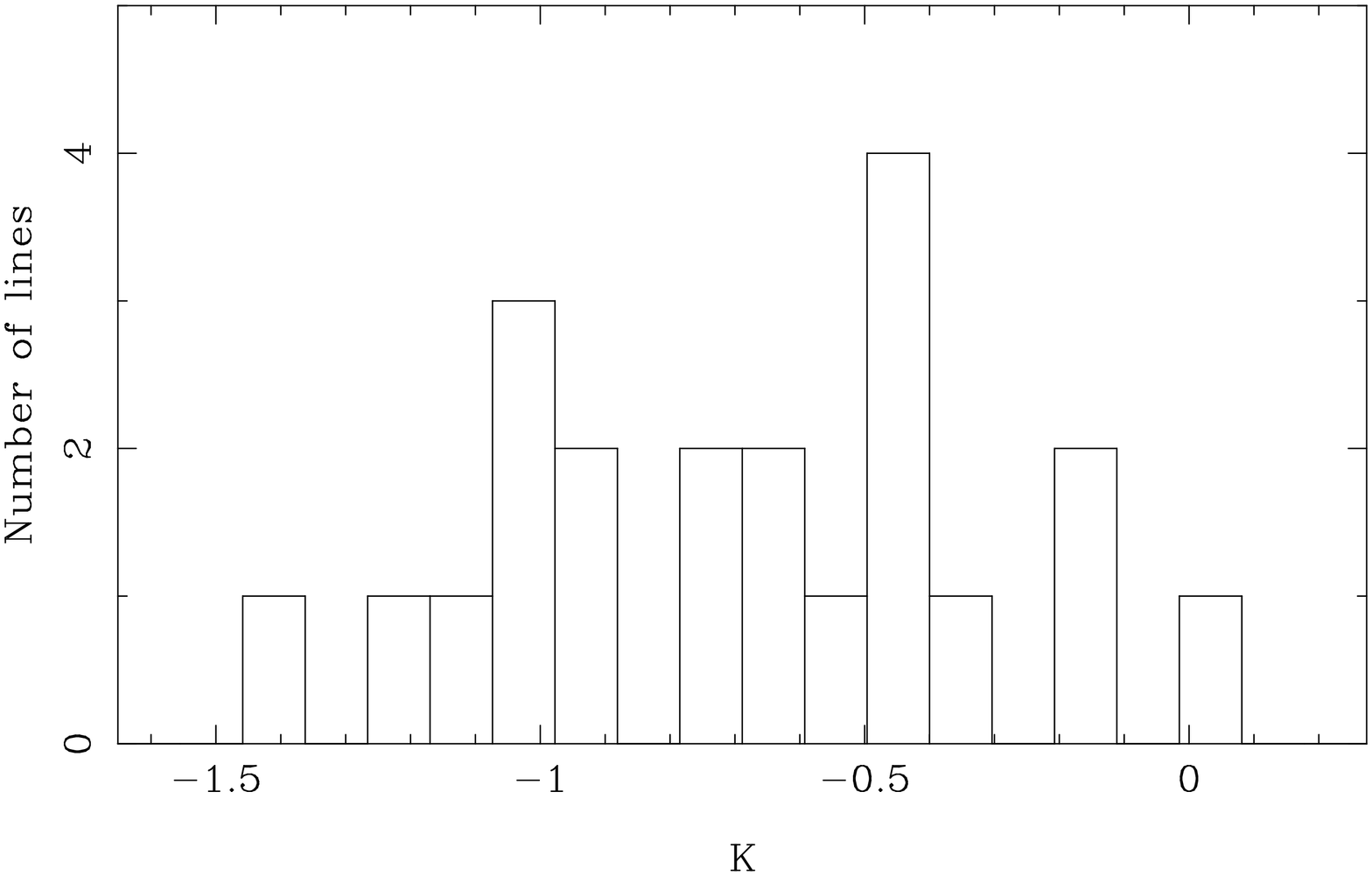}\\
\vspace{2mm}
\includegraphics[angle=-90,width=0.90\columnwidth]{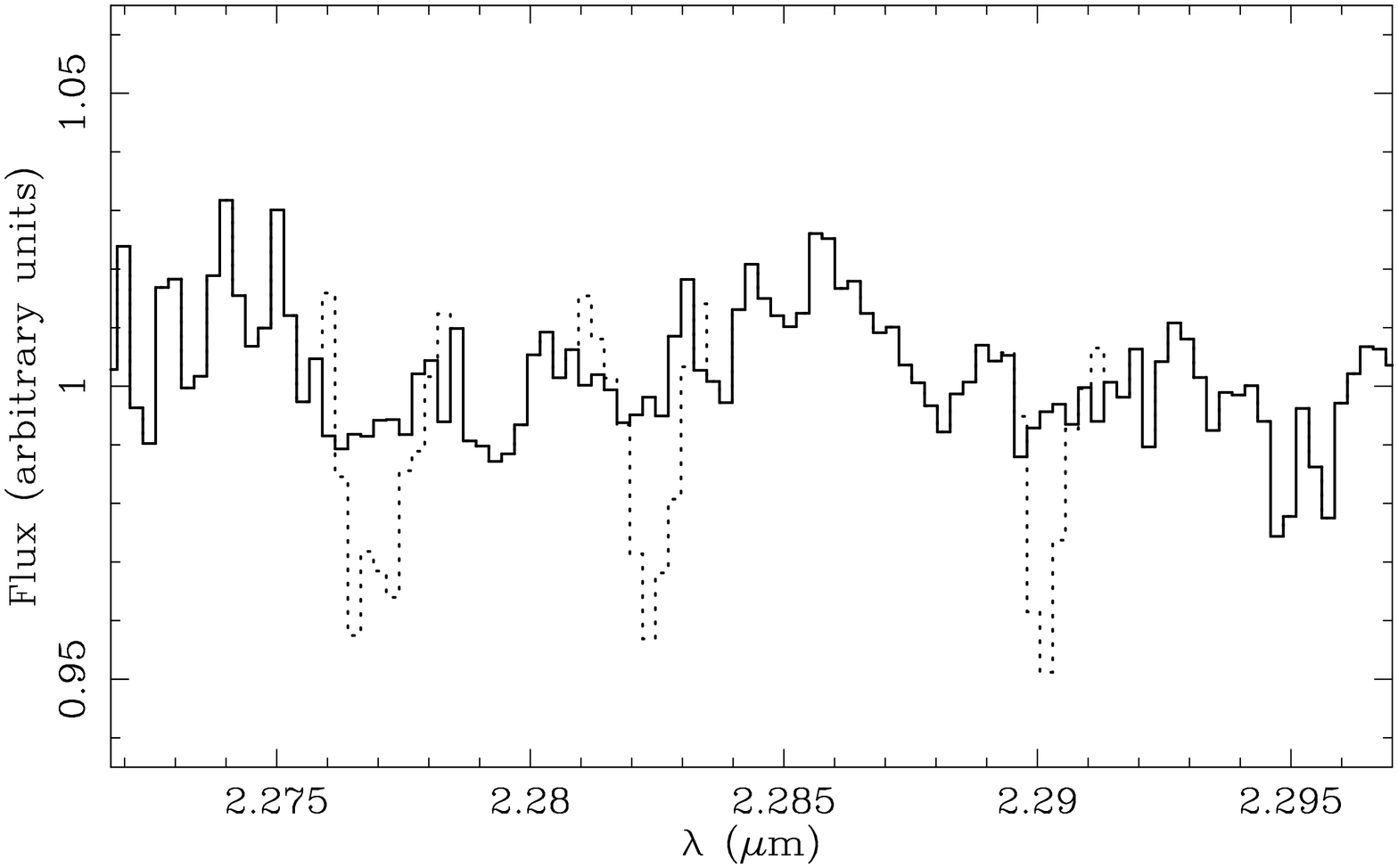}\\
\caption{{\it Upper panel}: Histogram with the different $K$ factors (see
Eq.~\ref{eq-telluric-correction}) employed to correct a particular stellar
spectrum from telluric absorptions. {\it Lower panel}: Example of a flux
standard spectrum before (dotted line) and after (solid line) telluric
correction (see details in the text).}
\label{telluric}
\end{figure}

There are two ways of flux calibrating infrared spectra. The first method
consists of observing a solar type star close to the star to be calibrated. The
solar type star is reduced as usual, dividing the final spectrum by the solar
spectrum \citep{SUN} degraded to the same spectral resolution as the problem
star. In this way, a spectrum with the information about the response curve and
the telluric lines is obtained. This spectrum is then rectified by the ratio
between the blackbody spectra at the temperature corresponding to the solar
type star and the solar temperature in order to obtain the correct continuum.
This final spectrum is used to (relative) flux calibrate and carry out the
telluric correction in the stars to be calibrated. The advantage of this method
is that it is easy to find a star of this type for each observation. 

A second method consists in the observation of Vega type stars at different
airmasses during the night. The main reason for choosing these stars is that
they are known to have no relevant features in our observational window, except
the Br\,$\gamma$ line. After wavelength calibration, Vega type stars are
divided by the well known theoretical Vega spectrum in our spectral range. In
that way, a spectrum with both the response curve and the telluric corrections
is obtained. The final stellar spectrum is then obtained after dividing each
star by this spectrum. In this work, we have used this second approach.

\subsubsection{Second order telluric correction}
As we mentioned in the previous section, we used the flux standard star in
order, not only to flux calibrate the stellar spectra, but also to correct
simultaneously for the telluric absorption lines. Due to the variability of the
observing conditions during the night, some telluric lines are badly corrected
by applying the response curve derived from the flux standard star. For that
reason an extra correction was necessary. First of all, we computed a reference
spectrum with the information of the telluric lines. For observations at CAHA,
we checked the response curves for each night looking for the spectrum with the
best removal of telluric features. The ratio between each response curve and
the previous spectrum free from telluric contamination provides the telluric
spectra that we used to correct all the stellar spectra. In the case of the TNG
observations, the telluric spectrum was obtained by dividing the flux standard
spectra at high and low airmasses, observed during each night. The telluric
spectrum in both observatories, obtained as explained above, consists mainly of
differences in the strength of the telluric absorption lines. To correct for
these lines, we modified their intensity by multiplying by an adjustable factor
$K$, i.e.,

\begin{equation} 
S_K = \left(S_0 - 1\right) \times K + 1, 
\label{eq-telluric-correction}
\end{equation} 
where $S_0$ is the telluric spectrum, and $S_K$ is the telluric spectrum
adjusted to correct a specific stellar spectrum. We divided the latter stellar
spectrum by $S_K$ and computed the r.m.s.\ in the corrected spectrum. The best
correction factor, $K$, is the one which minimizes the r.m.s. This method was
applied to different identified telluric lines separately, since they do not
vary in the same way. This effect can be seen in the histogram of
Fig~\ref{telluric}, top panel, where the number of telluric lines corrected by
a factor $K$ for a given spectrum is represented. In Fig.~\ref{telluric},
bottom panel, we present an example of the telluric lines correction for a
given flux standard spectrum. Notice that the telluric absorption lines can be
present even after flux calibration and it is important to correct them in
infrared spectroscopy.


\section{Index definitions for the CO band at 2.3~$\mu$m}
\label{sec-index-definition}

\subsection{The K band region}
The most prominent features in the K band are due to the rotational-vibrational
transitions of the CO molecule around 2.3~$\mu$m. Important absorptions are
also produced by other metallic species, such as \ion{Na}{I}, \ion{Fe}{I},
\ion{Ca}{I} and \ion{Mg}{I} (KH86, see Table~\ref{tabla_lineas}). The only
hydrogen line in this spectral range, Br\,$\gamma$, is present generally in
absorption in O and B stars, going into emission for high-luminosity stars
\citep[for a detailed study of this kind of stars, see][]{Han}. 

In this paper, we focus our study on the first CO bandhead at 2.29~$\mu$m.
Contrary to the \ion{Ca}{I} and \ion{Na}{I}, the contribution of other species
to the CO absorption is almost negligible \citep[see][for a further
discussion]{WH96,Ram}.

\begin{table}
\caption{Main spectroscopic features in the K band \citep[from][]{KH86}.}
\label{tabla_lineas}
\centering
\begin{tabular}{cccc}
\hline\hline
Species                 & $\lambda$ &Transition                         & Lower state \\
                        & ($\mu$m)  &                                   & energy (eV) \\
\hline            
\ion{H}{I} Br\,$\gamma$ &   2.1661  & $4^2F^0 -7^2G$                    & 12.70\\
\ion{Na}{I}             &   2.2062  & $4s^2S_{1/2} - 4p^2P^0_{1/2}$     & 3.19 \\
\ion{Na}{I}             &   2.2090  & $4s^2S_{1/2} - 4p^2P^0_{3/2}$     & 3.19 \\
\ion{Fe}{I}             &   2.2263  & $x^5F^0_4 - e^5D_3$               & 5.07 \\
\ion{Fe}{I}             &   2.2387  & $x^5F^0_3 - e^5D_2$               & 5.04 \\
\ion{Ca}{I}             &   2.2614  & $4d^3D_{3,2,1} - 4f^3F^0_4$       & 4.68 \\
\ion{Ca}{I}             &   2.2631  & $4d^3D_{3,2,1} - 4f^3F^0_3$       & 4.68 \\
\ion{Ca}{I}             &   2.2657  & $4d^3D_{3,2,1} - 4f^3F^0_2$       & 4.68 \\
\ion{Mg}{I}             &   2.2814  & $4d^3D_{3,2,1} - 6f^3F^0_{2,3,4}$ & 6.72 \\
\element[][12]{CO}(2,0) &   2.2935  & (2,0) bandhead                    & 0.62 \\
\element[][12]{CO}(3,1) &   2.3226  & (3,1) bandhead                    & 0.86 \\
\element[][13]{CO}(2,0) &   2.3448  & (2,0) bandhead                    & 0.32 \\
\element[][12]{CO}(4,2) &   2.3524  & (4,2) bandhead                    & 1.12 \\
\hline
\end{tabular}
\end{table}


\subsection{Previous definitions}\label{previous_definitions}

In order to measure the CO absorption at 2.3~$\mu$m in an objective way,
several authors have proposed different index definitions.
\citet{1973ApJ...184..427B} suggested a photometric system to measure the CO
features based on two narrow filters ($\Delta\lambda=0.10\,\mu$m) centered at
2.30~$\mu$m and at 2.20~$\mu$m for the CO absorption and the continuum,
respectively. The CO index was defined as the difference of the two filters
relative to the values obtained for $\alpha$ Lyrae, in magnitudes.  Following
this idea, \citet{1978ApJ...220...75F} defined the most used photometric CO
index (CO$_{\rm phot}$), with slightly different filter parameters
($\Delta\lambda=0.08\,\mu$m, for the CO filter centered at 2.36~$\mu$m, and
$\Delta\lambda=0.11\,\mu$m for the filter centered at 2.20~$\mu$m for the
continuum estimate).

The first spectroscopic CO index (${\rm CO}_{\rm KH}^{\rm mag}$) for the
CO(2,0) bandhead at 2.3~$\mu$m was defined by  KH86 as 

\begin{equation}
{\rm CO_{KH}^{mag}} = -2.5 \log\; {\rm CO_{KH}} =
-2.5 \log\;\frac{{\cal F}_a}{{\cal F}_c},
\label{eq-CO-KH}
\end{equation}
where ${\rm CO_{KH}}={\cal F}_a/{\cal F}_c$ is the ratio between the fluxes
integrated over narrow wavelength ranges centered in the absorption line
(\mbox{$\lambda\lambda2.29305-2.29832~\mu$m}) and the nearby continuum
(\mbox{$\lambda\lambda2.28728-2.29252~\mu$m}), measured in magnitudes. These
band limits have been used to measure the index as an equivalent width
\citep[e.g.][]{1993A&A...280..536O}. Both measurements can be converted using
\citep{2000A&A...357...61O} 

\begin{equation} 
{\rm CO_{KH}^{mag}} =
-2.5 \log \left(1-\frac{{\rm W}_\lambda\left(2.29\right)}{53 \AA}\right), 
\end{equation} 
where ${\rm CO_{KH}^{mag}}$ is the spectroscopic index initially defined by
KH86 measured in magnitudes, and ${\rm W_\lambda\left(2.29\right)}$ is the same
index measured as an equivalent width.

\citet{1994ApJ...421..101D} studied the behaviour of CO$_{\rm phot}$ and
indicated several reasons to introduce their new spectroscopic definition 

\begin{equation} 
{\rm CO_{sp}} = -2.5 \  \log <R_{2.36}>, 
\end{equation} 
where $<R_{2.36}>$ is the mean value of the rectified spectrum (normalized in
the continuum) in the $2.31-2.40~\mu$m range. This rectified spectrum is
obtained by fitting the continuum in the $2.00-2.29~\mu$m range with a power
law (F$_\lambda\varpropto\lambda^{-\alpha}$), due to the similarity of the
stellar spectrum in the K band to a Rayleigh--Jeans law. As
\citet{2000A&A...357...61O} indicated, this index is just the equivalent width
over the CO range relative to a continuum which is extrapolated from shorter
wavelengths. As a main advantage, this definition allows measurements of the CO
even from poor quality spectra.

Other authors \citep[for example,][]{LR92,Ram,For} proposed their own
definitions, adopting the bandpasses for the absorption and the continuum
without considering the use of those definitions in general situations.  
Recently, \citet{Riffel07} measured the CO absorption at 2.3~$\mu$m as an
equivalent width between $\lambda\lambda2.2860-2.3100$~$\mu$m, computing a
continuum defined as a spline using points free from emission/absorption lines
in the broad interval $\lambda\lambda2.2350-2.3690$~$\mu$m.

After a study of different band limits for the CO index measured in terms of
equivalent widths for stars of different spectral type, \citet{Puxley} proposed
to extend the absorption band of KH86 up to the end of the CO (2,0) band
(2.320~$\mu$m), and the use of three different bands to estimate the continuum
($\lambda\lambda 2.253-2.261~\mu$m, $2.270-2.278~\mu$m and $2.285-2.291~\mu$m).
Note that this type of index definition is what \citet{Cen2001a} called a {\it
generic index}. \citet{Puxley} adopted this definition because it allows to
distinguish between giant and supergiant stars and the correction for velocity
dispersion is smaller than with the other definitions.

An additional definition was introduced by \citet{Frogel2001}, in which the CO
absorption feature is measured using multiple bandpasses to estimate the
pseudo-continuum level. 

\begin{table}
\caption{Spectroscopic CO index definitions. References are 
CO$_{\rm KH}^{\rm mag}$ \citep{KH86}, I$_{\rm Puxley}$ \citep{Puxley} and 
I$_{\rm Frogel}$ \citep{Frogel2001}. Wavelengths are in vacuum.} 
\label{definitions}
\centering
\begin{tabular}{lccl}
\hline\hline
Index             & Continuum        & Absorption        & Comments \\
                  & bands ($\mu$m)& bands ($\mu$m) &        \\
\hline
CO$_{\rm KH}^{\rm mag}$ & 2.2873--2.2925 & 2.2931--2.2983 & Color-like index\\
I$_{\rm Puxley}$  & 2.2530--2.2610 & 2.2931--2.3200 & Generic index\\
                  & 2.2700--2.2780 &                & \\
                  & 2.2850--2.2910 &                & \\
I$_{\rm Frogel}$  & 2.2300--2.2370 & 2.2910--2.3020 & Generic index\\
                  & 2.2420--2.2580 &                & \\
                  & 2.2680--2.2790 &                & \\
                  & 2.2840--2.2910 &                & \\
D$_{\rm CO}$      & 2.2460--2.2550 & 2.2880--2.3010 & Generic\\
                  & 2.2710--2.2770 &                & discontinuity\\
\hline
\end{tabular}
\end{table}


\subsection{New index definition}\label{new_index}

\begin{figure}
\centering
\includegraphics[angle=-90,width=1.00\columnwidth]{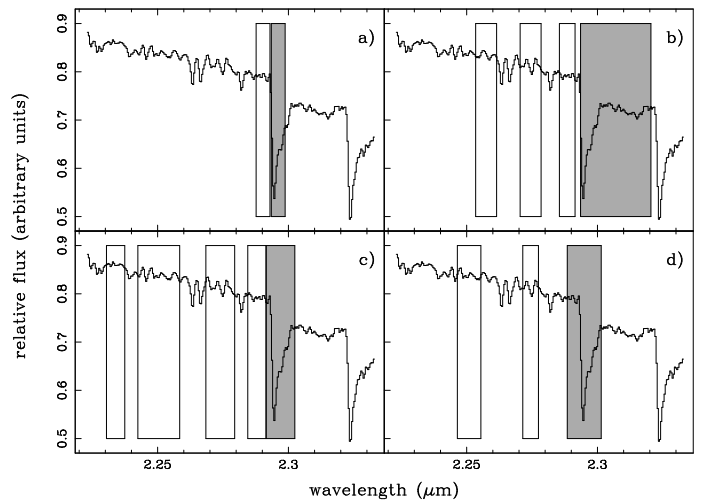}
\caption{Limits of the bandpasses in the definitions for the CO index proposed
by: a) KH86; b) \citet{Puxley}; c) \citet{Frogel2001}; and d) the new CO index
presented in this work. Grey and open bands represent absorption and continuum
bandpasses, respectively, for each index definition, superimposed on the
spectrum of HD137704.}
\label{index_bands}
\end{figure}

Even though the number of different CO index definitions is large, we have
explored in detail whether any of these is  actually well suited  for a
practical study of this spectroscopic feature in the integrated spectra of
galaxies. Curiously, from the list of previous index definitions only the one
presented by \citet{Puxley} (based on the previous definition by KH86) was
designed taking into account the variations of the index with radial velocity
and velocity dispersion, both of them very important in the study of galaxies.
In this work we have spent an additional effort to investigate the possibility
of finding an optimal CO index definition that could improve all the previous
definitions. 

In order to carry out this task, we have focused our efforts on defining a CO
index which is less sensitive to low signal-to-noise ratios, degradation due to
spectral resolution and/or velocity dispersion, and errors in wavelength
calibration (or errors in radial velocity), and in relative flux calibration
(see the next sections for a further study of each case).

After exploring different possibilities for the definition of the new index, we
propose to measure the CO at 2.29~$\mu$m as a {\it generic discontinuity},
i.e., as the ratio between the average fluxes in the continuum and in the
absorption bands

\begin{equation}
  {\rm D_{generic}} \equiv 
\frac{  
 \frac{\displaystyle\sum_{i=1}^{n_c}{\int_{\lambda_{c,i_1}}^{\lambda_{c,i_2}}F_{c,i}\left(\lambda\right)\;\mbox{d}\lambda}}
 {\displaystyle\sum_{i=1}^{n_c}{(\lambda_{c,i_2}-\lambda_{c,i_1})}
 }
 }
 {
 \frac{\displaystyle\sum_{i=1}^{n_a}\int_{\lambda_{a,i_1}}^{\lambda_{a,i_2}}F_{a,i}\left(\lambda\right)\;\mbox{d}\lambda}
 {\displaystyle\sum_{i=1}^{n_a}{(\lambda_{a,i_2}-\lambda_{a,i_1})}
 }
 }
\end{equation}
where ${\rm D_{generic}}$ is the generic discontinuity, and
$F_{a,i}\left(\lambda\right)$ and $F_{c,i}\left(\lambda\right)$ are the flux in
the $n_a$ absorption bands and $n_c$ continuum bands, respectively. Finally,
$\lambda_{x,i_1}$ and $\lambda_{x,i_2}$ are the lower and upper wavelength
limits of the $i^{\rm th}$ band $x$ (where $x$ is $a$ or $c$). This new
definition is similar to the B4000 index defined by \citet{1999A&AS..139...29G}
but using more than one bandpass to define the continuum and the absorption
regions.

Here we propose to measure the CO feature at $2.3 \mu$m as a generic
discontinuity, D$_{\rm CO}$, using two bandpasses for the continuum ($n_c=2$)
and one bandpass for the absorption region ($n_a=1$). The limits of these bands
are also listed in Table~\ref{definitions}. We selected the number of
bandpasses and their location taking into account several factors. Concerning
the continuum bandpasses, we have eluded the \ion{Ca}{I} and \ion{Mg}{I}
features, trying not to extend too far towards shorter wavelengths in order to
avoid potential systematic effects arising in the flux calibration of wide
line-strength indices. In the case of the absorption region, one single
bandpass is enough to cover the first CO bandhead. Compared with previous
definitions, we decided to shift slightly the blue continuum bandpass limit to
obtain an index which is more stable with velocity dispersion (i.e.\ spectral
resolution) and with radial velocity uncertainties.

Following \citet{1998A&AS..127..597C}, it is not difficult to show that the
expected variance in a D$_{\rm generic}$ index can be computed as

\begin{equation}
\sigma^2[{\rm D}_{\rm generic}] = 
  \frac{
    {\cal F}_c^2 \sigma_{{\cal F}_a}^2+
    {\cal F}_a^2 \sigma_{{\cal F}_c}^2
  }
  {
    {\cal F}_a^4
  },
\end{equation}
where ${\cal F}_x$ is the total flux per wavelength unit in the continuum
($x=c$) and the absorption ($x=a$) region, determined from the coaddition of
the flux in all the corresponding bandpasses, i.e.,

\begin{equation}
{\cal F}_x \equiv 
  \Theta \;
\frac{  
  \displaystyle\sum_{i=1}^{n_x} 
  \sum_{k=1}^{N_{\rm pixels}^i} F_{x,i}\left(\lambda_k\right)}
  {{\displaystyle\sum_{i=1}^{n_x}{(\lambda_{x,i_2}-\lambda_{x,i_1})}}},
\end{equation}
being $n_x$ the number of bandpasses in either the continuum or the absorption
region, $\Theta$ is the linear dispersion (in \AA/pixel), $N_{\rm pixels}^i$ is
the number of pixels covered by the $i^{\rm th}$ bandpass of the $x$ region
(with $x$ equal to $c$ or $a$), and $\lambda_k$ is the central wavelength of
the $k^{\rm th}$ pixel. The variance in these total fluxes are simply computed
as the quadratic sum of the individual variances in each pixel, i.e.,

\begin{equation}
\sigma^2_{{\cal F}_x} =
  \Theta^2
\frac{  
  \displaystyle\sum_{i=1}^{n_x} 
  \sum_{k=1}^{N_{\rm pixels}^i} \sigma^2_{F_{x,i}}\left(\lambda_k\right)}
  {\left[{\displaystyle\sum_{i=1}^{n_x}{(\lambda_{x,i_2}-\lambda_{x,i_1})}}\right]^2},
\end{equation}
where, in particular, $\sigma^2_{F_{x,i}}\left(\lambda_k\right)$ is the
variance corresponding to the random error in the $k^{\rm th}$ pixel. It is
important to highlight that in these expressions we are assuming that the
random errors in each pixel are not correlated.


\subsection{Sensitivities of the indices to different effects} 
\label{subsec-sensitivities}

\begin{figure*}
\centering
\includegraphics[width=0.75\textwidth]{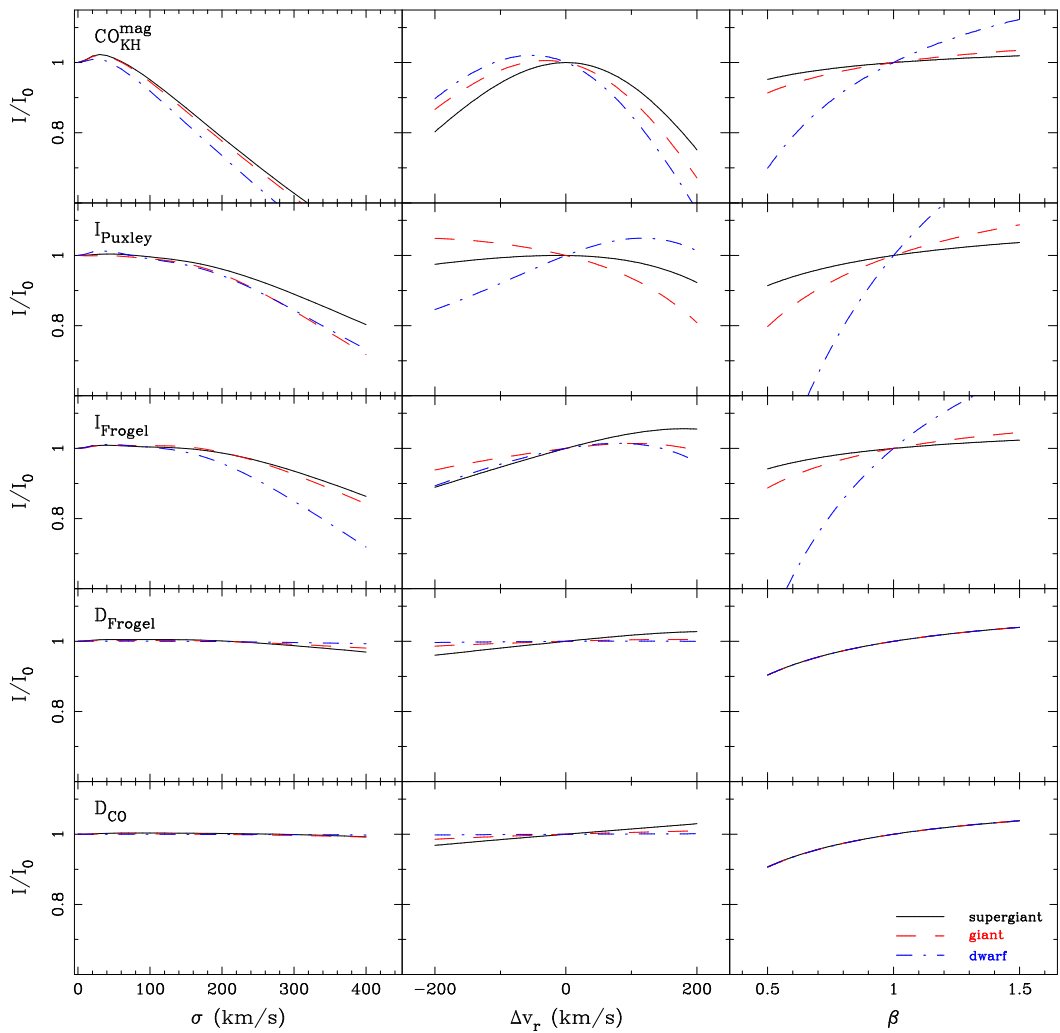}
\caption{Study of the sensitivity of the different CO index definitions (from
top to bottom, CO$_{\rm KH}^{\rm mag}$,  I$_{\rm Puxley}$, I$_{\rm Frogel}$,
D$_{\rm Frogel}$ and D$_{\rm CO}$) to several relevant parameters.  {\it Left
column:\/} ratio between the index I measured on the spectra broadened with a
given $\sigma$, and the index I$_0$ measured on the initial spectra
\citep[$\sigma_0=3$~km/s, corresponding to the resolution of the stellar
library of][]{WH96}. {\it Central column:\/} ratio between the index I for a
given v$_{\rm r}$ and the index I$_0$ in the original spectrum (v$_{\rm r} = 0$
km/s).  {\it Right column:\/} ratio between the index I measured on the
spectrum multiplied by a curved spectrum parametrized by $\beta$ (see
\S~\ref{study_flux} for details) and the index measured over the original
spectrum, I$_0$, as a function of the curvature parameter $\beta$.}
\label{study_index}
\end{figure*}

In this section, we discuss the sensitivity of previous spectroscopic indices
defined by KH86 (CO$_{\rm KH}^{\rm mag}$), \citet{Puxley} (I$_{\rm Puxley}$)
and \citet{Frogel2001} (I$_{\rm Frogel}$), and the new CO index (D$_{\rm CO}$)
to velocity dispersion (or spectral resolution), wavelength calibration (radial
velocity), relative flux calibration and signal-to-noise (S/N) ratio. A fifth
index, D$_{\rm Frogel}$, is considered: a generic discontinuity based on the
same bands proposed by \citet{Frogel2001}. In Fig.~\ref{index_bands} we show
the bandpasses for these index definitions.  For this study, we selected from
the high resolution library of \citet{WH96} three stars with similar spectral
type (M2--5; chosen because of their strong CO features) and different
luminosity class (supergiant, giant and dwarf) in order to account for
differences depending on the type of star. The resolution of the spectra is
0.54~\AA (FHWM) and they are all shifted to rest-frame.

\subsubsection{Spectral resolution and velocity dispersion broadening}
In order to study the sensitivity of the spectroscopic CO indices to the
spectral resolution or velocity dispersion broadening ($\sigma$), we broadened
the stellar spectra of the selected stars with additional $\sigma$'s from the
initial $\sigma_0$ up to $\sigma=400$ km/s (in steps of 10 km/s). The different
indices were measured on all these broadened spectra and we computed the ratio
between the index (I) at each $\sigma$ and the index measured on the original
spectrum (I$_0$). Fig.~\ref{study_index} (left column) shows this ratio as a
function of the velocity dispersion for the definitions we are studying.
Compared to the other index definitions, the two generic discontinuities
(D$_{\rm Frogel}$ and D$_{\rm CO}$) are clearly the less sensitive to velocity
dispersion broadening.

\subsubsection{Wavelength calibration}
Sometimes, errors in the wavelength calibration arise in the spectra even after
a very careful reduction or due to an inaccurate radial velocity (v$_{\rm r}$)
estimate of the studied object. Because of that, it is important to define
indices with the least possible sensitivity to this kind of uncertainties.  To
quantify this effect, we measured the CO absorption with the different index
definitions in the stellar spectra shifted from $-200$ to $+200$ km/s with
steps of 4~km/s in radial velocity. In Fig.~\ref{study_index} (central column)
we present the ratio between the index, I, measured at v$_{\rm r}$ and the
initial value I$_0$ (assumed v$_{\rm r} = 0$ km/s) as a function of the
considered v$_{\rm r}$ for different types of stars. It is apparent from the
figures that the indices CO$_{\rm KH}^{\rm mag}$, I$_{\rm Puxley}$ and I$_{\rm
Frogel}$ are very sensitive to radial velocity uncertainties, while D$_{\rm
Frogel}$ and the new index definition D$_{\rm CO}$ are more robust to this
effect.

\subsubsection{Flux calibration}\label{study_flux}
As we explained in \S~\ref{flux}, it is common to use theoretical spectra to
recover the real shape of the continuum. This practise implies the knowledge of
the temperatures of the standard stars. For that reason, we have studied the
impact, during flux calibration, of an error in the temperature estimate of the
standard stars. To analyse the impact when solar-type stars are used as flux
standards, we computed the blackbody spectrum in the interval $5600\leq
T_\mathrm{eff}\leq6300$~K, and derived the ratio between these spectra and the
blackbody, at solar temperature. To study the effect when Vega type stars are
used as calibrators, we analysed the differences from the theoretical spectrum
of Vega ($T_\mathrm{eff}\sim9400$~K) and the real temperature of the Vega type
stars (from 8400 to 14400~K for our study). In both cases, we found that the
changes in the continuum produced by differences in the assumed temperature of
standard stars produce negligible differences in the measured indices.

\begin{figure}
\centering
\includegraphics[angle=-90,width=\columnwidth]{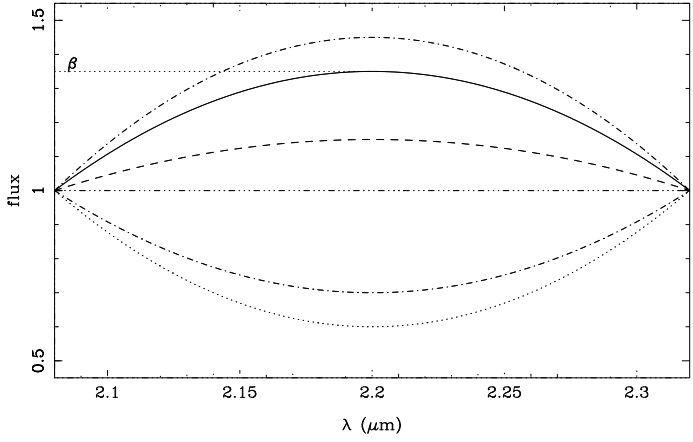}
\caption{Examples of the curved spectra used in the study of the sensitivity
of the different index definition to a wrong curvature of the spectrum. The
parameter $\beta$ determines the distance to the maximum/minimum in the center
of the spectrum (shown here for the polynomial displayed with a solid line).}
\label{curved_sp}
\end{figure}

Finally, we studied the impact of a wrong curvature in the response curve,
which is a typical source of systematic error. In order to obtain an estimate
of this effect, as a first order approach we have artificially modified the
continuum shape of the original spectra by multiplying them by a second order
polynomial. This polynomial was chosen to pass through 3 fixed points, two at
the borders of the wavelength range (where the polynomial were forced to be
equal to 1.0), and another point at the center of that range (where the
polynomial was set to a variable parameter $\beta$ ranging from 0.5 to 1.5). In
Fig.~\ref{curved_sp} we show different examples of these polynomials for
distinct values of $\beta$. Note that, with this exercise, we simply studied
the effect of a low frequency error in flux calibration.

In Fig~\ref{study_index} (right column) we present the ratio between the
measured index in the stellar spectrum multiplied by the polynomial for a given
value of $\beta$, I, and the original one, i.e., I$_0$ for $\beta = 1.0$ (no
additional curvature), as a function of the parameter $\beta$. The sensitivity
of each index definition to the $\beta$ parameter is, not surprisingly,
dependent on the location and wavelength coverage of the index bandpasses, and
also depends on the way the pseudo-continuum is determined, and the absolute
value of the index. For these reasons, CO$_{\rm KH}^{\rm mag}$, I$_{\rm
Puxley}$ and I$_{\rm Frogel}$ are the most sensitive to an error in the
response curve.  In particular, the CO$_{\rm KH}^{\rm mag}$ depends strongly on
the strength of the CO absorption. On the other hand, I$_{\rm Puxley}$ and
I$_{\rm Frogel}$, defined as generic indices, extrapolate the continuum value
into the absorption band and, in that way, the wrong curvature too.  In
addition, the generic discontinuities D$_{\rm Frogel}$ and D$_{\rm CO}$,
computed as the averaged flux in the continuum and absorption bands, exhibit no
differences between dwarf, giant and supergiant stars. 

\subsubsection{S/N ratio}

\begin{figure}
\centering
\includegraphics[angle=-90,width=1.00\columnwidth]{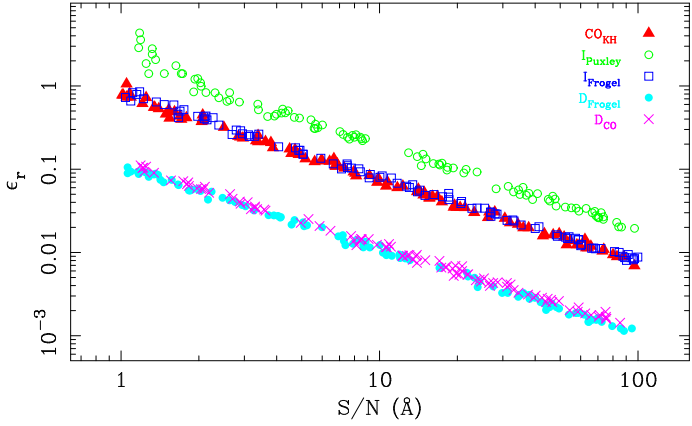}
\caption{Example of the study of the dependence of the relative error in the
measure of the first CO bandhead using different index definitions at a given
S/N ratio per \AA. We represent the relative errors vs.\ S/N ratio (in
logarithmic scale) measured on the simulated spectra of a giant star (the
results are independent of the type of star). Different symbols are used to
indicate different index definitions, as shown in the key. See text for
details.}
\label{SN}
\end{figure}

One important issue to take into account in the definition of a new index is
the dependence of the relative error of the measurements on the S/N ratio. In
this sense, the aim is to find an index definition which provides the lowest
relative error in the measurements with the lowest signal-to-noise ratio in the
spectra. For that reason, we have studied the behaviour of relative errors
measured with previously analyzed CO index definitions as a function of the S/N
ratio. Using a particular stellar spectrum, we have simulated a set of one
hundred spectra (and their associated error spectra) with random S/N(\AA)
ratios in the range 1.0--100.0. For this task we have used the program {\tt
indexf}\footnote{http://www.ucm.es/info/Astrof/software/indexf/}
\citep{indexf}. In Fig~\ref{SN} we compare the results obtained for a giant
star (same results are obtained for supergiant and dwarf stars). Not
surprisingly, the relative errors in all the definitions follow

\begin{equation} 
\varepsilon_{\rm r} = \frac{c}{\rm S/N(\AA)}, 
\label{eq-sn}
\end{equation} 
where $c$ is a constant that depends on the particular index.
This result was already found for atomic and molecular indices
\citep[][]{1998A&AS..127..597C}, and for generic indices \citep[][]{Cen2001a}.
It is clear from Fig.~\ref{SN} that the same holds for generic discontinuities.
Considering Eq.~\ref{eq-sn}, it is evident that, at a given S/N ratio, the
lower relative errors correspond to the index definitions with lower $c$
values. Table~\ref{table-ctes-sn} list these values for the five index
definitions under study. From these numbers and the data displayed in
Fig.~\ref{SN}, it is obvious that D$_{\rm CO}$ is comparable to D$_{\rm
Frogel}$, while CO$_{\rm KH}^{\rm mag}$, I$_{\rm Puxley}$ and I$_{\rm Frogel}$
provide larger relative errors for a given S/N ratio.

\begin{table}
\caption{Values of the constant $c$ of Eq.~\ref{eq-sn} for the different index
definitions analyzed in \S~\ref{subsec-sensitivities}.}
\label{table-ctes-sn}
\centering
\begin{tabular}{lc}
\hline\hline
Index                   &    $c$    \\ 
\hline 
CO$_{\rm KH}^{\rm mag}$ & $0.7537$ \\
I$_{\rm Puxley}$        & $2.0258$ \\
I$_{\rm Frogel}$        & $0.8123$ \\
D$_{\rm Frogel}$        & $0.1075$ \\
D$_{\rm CO}$            & $0.1198$ \\
\hline
\end{tabular}
\end{table}


\subsubsection{The best index definition}

Once we have studied the behaviour of the different CO index definitions as a
function of all the relevant parameters, we can conclude that the D$_{\rm CO}$
index definition is, in general, preferable.

On one hand, CO$_{\rm KH}^{\rm mag}$, I$_{\rm Frogel}$ and I$_{\rm Puxley}$ are
too sensitive of spectral resolution and errors in wavelength calibration and
radial velocity. In addition, the behaviour of CO$_{\rm KH}^{\rm mag}$, I$_{\rm
Frogel}$ and I$_{\rm Puxley}$ are also too sensitive to uncertainties in the
spectrophotometric system (i.e., flux calibration).

When the sensitivity to S/N is included in the comparison, it is clear that the
best definitions are the two generic discontinuities, namely D$_{\rm Frogel}$
and D$_{\rm CO}$. Since the use of generic discontinuities for measuring the CO
absorption is introduced for the first time in this paper, and considering that
the D$_{\rm CO}$ is practically insensitive to spectral resolution (or velocity
dispersion broadening) up to $\sigma\sim 400$~km/s, we propose to use the new
definition, especially for the future analysis of integrated spectra.

\subsection{Conversion between different CO index systems}
\label{subsec-conversion}

\begin{figure*}
\centering
\includegraphics[angle=-90,width=0.40\textwidth]{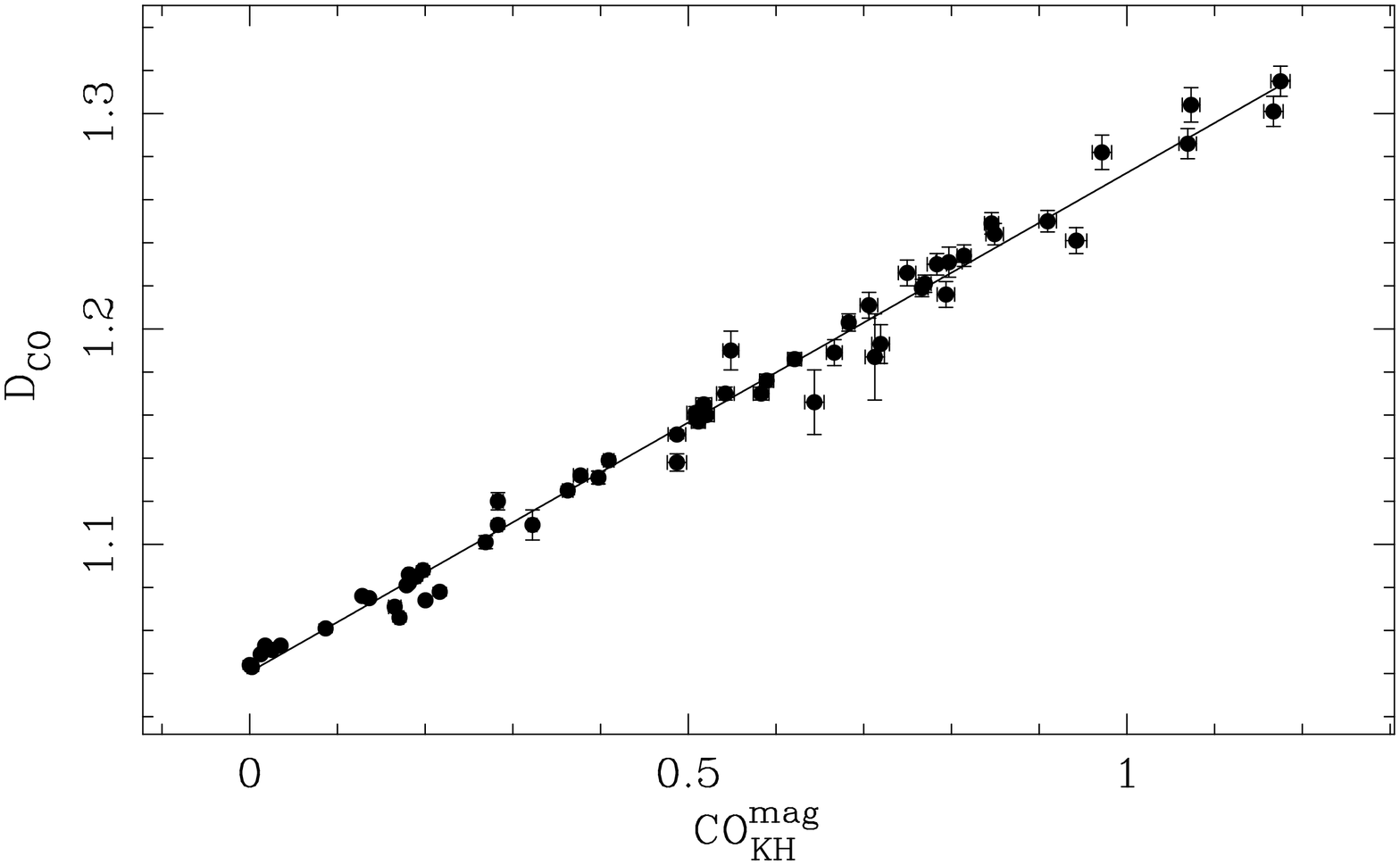}
\hspace{4.0mm}
\includegraphics[angle=-90,width=0.40\textwidth]{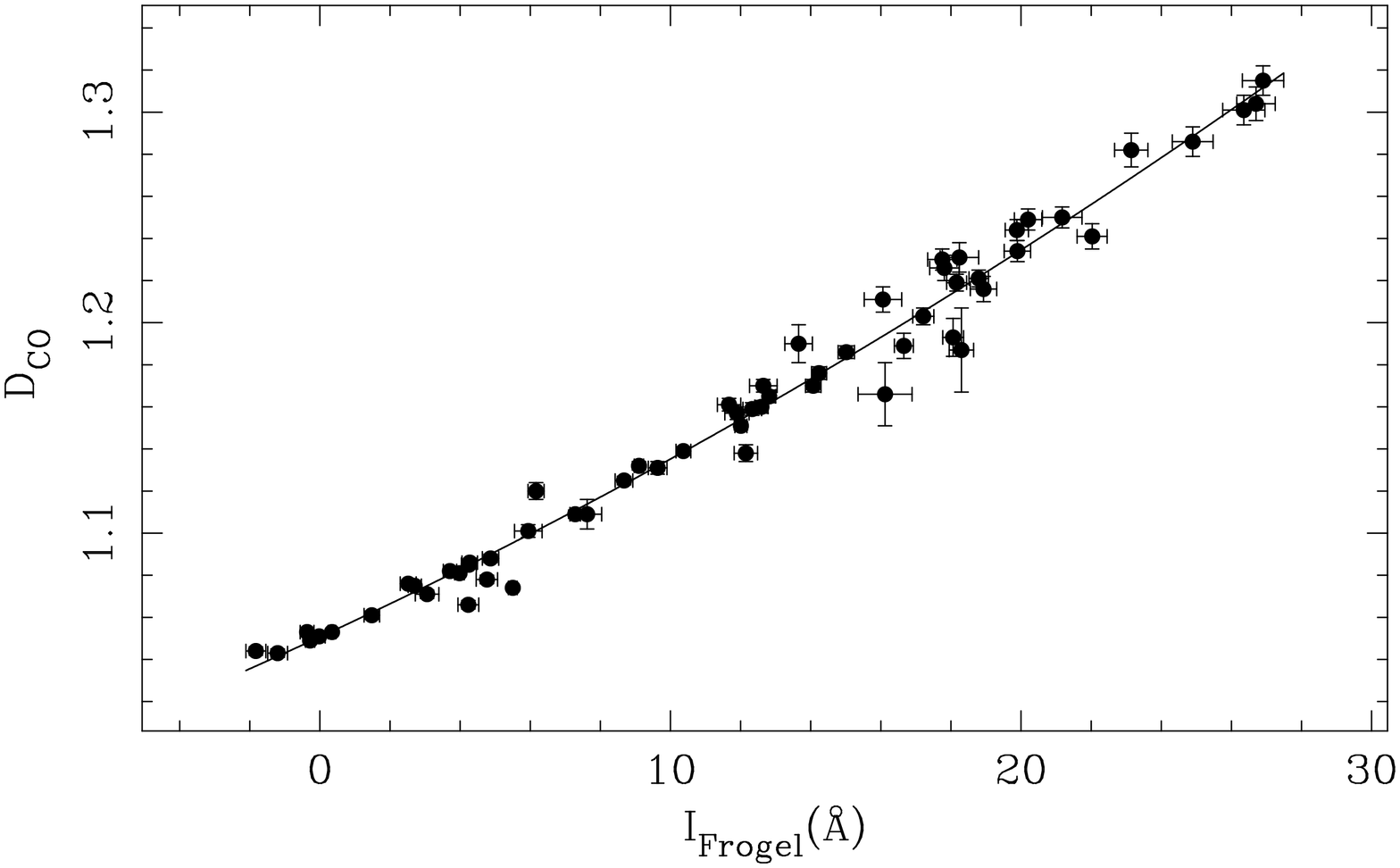}
\vspace{4.0mm}

\includegraphics[angle=-90,width=0.40\textwidth]{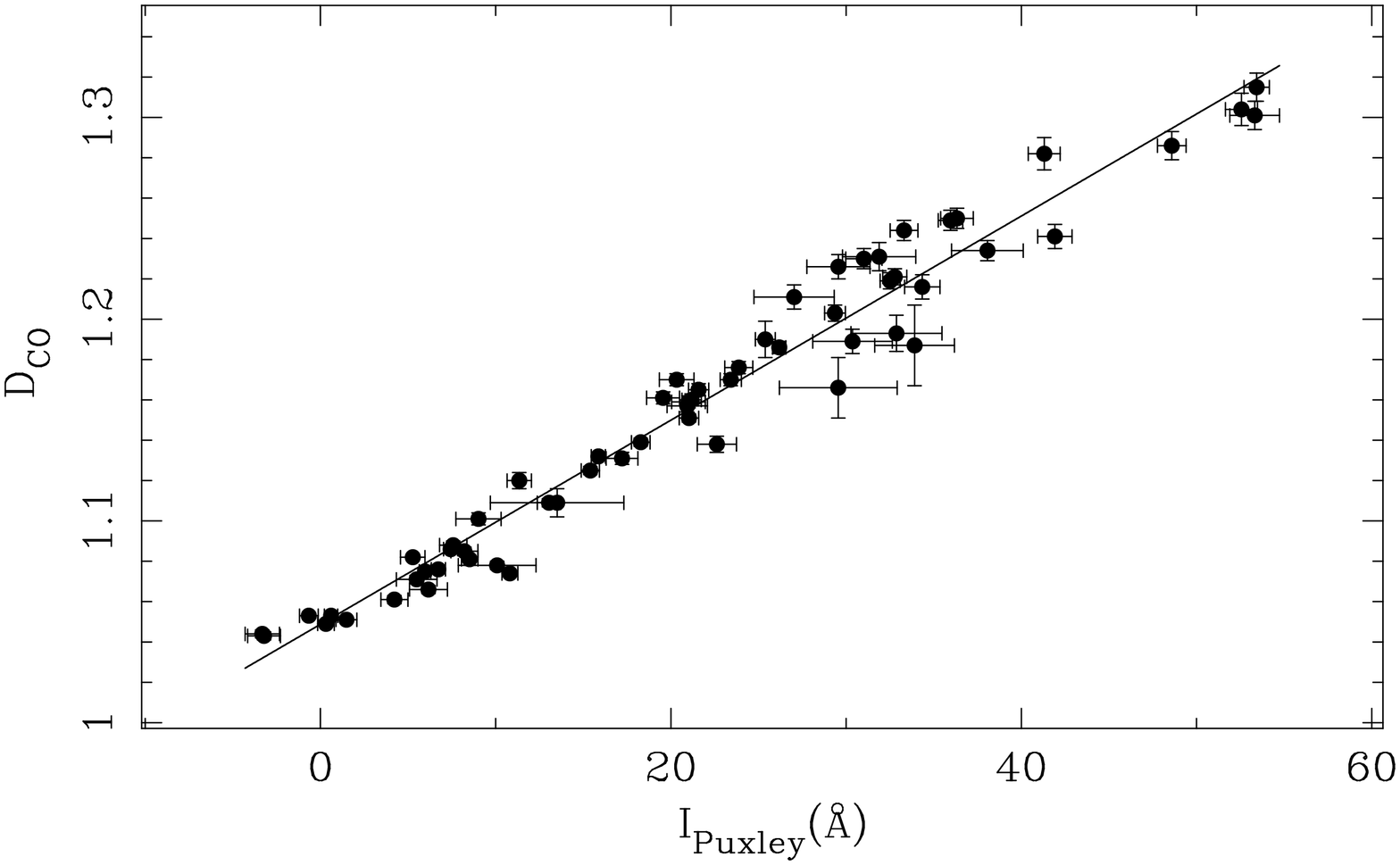}
\hspace{4.0mm}
\includegraphics[angle=-90,width=0.40\textwidth]{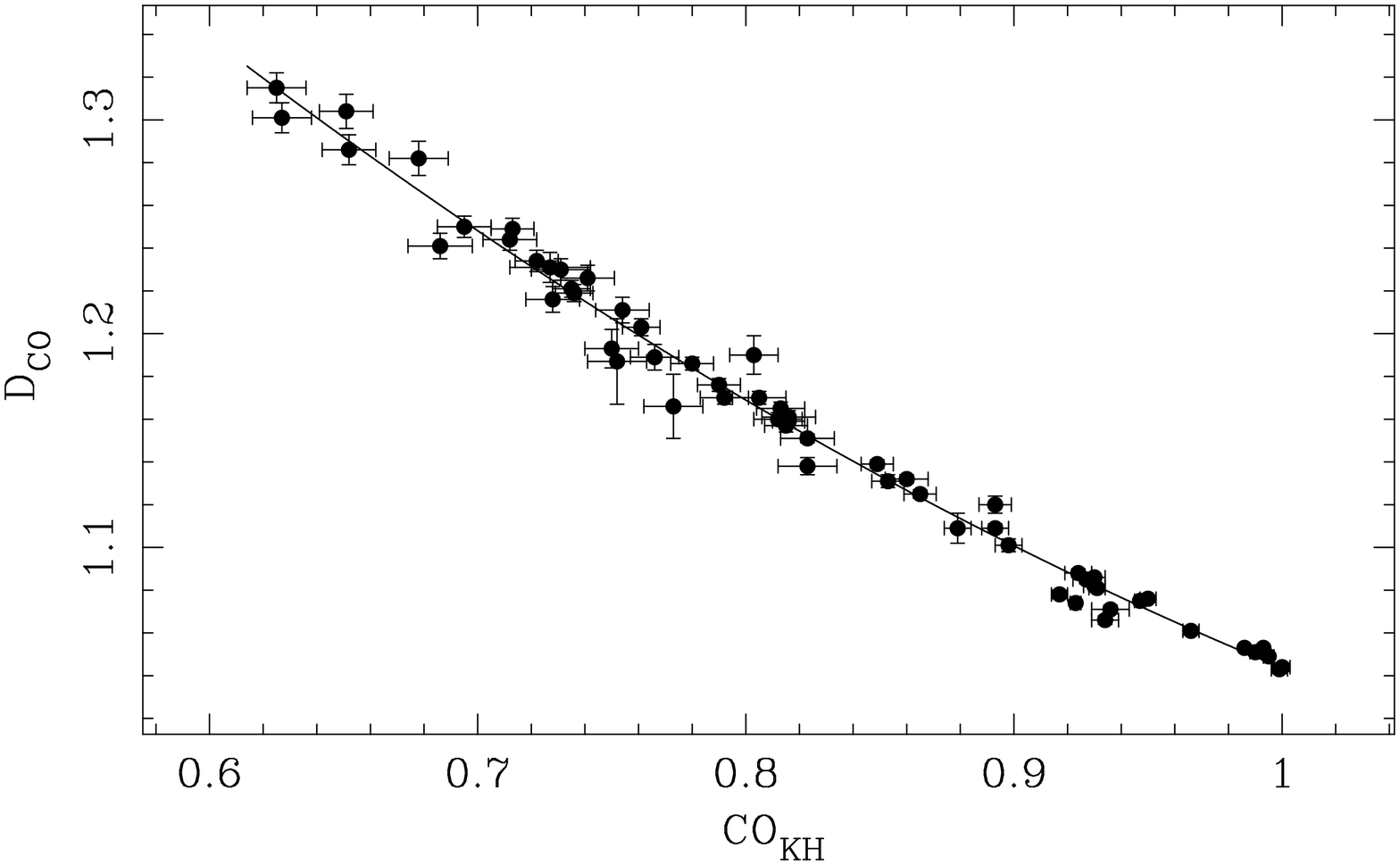}
\caption{Comparison between measurements on the TNG subsample of the first CO
bandhead using different index definitions. The panels show the transformation
between the new index D$_{\rm CO}$ and CO$_{\rm KH}^{\rm mag}$, I$_{\rm
Frogel}$, I$_{\rm Puxley}$ and CO$_{\rm KH}$ (from top to bottom, and from left
to right). The solid lines are least squares fits to the data, and correspond
to the transformations given in Eqs.~\ref{KHmag_to_new}--\ref{KH_to_new}.}
\label{fig-conversions} 
\end{figure*}

In this section we give the calibrations to convert between the new CO index
definition and the CO indices defined by KH86, \citet{Puxley} and
\citet{Frogel2001}. In order to obtain these conversions, we have measured the
indices on the subsample of stars observed at the TNG ($3200 \le T_\mathrm{eff}
\le 9625$~K, $0.00 \le \log g \le 5.00$, $-1.73 \le [{\rm Fe/H}] \le +0.36$).
The calibrations were computed by deriving a least squares fit to the data. The
fits are completely compatible with index measurements on the KH86 and
\citet{WH97} spectra which are on the same spectrophotometric system.  Just six
stars from \citet{WH97} deviate more than $3\sigma$ from the fitted relation
due to a problems in the continuum and the telluric correction of those
spectra. In Fig.~\ref{fig-conversions} we show all these fits and the data used
to compute them. 

The conversion between the index defined by KH86, ${\rm CO_{KH}^{mag}}$, and
the new index D$_{\rm CO}$ is given by

\begin{equation}\label{KHmag_to_new}
{\rm D_{CO}} = 1.0407 \left(\pm 0.0021\right)+0.2317 \left(\pm 0.0035\right)\;{\rm CO_{KH}^{mag}}
\end{equation}
with $r^2 = 0.9863$. 

The expression to compute D$_{\rm CO}$ from the index defined by
\citet{Frogel2001} is

\begin{eqnarray}\label{Frogel2new}
{\rm D_{CO}} & = & 1.0507 \left(\pm 0.0031\right)+0.0077 \left(\pm 0.0005\right)\;{\rm
I_{Frogel}}+ \nonumber \\
& & +0.00007 \left(\pm 0.00002\right)\;{\rm I_{Frogel}^2} 
\end{eqnarray}
with $r^2 = 0.9802$, where ${\rm I_{Frogel}}$ is measured as an equivalent
width (\AA).

As we mentioned before, ${\rm I_{Puxley}}$ is also measured as an equivalent
width. The expression to compute the D$_{\rm CO}$ index from I$_{\rm Puxley}$
is

\begin{equation}
{\rm D_{CO}} = 1.0488 \left(\pm 0.0033\right)+0.0051 \left(\pm 0.0001\right)\;{\rm I_{Puxley}}
\end{equation}
with $r^2 = 0.9629$.

Finally, we have also computed the conversion between the ratio CO$_{\rm KH}$
(not to be confused with CO$_{\rm KH}^{\rm mag}$; see Eq.~\ref{eq-CO-KH}) and
the new CO index

\begin{eqnarray}\label{KH_to_new}
{\rm D_{CO}} & = & 2.1119 \left(\pm 0.0724\right)-1.6205 \left(\pm 0.1772\right)\;{\rm CO_{KH}}+
\nonumber \\
& & +0.5521 \left(\pm 0.1071\right)\;{\rm CO_{KH}^2}
\end{eqnarray}
with $r^2 = 0.9867$. This last transformation will be used in next section.


\section{Measurements of the CO absorption for the stellar library and error
estimates}
\label{sec-CO-measurements}

\subsection{Index measurements}

\begin{figure*}
\centering
\includegraphics[angle=-90,width=0.40\textwidth]{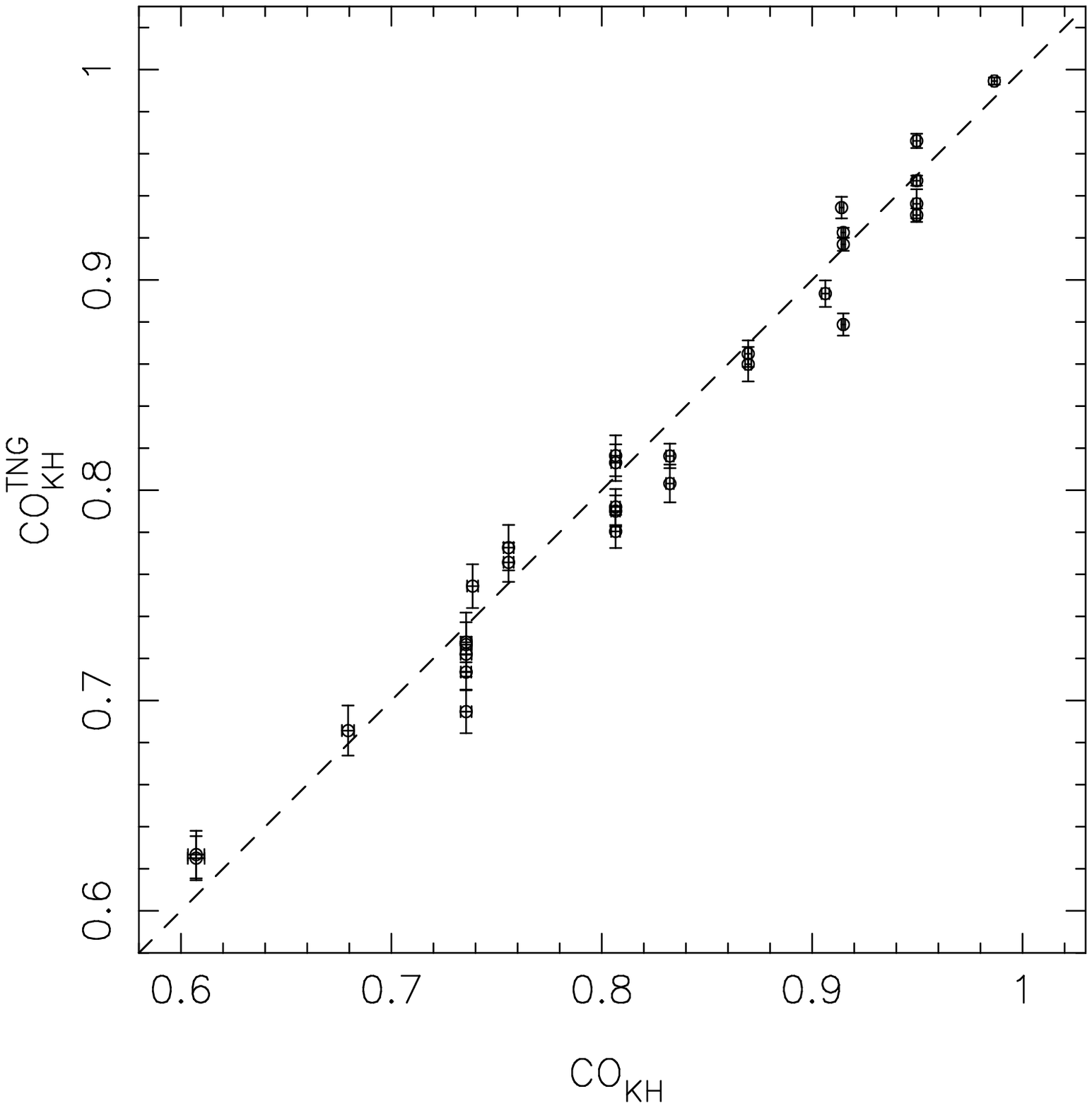}
\hspace{5mm}
\includegraphics[angle=-90,width=0.40\textwidth]{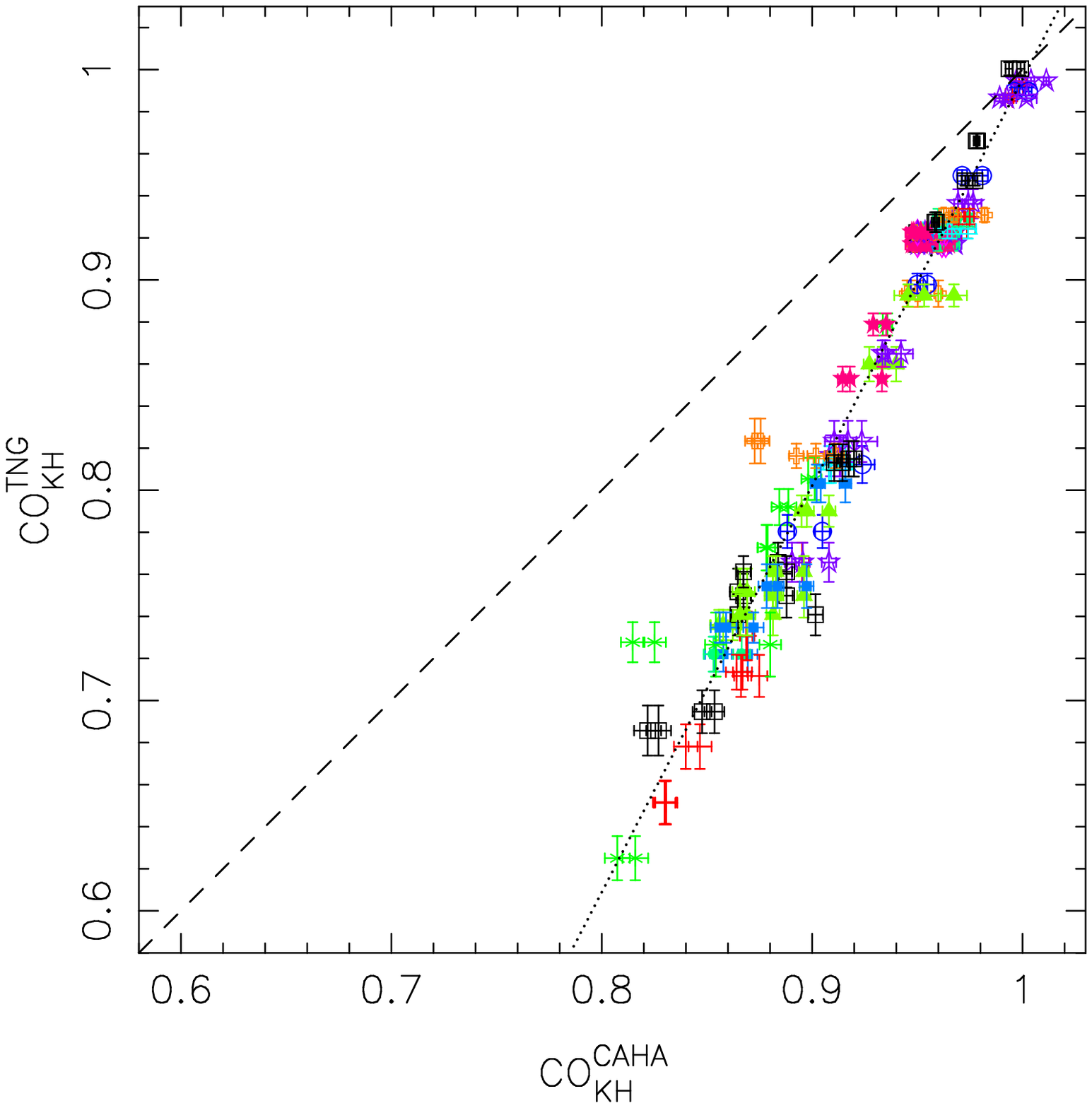}
\caption{{\it Left panel}: CO$_{\rm KH}$ index measured on the stars with solar
metallicity observed at the TNG (CO$_{\rm KH}^{\rm TNG}$) vs.\ the same
measurements in the most similar stars from the KH86 library (CO$_{\rm KH}$).
{\it Right panel}: CO$_{\rm KH}$ index measured on stars observed in
common at TNG (CO$_{\rm KH}^{\rm TNG}$) and CAHA (CO$_{\rm KH}^{\rm
CAHA}$).  Different symbols are used for distinct observing runs at CAHA,
although no segregation of the data between runs is apparent. The dotted line
shows the empirical correction for the CAHA measurements. In both panels, the
dashed line shows the one-to-one relation.}
\label{corrections}
\end{figure*}

A detailed study of the continuum in spectra observed at CAHA compared with the
the spectra published by KH86 have revealed some problems with the flux
calibration of the CAHA data. The shape of the continuum in these spectra
showed a spurious and non-reproducible high-frequency structure superimposed to
the real continuum, which was affecting not only the shape of the continuum but
also the final index measurements. During the reduction of the data it was
neither possible to identify nor correct this additional source of noise.

In order to handle, at least in an empirical way, the spectrophotometric
calibration of the CAHA spectra, we re-observed a good subsample of the stellar
library at the TNG. To guarantee that the TNG data were on the appropriate
spectrophotometric system, the CO measurement of each star observed at the TNG
was compared with the measurement of the most similar spectrum (in
$T_\mathrm{eff}$ and luminosity class) available in the KH86 library at the
same spectral resolution.

Since the bandpasses for the new index D$_{\rm CO}$ encompass a wide range in
wavelength, the strange behaviour of the continuum shape in the CAHA spectra
has a large impact on the index measurements. Luckily, this is not such a big
issue for the CO$_{\rm KH}$ ratio, since both continuum and absorption
bandpasses are very close in this definition.  For that reason, we decided to
measure the CO$_{\rm KH}$ index, transforming afterward the results into the
D$_{\rm CO}$ index using Eq.~\ref{KH_to_new} (which provides a good conversion
between both indices, as shown in the previous section). In more detail, the
method to derive the final spectrophotometric calibration can be summarized as
follows.

First of all, the CO$_{\rm KH}$ measurements of the subsample of stars with
solar metallicity re-observed at the TNG were compared with the corresponding
star in the KH86 library, as explained before. In Fig.~\ref{corrections}, left
panel, the results of this comparison are shown. A least squares fit to the
one-to-one relation was computed, providing $r^2=0.9768$. Although some points
in this figure appear relatively far from the 1:1 relation (considering their
error bars), it is important to highlight that we are not comparing the same
stars, and that, in any case, the determination coefficient $r^2$ is high
enough to guarantee the quality of the transformation.

Finally, we used the stars in common between the TNG and CAHA to empirically
correct the measured indices on the CAHA spectra sampled at the spectral
resolution of the TNG spectra. Fig.~\ref{corrections}, right panel, presents
the relation between the CO$_{\rm KH}$ ratio in common stars observed in both
observatories.  A least squares fit to a straight line gives $r^2=0.9539$. We
have checked that a unique empirical correction is valid for all the observing
runs at CAHA (within the statistical errors). For this study, we have used all
the individual measurements for each star instead of averaging all the
observations along the slit. After transforming the CAHA CO$_{\rm KH}$
measurements onto the correct spectrophotometric system, we applied
Eq.~\ref{KH_to_new} to obtain the new D$_{\rm CO}$ measurements, which will be
used later to derive the empirical fitting functions.


\subsection{Random errors and systematic effects}\label{sub-sec-lib-errors}

\subsubsection{Known sources of random errors}

We have considered several sources of random errors: photon statistic and
read-out noise, errors due to the detector cosmetic, the combined effect of
wavelength calibration and radial velocity uncertainties, and flux calibration
uncertainties.

({\it i}) {\it Photon statistics and read-out noise}. 
With the aim of tracing the propagation of photon statistic and read-out noise,
we followed a parallel reduction of data and error frames with the reduction
package \reduceme\,, which creates error frames at the beginning of the
reduction procedure and translates into them, by following the law of
combination of errors, all the manipulations performed over the data frames. In
this way the most problematic reduction steps (flat-fielding and distortion
corrections, wavelength calibration, etc.) are taken into account and, finally,
each data spectrum has its corresponding error spectrum which can be used to
derive reliable photon errors in the index ($\sigma_{\rm photon}$). A detailed
description of the estimate of random errors in the measurement of classical
line-strength indices and the impact of their interpretation are shown in
\citet{1998A&AS..127..597C,Cardiel03}. The new CO index is defined in this
paper as a {\it generic discontinuity} and follows the expressions given in
\S~\ref{new_index} for the errors.

\begin{figure}
\centering
\includegraphics[angle=-90,width=\columnwidth,clip=]{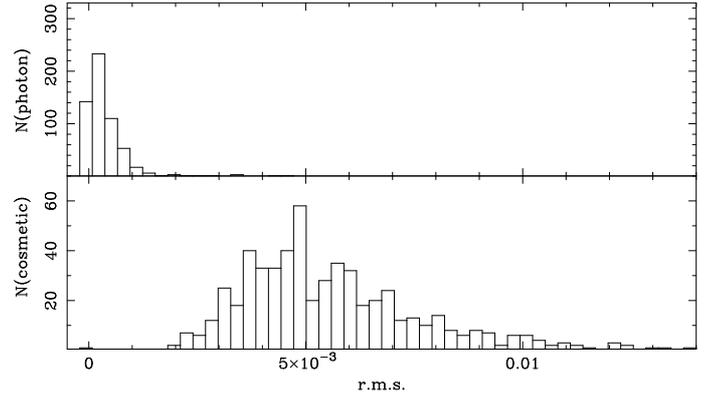}
\caption{{\it Upper panel}: Histogram with the photon statistic and read-out
noise error in the D$_{\rm CO}$ for the stars in the stellar library. {\it
Lower panel}: Similar histogram with the errors derived from measurements of the
r.m.s.\ on two spectral regions with no apparent absorption features for the
stars in the stellar library.}
\label{cosmetic_errors}
\end{figure}

({\it ii}) {\it Errors due to the detector cosmetic}. 
In the case of infrared detectors, the errors due to the detector cosmetic are
very important. We measured the r.m.s. ({\it root-mean-squared}) on two
different regions of the spectra with no apparent absorption features
($\lambda\lambda2.2685-2.2790$~$\mu$m and
$\lambda\lambda2.2825-2.2890$~$\mu$m), in order to obtain an estimation of
photon statistics and read-out noise errors together with the errors due to
imperfections present in the images even after flat-fielding correction. In
Fig.~\ref{cosmetic_errors} we compare the errors due to photon statistic and
read-out noise errors (upper panel), derived from first principles (using the
parallel reduction of data and error frames), with the errors from direct
measurement of the r.m.s in the spectra (lower panel).  As it can be seen, we
are underestimating the random errors if we do not consider the errors due to
the detector cosmetic.

({\it iii}) {\it Wavelength calibration and radial velocity errors}. 
The combined effect of wavelength calibration and radial velocity correction is
another random source of error. We considered as an upper limit for the
wavelength calibration error the projection on the spectral direction of half
of the slit width. These values were measured from the FWHM of the arc lines,
providing errors of 45 km/s for the observations at CAHA and 85 km/s for the
TNG images.  These numbers translate into typical errors of $\sigma[{\rm
D_{CO}}]=0.002$ and $\sigma[{\rm D_{CO}}]=0.004$, for CAHA and TNG
respectively. 

Radial velocity for  most of the stars in the stellar library were taken from
the Hipparcos Input Catalogue \citep{Hipparcos}, which in the worst cases are
given with mean probable errors of 5 km/s. For the missing stars in this
catalogue, we adopted radial velocities from the SIMBAD database, which
presents typical errors of a few km/s. These errors are completely negligible
in comparison with the wavelength calibration errors already mentioned.

({\it iv}) {\it Flux calibration errors}. 
In order to check for possible random errors in the index measurements due to
the flux calibration method, we should have observed several spectrophotometric
reference spectra each observing night.  Since this is a highly time-consuming
approach, we did not observe them.  Alternatively, we have decided to check the
impact of this kind of error through the comparison of common stars between
different nights and runs and we will discuss it in
\S~\ref{systematic-effects}.

\subsubsection{Additional sources of random errors}

Expected random errors for each star can be computed by adding quadratically
the random errors derived from the known sources previously discussed, i.e.,

\begin{equation}
\sigma^2 [{\rm CO}]_{\rm initial}=
\end{equation}
\[
= \sigma^2 [{\rm CO}]_{\rm photon+cosmetic}+ 
\sigma^2 [{\rm CO}]_{\rm wavelength}.
\]

However, additional (but unknown) sources of random errors may still be present
in the data. Using the multiple observations available for each particular
star, we compared the standard deviation of the different D$_{\rm CO}$
measurements ($\sigma [{\rm CO}]_{\rm rms}$) with the initial random error
($\sigma^2 [{\rm D}_{\rm CO}]_{\rm initial}$) for that star. In the cases in
which the standard deviation was significantly larger than the expected error
(using a $\chi^2$-test of variances), a residual random error $\sigma[{\rm
D}_{\rm CO}]_{\rm residual}$ was derived as

\begin{equation}
\sigma^2 [{\rm CO}]_{\rm residual}= 
\sigma^2 [{\rm CO}]_{\rm rms}- 
\sigma^2 [{\rm CO}]_{\rm initial},
\end{equation}
and quadratically added to the initial random errors to obtain the final random
errors, i.e.,

\begin{equation}
\sigma^2 [{\rm CO}]_{\rm final}= 
\sigma^2 [{\rm CO}]_{\rm initial}+ 
\sigma^2 [{\rm CO}]_{\rm residual}.
\end{equation}

It is worth noting that this additional error was only necessary for a few
stars, since most of them presented consistent errors between different
measurements.

\subsubsection{Systematic effects}
\label{systematic-effects}

Due to the large number of runs needed to complete the whole library,
systematic errors can arise because of possible differences between the
spectrophotometric systems of each run at both the CAHA and TNG telescopes. To
guarantee that the whole dataset is on the same system, we observed stars in
common between different runs at each telescope. We compared the index
measurements of these stars by applying a $t$-test to study whether the
differences of the index measurements of the common stars were statistically
larger than zero, finding no systematic effects between different nights within
a given observing run, and between different runs. For that reason we have
considered that an additional error exclusively due to flux calibration was not
required (i.e., the actual flux calibration errors are within the already
computed random errors).

It is important to highlight that there is not guarantee that our data are on a
{\it true} spectrophotometric system, so we encourage readers interested in
using our results to include in their observing plan stars in common with the
stellar library to ensure a proper comparison.

\subsection{Additional measurements of the CO absorption: globular cluster
stars}\label{cluster_stars}
To improve the stellar parameter coverage of our stellar library, additional
stars were included for the computation of empirical fitting functions for the
D$_{\rm CO}$ (see \S~\ref{sec-fitting-functions}).  \citet{Frogel2001} and
\citet{2004AJ....127..925S} presented a sample of globular cluster giant stars
($R\sim 1500$, $\sigma \sim 85$~km/s), characterized by their low metallicity,
with measurements of the CO absorption at 2.3~$\mu$m using the definition of
\citet{Frogel2001}. Since there is not dependence of I$_{\rm Frogel}$ at the
spectral resolution of these data (see Fig~\ref{study_index}), we transformed
their CO measurements to the new index by applying Eq.~\ref{Frogel2new}. The
stellar atmospheric parameters of these stars were determined from $J$ and $K$
photometry, as it is explained in \S~\ref{parameter_cluster}. Finally, we
considered 80 stars from \citet{Frogel2001} and 14 stars from
\citet{2004AJ....127..925S}, which, together with the stellar library presented
in this work, will be used to parametrize the behaviour of the CO index as a
function of the stellar atmospheric parameters.


\section{Stellar atmospheric parameters}
\label{sec-atmospheric-parameters}

\begin{figure*}
\centering
\includegraphics[angle=-90,width=0.90\textwidth]{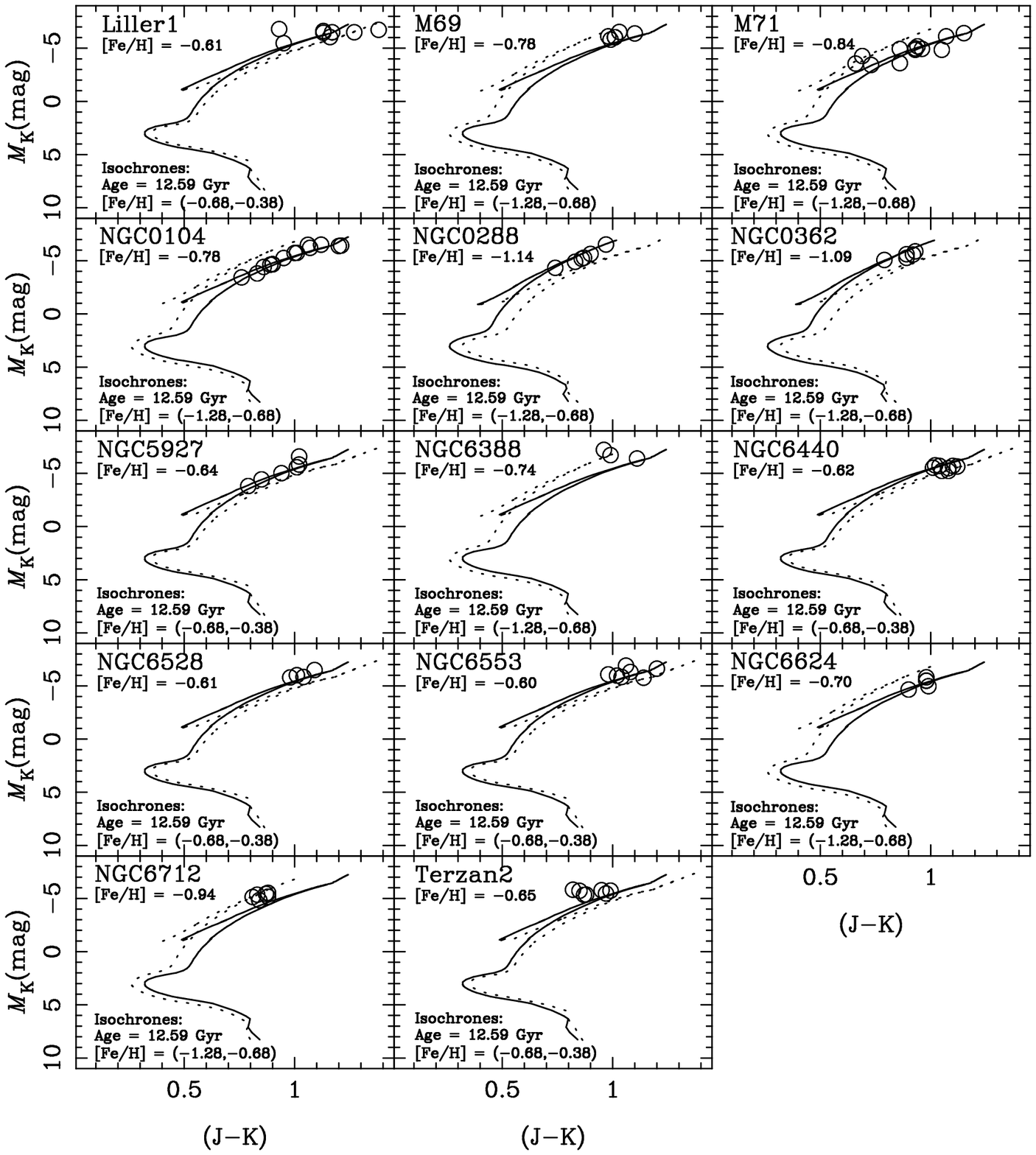}
\caption{Pseudo-HR diagrams for the cluster stars together with
adequate isochrones
\citep{2000A&AS..141..371G,1997A&AS..125..229L,1998A&AS..130...65L} for each
individual cluster. Open circles are used for individual stars in the clusters.
Solid and dashed lines illustrate isochrones of 12.59 Gyr and two metallicity
values enclosing the one of the cluster (as shown in the labels). Adopted
metallicities for the clusters are labeled in the figure. In all cases,
the solid line is employed to indicate the isochrone whose metallicity is
closer to that of the cluster.  Surface gravity for each star was estimated by
comparing to the predicted $M_K$ as explained in \S~\ref{parameter_cluster}.
Final atmospheric parameters for each cluster star, and their
corresponding errors, are given in Table~\ref{table_cluster_stars}.} 
\label{isochrones} 
\end{figure*}

In this section, we present the atmospheric parameters for the stars considered
in the computation of the empirical fitting functions for the D$_{\rm CO}$
index presented in \S~\ref{sec-fitting-functions}.

\begin{figure*}
\centering
\includegraphics[angle=-90,width=0.70\textwidth]{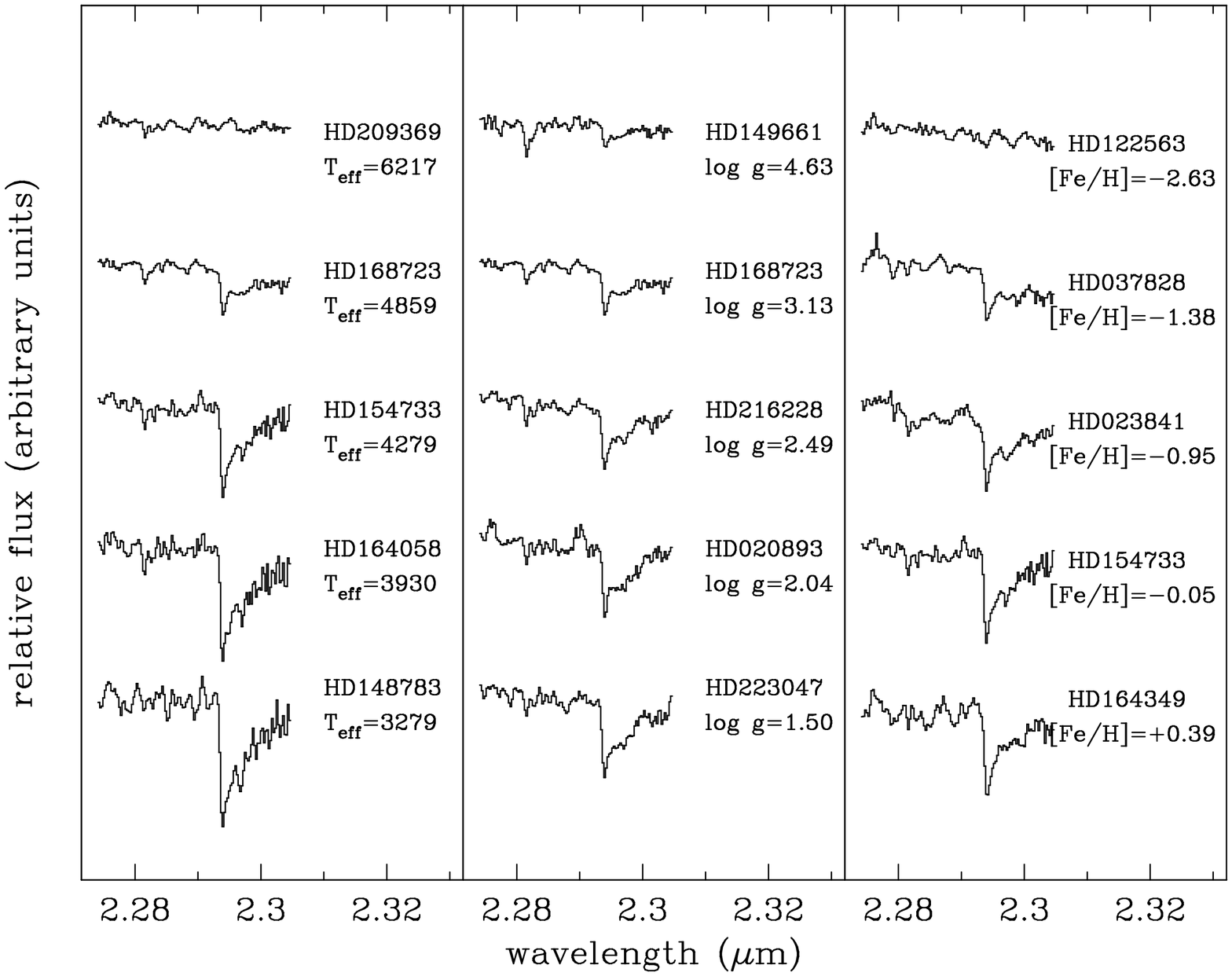}
\caption{Behaviour of the first CO bandhead with T$_{\rm eff}$, $\log g$ and
[Fe/H] (from left to right). In each case, the displayed stars have been
selected to exhibit a range in a given atmospheric parameter (as shown in the
labels) with similar values for the other two parameters.}
\label{behaviour_CO}
\end{figure*}

\subsection{Atmospheric parameters for the observed stellar library sample}
As we mentioned before, the stellar library presented in this work is a
subsample of the MILES library \citep{MILES}. Following the previous work by
\citet{Cen2001b}, a reliable and highly homogeneous data set of atmospheric
stellar parameters for the stars in MILES library was derived by
\citet{MILES_PARAM} as the result of a previous, extensive compilation from the
literature and a carefully calibration and correction by bootstrapping of the
data on to a reference system. In short, the method can be itemized in the
following steps (see \citealt{Cen2001b} and \citealt{MILES_PARAM} for a more
detailed explanation of the working procedure): (i) selection of a
high-quality, standard reference of atmospheric parameters, (ii) bibliographic
compilation of atmospheric parameters of the stars in the library, (iii)
calibration and correction of systematic differences between the different
sources and the standard reference system, and (iv) determination of averaged,
final atmospheric parameters for the library stars from all those references
corrected on to the reference system. Together with the atmospheric parameters,
an estimation of the uncertainties in their determination is given for each
star.  In that way, stellar atmospheric parameters (and their uncertainties)
are perfectly well known for the stars in the new stellar library.

\subsection{Atmospheric parameters for the additional globular cluster stars}
\label{parameter_cluster}

For the subsample of 94 red giant branch (RGB) stars from Galactic globular
clusters, effective temperatures ($T_\mathrm{eff}$) and surface gravities
($\log g$) were derived from $J$ and $K$ photometry, following a technique
similar to that explained in \citet{MILES_PARAM} \citep[see also][]{Cen2001b}.
Absolute, reddening corrected photometry for all clusters was taken from
\citet{Frogel2001} and \citet{2004AJ....127..925S}.

The surface gravity for each cluster star was estimated by matching its
location in a $M_K$--$\left(J-K\right)$ diagram to the appropriate isochrones
from \citet{2000A&AS..141..371G} (see Fig.~\ref{isochrones}), whose colors and
magnitudes were previously transformed to the observational plane using the
empirical colour-temperature relations for giant stars from
\citet{1999A&AS..140..261A} and
\citet{1997A&AS..125..229L,1998A&AS..130...65L}, with the latter being for
giants with $T_\mathrm{eff} \leq 3500$\,K \citep[see more details
in][]{2003MNRAS.340.1317V}. 

If available, the metallicity for each cluster was taken from the work by
\citet{RHS97} (hereafter RHS97), which provides metallicity estimates for
Galactic globular cluster on the \citet{CG97} (hereafter CG97) scale based on
the \ion{Ca}{II} triplet strengths of their RGB stars. This was the case for
\object{NGC0104}, \object{NGC0288}, \object{NGC0362}, \object{NGC5927}, 
\object{NGC6553}, \object{NGC6624}, \object{NGC6712}, and \object{M69}. For
\object{M71}, however, we kept the value in the CG97 scale inferred by
\citet{2002MNRAS.329..863C} according to the CaT indices of their stars. If not
available in RHS97, metallicity values in the \citep[][hereafter ZW84]{ZW84}
scale were transformed into the CG97 scale using Equation~7 in CG97. This was
the case for the rest of our GC sample. For \object{NGC6388}, \object{NGC6440}, 
\object{Liller1}, and \object{Terzan2}, ZW84 metallicity values were taken 
from Table~6 in that paper. For
\object{NGC6528}, the value given in the 2003 revised version of the catalog of
parameters for Milky Way globular clusters \citep[][]{Harris96} was employed.
It is important to note that there is a double reason for using the CG97
metallicity scale in this work. First, since, as compared to the \citet{ZW84}
metallicity scale, the agreement between the locus of the cluster RGB stars and
the corresponding isochrones in the $M_K$--$\left(J-K\right)$ plane is much
better when using CG97 values, particularly in the high metallicity regime.
Also, because the metallicities of the rest of cluster stars in this paper
\citep[as taken from either][]{Cen2001b,MILES_PARAM} are already on the CG97
scale, thus guaranteeing full consistency among the metallicities of the whole
star sample, and minimizing systematics in the computation of the fitting
functions.

Assuming that all the galactic globular cluster under study are similarly old,
and taking into account the above metallicities, for all the stars in each
cluster we used two of the Girardi's isochrones of 12.6\,Gyr with metallicities
enclosing the corresponding value of the cluster (solid and dashed lines in
Fig.~\ref{isochrones}). For each isochrone, a $\log g$ value for each star was
estimated by interpolating in $M_K$. Final $\log g$ values were computed as
weighted means of the single values derived from the two different metallicity
isochrones, with weights accounting for the differences between the adopted
cluster metallicity and the isochrone values.

Effective temperatures for all the cluster giants were computed on the basis of
the $\left(J-K\right)$--$T_\mathrm{eff}$ relations given in
\citet{1999A&AS..140..261A} (for $T_\mathrm{eff} > 3500$ K) and
\citet{1997A&AS..125..229L,1998A&AS..130...65L} (for $T_\mathrm{eff} \leq
3500$\,K). Interestingly, since Alonso's relation for $\left(J-K\right)$ does
not depend on either metallicity and $\log g$, $T_\mathrm{eff}$ can be directly
determined from $\left(J-K\right)$. This indeed helps to minimize the sources
of uncertainties in the final temperatures of most globular cluster giants. As
a matter of fact, only a few temperatures for very cold stars were computed
using Lejeune's relations.

Errors for the derived atmospheric parameters of the cluster stars were
estimated from input uncertainties in [Fe/H] and in the photometric data
employed in this technique. For those [Fe/H] values taken from RHS97 and
\citet{2002MNRAS.329..863C}, the uncertainties given therein were assumed. For
the rest of the clusters, metallicity errors were computed from those given in
ZW84 through standard error propagation in Equation~7 of CG97. Since most stars
in \citet{Frogel2001} and \citet{2004AJ....127..925S} were selected from the
bright ends of the globular cluster luminosity functions, photometric
observational errors in $J$ and $K$ bands turned out to be very small
\citep[$<< 0.06\,$mag; see
e.g.][]{1995AJ....109.1154F,1995AJ....109.1131K,1995AJ....110.2844K} as
compared to typical uncertainties in the assumed distance moduli and reddening
corrections. Errors in $M_K$, and $\left(J-K\right)$ are therefore strongly
dominated by the above effects. For this reason, typical errors of 0.20\,mag
and 0.06\,mag for $M_K$ and $\left(J-K\right)$, respectively, have been
considered for all the cluster stars.

Final values of $T_\mathrm{eff}$, $\log g$, [Fe/H], and their corresponding
errors for the whole sample of cluster stars are presented in
Table~\ref{table_cluster_stars}.


\section{Empirical calibration of the new CO index}
\label{sec-fitting-functions}

In this section we parametrize the behaviour of the new CO index in terms of
the stellar atmospheric parameters (T$_{\rm eff}$, $\log g$ and
[Fe/H]. For that purpose, we
use the D$_{\rm CO}$ index measurements of the stars of the new stellar library
and on the sample of globular cluster stars from \citet{Frogel2001} and
\citet{2004AJ....127..925S} (see \S~\ref{cluster_stars}). This paper shows
a detailed quantitative metallicity dependence of the CO feature at
2.3~$\mu$m.

\begin{figure} \centering
\includegraphics[angle=-90,width=\columnwidth]{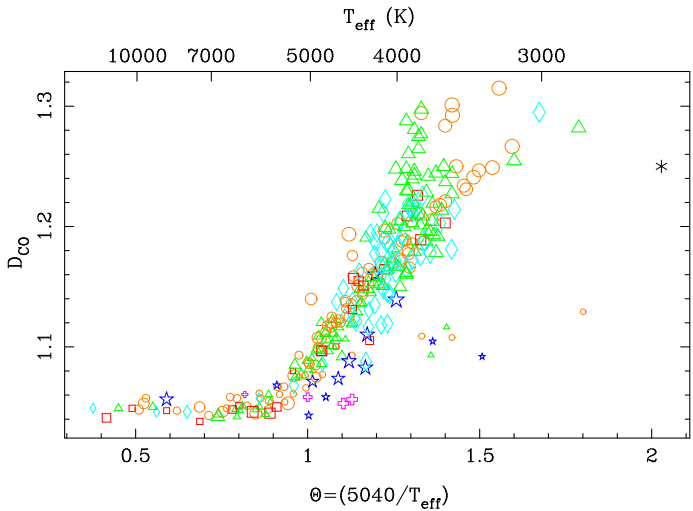}
\caption{D$_{\rm CO}$ index as a function of $\theta\,
\left(=5040/T_\mathrm{eff}\right)$ for the stars of the new library plus the
globular cluster stars from \citet{Frogel2001} and \citet{2004AJ....127..925S}
used in this study. Symbols mean metallicity, as explained in
Fig.~\ref{hr_diagram}. Relative symbol sizes indicate gravity ranges (small
symbols for dwarfs, increasing symbol size for decreasing gravity).}
\label{CO_vs_T} \end{figure}


\subsection{Behaviour of the CO index as a function of stellar atmospheric
parameters}\label{qualitative}

The dependence of the strong CO absorption bands as a function of the effective
temperature and surface gravity is well known from the first studies in the
infrared \citep[KH86,][]
{Joh,1978ApJ...220...75F,LR92,1994ApJ...421..101D,WH97,Ram,For,Lan,Ivanov,Fornax_red}.
The CO absorptions turn deeper as the stars are cooler, while hot stars show no
trace of CO features. On the other hand, as surface gravity decreases, the CO
absorptions become prominent, i.e., supergiant stars present more important CO
absorption than dwarfs. All the previous works in the K band highlighted both
the tight correlation of their defined CO indices with temperature (spectral
type or $J-K$ color in the first papers), and the dependence of the CO
absorption with surface gravity. In addition, a few theoretical studies
\citep[][]{1984PASP...96..882M,1993A&A...280..536O} indicate that these
spectroscopic features should be metallicity dependent.
\citet{1991ApJ...378..742T} showed that Baade's window stars have deep
2.3~$\mu$m bands, what they interpreted as these stars probably having a high
metallicity. Model predictions by \citet{Maraston05} also showed the dependence
of the CO absorptions in the K band with metallicity. Observations of composite
stellar population \citep[e.g.][]{2000A&A...357...61O,Riffel07} give support to
this dependence. Unfortunately, the lack of empirical stellar libraries with an
appropriate coverage in metallicity had prevented, until now, a detailed
quantitative description of this dependence.  Fig.~\ref{behaviour_CO} shows,
graphically, how the first CO bandhead at 2.3~$\mu$m varies with the stellar
atmospheric parameters ($T_\mathrm{eff}$, $\log g$ and [Fe/H]).

Several authors have partially parametrized the described behaviour of the
first CO bandhead. \citet{LR92} computed a relation of their CO index with the
color temperature of giant stars. \citet{1994ApJ...421..101D} parametrized the
behaviour of their CO index with the effective temperature for supergiant,
giant and dwarf stars, separately. More recently, \citet{Ram} used their CO
index to obtain effective temperatures for giants. They applied the same method
to dwarf stars from \citet{Ali}. However, there is no systematic study of the
dependence of the CO absorption as a function of the atmospheric stellar
parameters due to the poor stellar parameter coverage of previous libraries,
especially in metallicity. In the next subsection we compute an empirical
calibration for the D$_{\rm CO}$ index measured for the stars of our stellar
library which accounts for the previously described qualitative behaviour of
the CO absorption.

\subsection{Fitting functions}
To reproduce the behaviour exhibited by a given feature, empirical calibrations
of the corresponding line-strength index as a function of the stellar
atmospheric parameters were derived. These calibrations, called fitting
functions, are just a mathematical description of the observed behaviour and
there is not physical justification for the explanation of each of the
significant terms in such fitting functions. We use
$\theta=5040/T_\mathrm{eff}$ as temperature indicator, together with $\log g$
and [Fe/H] as parameters for gravity and metallicity. Following previous works
\citep{1993ApJS...86..153G, 1994ApJS...94..687W, 1999A&AS..139...29G,
2002MNRAS.329..863C}, two possible functional forms for the computation of the
fitting functions are typically considered, in particular 

\begin{equation}
I\left(\theta,\log g, \left[{\rm Fe/H}\right]\right)=
p\left(\theta,\log g, \left[{\rm Fe/H}\right]\right) 
\end{equation}
and 
\begin{equation} 
I\left(\theta,\log g, \left[{\rm Fe/H}\right]\right)={\rm constant}+
\exp[p\left(\theta,\log g, \left[{\rm Fe/H}\right]\right), 
\end{equation} 
where $I$ refers to a given index, and $p\left(\theta,\log g, \left[{\rm
Fe/H}\right]\right)$ is a polynomial with terms up to the third order,
including all possible cross-terms among the three parameters, i.e. 

\begin{equation}\label{polinomio}
p\left(\theta,\log g, \left[{\rm Fe/H}\right]\right)=
\sum_{0\leq i+j+k\leq 3} c_{i,j,k} \;\theta^i\;
\left(\log g\right)^j\;\left[{\rm Fe/H}\right]^k,
\end{equation}
with $0\leq i+j+k\leq 3$ and $0\leq i,j,k$.

The optimum fitting function is the one which minimizes the residuals of the
fits, i.e., when the differences between the measured index ($I_{\rm obs}$) and
the index predicted by the fitting function ($I_{\rm pred}$) are the lowest.

In general, not all the terms in Eq.~\ref{polinomio} are necessary for
reproducing the dependences of an index on the stellar parameters. To obtain
the significant terms for the best fitting functions, we follow two different
approaches \citep{2002MNRAS.329..863C}. Both of them consist of an iterative
and systematic procedure based on the computation of a general fit together
with the analysis of the residual variance of the fit and estimated errors of
each fitted coefficient, for a given set of stars with well known atmospheric
parameters, index measurement and error. The significance of each term
considered in the fit is calculated by means of a {\it t}-test (that is, using
the error in that coefficient to check whether it is significantly different
from zero). Typically, we consider a term as significant for a threshold value
of the significance level parameter $\alpha=0.10$. The first method consists of
computing the fit with all the terms in Eq.~\ref{polinomio} and eliminating,
during each iteration, the least significant term.  A parallel method consists
of computing an initial one-parameter fit and incorporating new terms only when
they are significant.  Together with this criterion, we study the residuals
derived for each set of terms. The final fit will be the one which minimizes
the sum of the residuals while having all the employed $c_{i,j,k}$ coefficients
statistically significant, that does not produce systematic deviations in the
residuals for stars of different types, e.g., stars of a given metallicity
range, and that reproduces the observed behaviour of the index.

\subsubsection{The general fitting procedure}
As a result of the large stellar parameter coverage of the library, there is
not a single fitting function able to reproduce the whole behaviour of the
D$_{\rm CO}$ index. For that reason, we have divided the stellar parameter
space into several ranges (boxes) where local fitting functions have been
computed following the methods explained in the previous section. The final
fitting function for the whole parameter space is then constructed by
considering the derived local fitting functions. In some ranges, an overlapping
region in a generic parameter $z$ has been allowed between two different boxes.
If $I_1\left(x,y,z\right)$ and $I_2\left(x,y,z\right)$ are two local fitting
functions defined respectively in the intervals ($z_{1,1},z_{1,2}$) and
($z_{2,1},z_{2,2}$) with $z_{2,1}<z_{1,2}$, the final predicted index
$I\left(x,y,z\right)$ in the overlapping region is obtained by interpolating
the index value in both boxes, i.e., 

\begin{equation}
I\left(x,y,z\right) = w I_1\left(x,y,z\right) + \left(1-w\right)
I_2\left(x,y,z\right),
\end{equation}
where the weight $w$ is modulated by the distance to the overlapping limits as  
\begin{equation}
w = \frac{z-z_{2,1}}{z_{1,2}-z_{2,1}},
\end{equation}
or
\begin{equation}
w = \cos \left[ \frac{\pi}{2} 
    \left( \frac{z-z_{2,1}}{z_{1,2}-z_{2,1}} \right) \right], 
\end{equation}
for a cosine-weighted mean, where $z_{2,1} \le z \le z_{1,2}$.

\subsubsection{Fitting functions for the D$_{\rm CO}$ index}\label{CO_fitting}

\begin{figure*}
\centering
\includegraphics[angle=-90,width=0.80\textwidth]{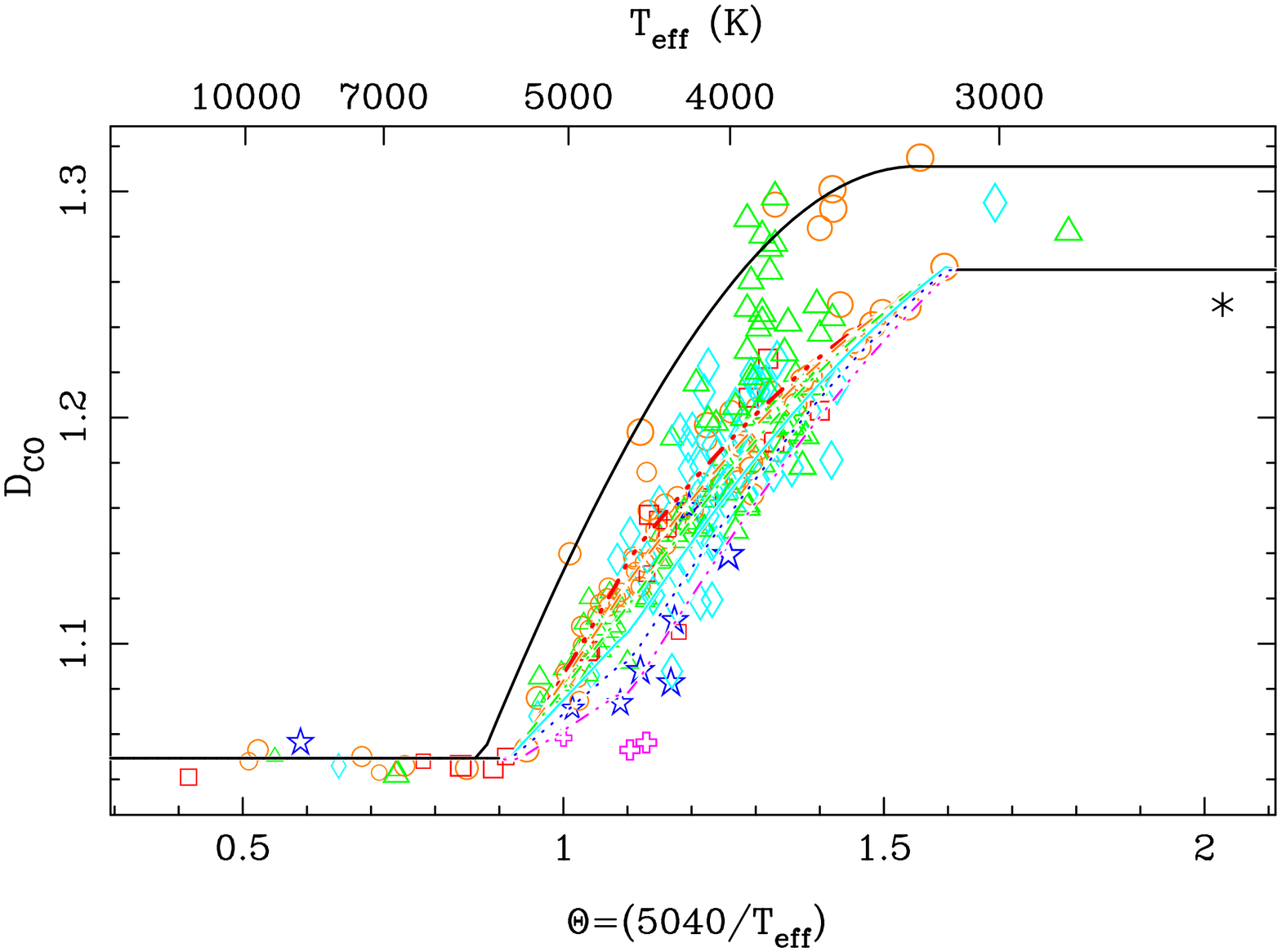}

\includegraphics[angle=-90,width=0.80\textwidth,clip=true]{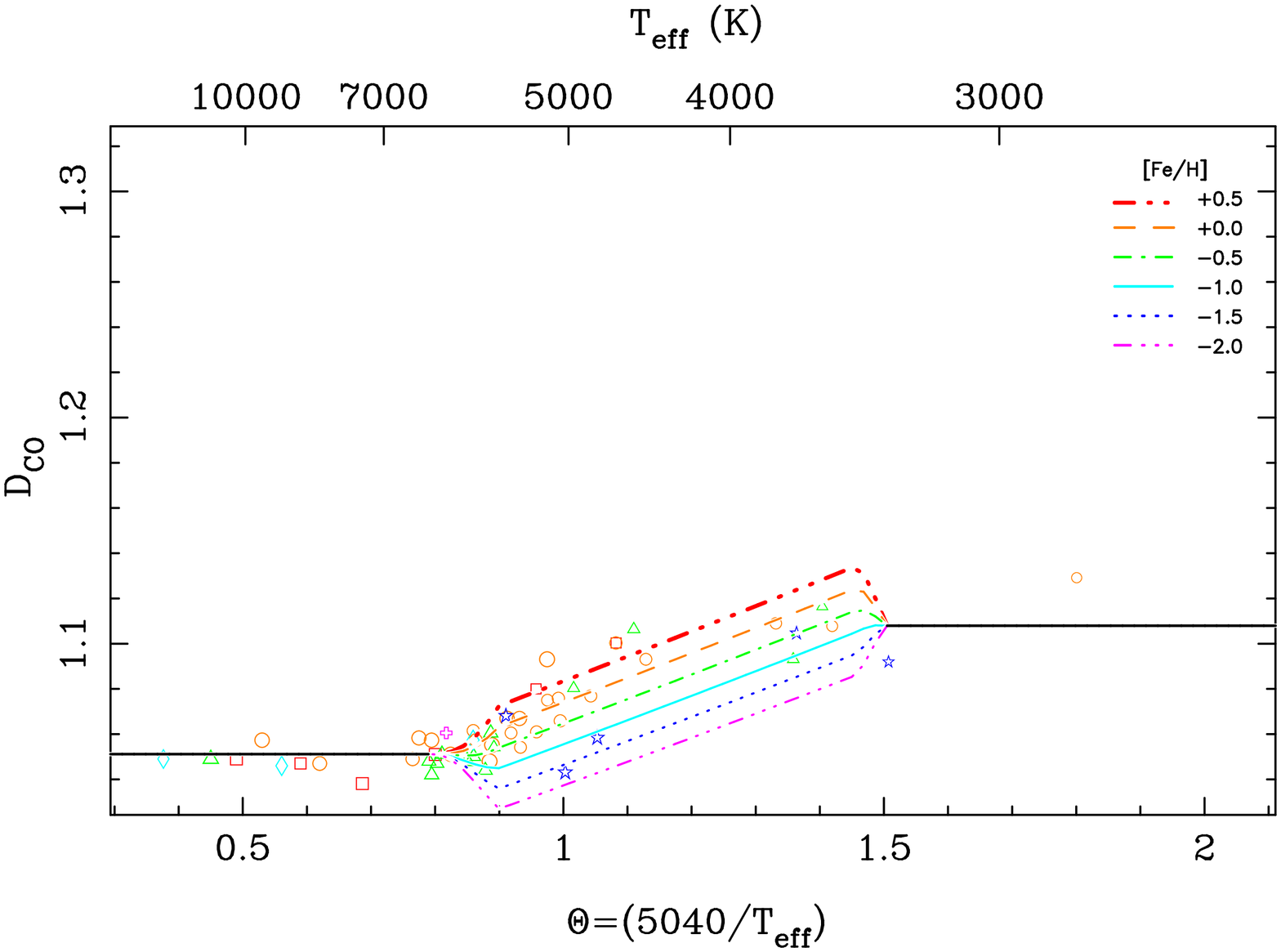}
\caption{Empirical fitting functions for the D$_{\rm CO}$ index for giants,
including the fit for AGB stars (top panel) and dwarfs (bottom panel), computed
as explained in the text. Symbol types and size mean metallicity and gravity,
the same than in previous figures. The different lines represent the
metallicities [Fe/H]$= +0.5,+0.0,-0.5,-1.0,-1.5, {\rm and} -2.0$, from top to
bottom. The reduced scatter for $\theta \le 0.9$ is due to the absence of CO
absorptions for hot stars, as explained in the text.}
\label{fitting_CO}
\end{figure*}

Fig.~\ref{CO_vs_T} shows the behaviour of the D$_{\rm CO}$ index versus
$\theta$ for the stars in the new library. There is a clear dichotomy in the
behaviour of stars depending on their gravity. For that reason, we have divided
the stellar atmospheric parameter space in two main groups: dwarf stars ($\log
g>3.5$ dex) and giant and supergiant stars ($\log g<3.5$ dex). As we explained
in \S~\ref{qualitative}, there is also a strong dependence of the CO absorption
with the effective temperature. That is why we have subdivided each gravity
group into different temperature ranges. Independently of their gravity, stars
with high effective temperature exhibit no traces of CO absorptions and their
index value tends to a constant (D$_{\rm CO} \simeq 1.05$). On the other hand,
due to the lack of very cold stars in both gravity regimes, we have computed a
constant value of the index for cold dwarf and giant stars. In short, we have
considered three temperature ranges for dwarf stars, while we have considered
four different ranges for the giants (see Table~\ref{coefficients}). The
boundaries of these ranges are shown in Fig.~\ref{hr_diagram}. In some cases,
an overlapping region has been considered.  After some experimentation in order
to obtain the smoothest fit, we have computed the final index in the
intersection region considering a different weight $w$ depending on the case
(see Table~\ref{coefficients}).

Besides the global behaviour described for giant stars, two different
trends are found for this type of stars around $\theta = 1.3-1.4$ in
Fig.~\ref{CO_vs_T}. After a carefully study of these stars with a higher CO
index, we found that they are stars in the asymptotic giant branch (AGB).
For that reason, we decided to compute an independent fit for these stars in
the range $\theta = 1.01-1.56$, shown Fig.~\ref{fitting_CO} (coefficient in
Table~\ref{coefficients}). Since there are no AGB stars for $\theta > 1.56$, 
we simply extrapolate the constant value of the CO index at $\theta = 1.56$.

Up to now, we have described the general method to compute fitting functions.
However, we have used a slightly different method for cold and warm giant
stars. First of all, we have obtained constant fits for hot ($\theta < 0.90$)
and cold stars ($\theta > 1.55$) as explained before. After this, we have
generated a set of fake stars with random gravity and metallicity, and
effective temperature of $\theta=5040/T_\mathrm{eff}=1.60$. Their index values
were evaluated from the fitting function computed for cold giant stars. For the
associated error of the indices of these fake stars we assumed the mean error
of the cold stars. We calculated the local fitting function of cool giants,
including these new stars. In this way, we were forcing the fitting function to
pass through these stars, i.e., approaching the constant value for cold giants.
In a similar way, for the computation of the fitting functions of warm giants,
we also created fake stars at $\theta=1.10$ and $\theta=1.13$, and we assigned
their index value according to the fitting function for cool stars. A new set
of random stars at $\theta=0.92$ was introduced with the constant index value
computed from hot stars. Finally, we used these fake stars together with the
real stars to obtain the final fit for the warm giants.

\begin{figure*}
\centering
\includegraphics[angle=-90,width=0.80\textwidth]{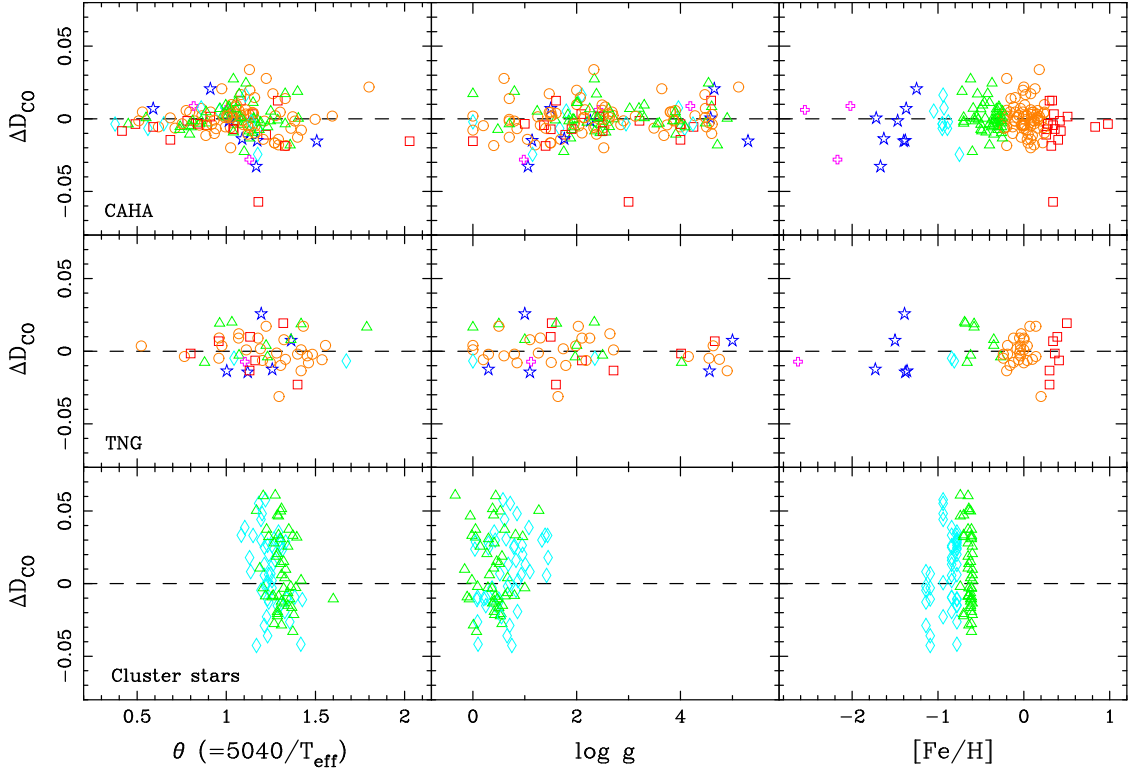}
\caption{Residuals of the derived fitting functions (observed minus derived)
against the three input stellar atmospheric parameters for the stars observed
at CAHA (upper panels) and TNG (middle panels). Cluster data from
\citet{Frogel2001} and \citet{2004AJ....127..925S} are presented in lower
panels. Codes and relative sizes of the star symbols are explained in
Fig.~\ref{hr_diagram}~and~\ref{CO_vs_T}.}
\label{residuals}
\end{figure*}

We derived the local fitting functions computing a weighted least squares fit
to the stars within each parameter range, with weights according to the inverse
of the squared uncertainties of the indices for each individual star. Here we
considered the nominal errors in the index measurements, using the
uncertainties derived in \S~\ref{sub-sec-lib-errors} for the library stars, and
the individual errors for each cluster quoted by \citet{Frogel2001} and
\citet{2004AJ....127..925S}. After the analysis of the residuals, it was
necessary to repeat the fitting procedure with the inclusion of additional
random uncertainties in some of the stars (see next subsection). The final
fitting functions for each stellar parameter range are presented in
Table~\ref{coefficients}. This table includes the functional forms of the fits,
the significant coefficients and their corresponding formal errors, the typical
index error for the $N$ stars employed in each interval
($\sigma^2_{typ}=N/\sum^N_{i=1}\sigma^{-1}_i$), the unbiased residual variance
of the fit ($\sigma^2_{\rm std}$) and the determination coefficient ($r^2$).
Also the intersection regions are indicated and the procedure used for
computing the index. We plot in Fig.~\ref{fitting_CO} the fitting
functions for giant and dwarf stars as a function of $\theta$ for several
metallicities, as well as the simple fit for AGB stars.  Fig.~\ref{residuals}
shows the final residuals ($\Delta{\rm D_{CO}} = {\rm D_{CO\, obs} - D_{CO\,
pred}}$) of the new CO index for the different groups of stars used for the
computation of the fitting functions (stars observed at CAHA, stars observed
at the TNG and cluster stars from previous works) as a function of the
atmospheric parameters. As we expected, the residuals for each group of stars
do not exhibit any systematic trend. 

Users interested in implementing these fitting functions into their population
synthesis codes can make use of the {\sc FORTRAN} subroutine available at the
URL address {\tt http://www.ucm.es/info/Astrof/ellipt/CO.html}.

\begin{table}
\caption{Coefficients and statistical data for the local empirical fitting
functions for the D$_{\rm CO}$ index in each range of stellar parameters.}
\label{coefficients}
\centering
\begin{tabular}{lrc}
\hline\hline
{\bf Hot dwarfs}      & $0.38 < \theta < 0.90$ & $3.50 < \log g < 5.50$ \\
exponential fit       & N = 28 stars           & $\sigma_{\rm typ} = 0.00913$ \\
$c_0$                 & $ 0.0499 \pm 0.0011 $  & $\sigma_{\rm std} = 0.00557$ \\
\hline
{\it Intersection}    & Cosine-weighted mean   & $0.80 < \theta < 0.90$ \\
\hline
{\bf Cool dwarfs}     & $0.80 < \theta < 1.50$ & $3.50 < \log g < 5.50$ \\
exponential fit       & N = 39 stars           & $\sigma_{\rm typ} = 0.00748$\\
$c_0$                 & $-0.0292 \pm 0.0330$   & $\sigma_{\rm std} = 0.01254$\\
$\theta$              & $ 0.1006 \pm 0.0329$   & $ r^2 = 0.765 $             \\
$[{\rm Fe/H}]$        & $ 0.0174 \pm 0.0067$   &                             \\
\hline
{\it Intersection}    & Cosine-weighted mean   & $1.45 < \theta < 1.50$ \\
\hline
{\bf Cold dwarfs}     & $1.33 < \theta < 1.80$ & $3.50 < \log g < 5.50$      \\
exponential fit       & N = 7 stars            & $\sigma_{\rm typ} = 0.00960$\\
$c_0$                 & $ 0.1025 \pm 0.0046 $  & $\sigma_{\rm std} = 0.01195$\smallskip\\
\hline\hline
{\bf Hot giants}      & $0.42 < \theta < 0.90$ & $-0.40< \log g < 3.50$ \\
exponential fit       & N = 15 stars           & $\sigma_{\rm typ} = 0.01954$\\
$c_0$                 & $ 0.0459 \pm 0.0010 $  & $\sigma_{\rm std} = 0.00398$\\
\hline
{\it Intersection}    & Cosine-weighted mean   & $0.90 < \theta < 0.93$ \\
\hline
{\bf Warm giants}     & $0.90 < \theta < 1.131$ & $ 0.13 \log g < 3.5$ \\
exponential fit       & N = 63 stars           & $\sigma_{\rm typ} = 0.00764$\\
$c_0$                 & $-0.3073 \pm 0.0046 $  & $\sigma_{\rm std} = 0.00428$\\
$\theta$              & $ 0.3876 \pm 0.0043 $  & $ r^2 = 0.982 $             \\
$[{\rm Fe/H}]$        & $-0.1016 \pm 0.0042 $  &                             \\
$\theta [{\rm Fe/H}]$ & $ 0.1072 \pm 0.0039 $  &                             \\
$[{\rm Fe/H}]^2$      & $-0.0023 \pm 0.0005 $  &                             \\
\hline
{\it Intersection}    & Distance-weighed mean  & $1.09 < \theta < 1.10$ \\
\hline
{\bf Cool giants}     & $1.10 < \theta < 1.60$ & $-0.34< \log g < 3.41$ \\
exponential fit       & N = 167 stars          & $\sigma_{\rm typ} = 0.01062$\\
$c_0$                 & $-0.5224 \pm 0.0970 $  & $\sigma_{\rm std} = 0.00890$\\
$\theta$              & $ 0.8257 \pm 0.1417 $  & $ r^2 = 0.958 $             \\
$[{\rm Fe/H}]$        & $ 0.0674 \pm 0.0101 $  &                             \\
$\theta [{\rm Fe/H}]$ & $-0.0444 \pm 0.0065 $  &                             \\
$\theta^2$            & $-0.2200 \pm 0.0509 $  &                             \\
$[{\rm Fe/H}]^2$      & $-0.0023 \pm 0.0014 $  &                             \\
\hline
{\it No intersection} & & \\
\hline
{\bf Cold giants}     & $1.55 < \theta < 2.03$ & $-0.07< \log g < 3.50$ \\
exponential fit       & N = 7 stars            & $\sigma_{\rm typ} = 0.02156$\\
$c_0$                 & $ 0.2397 \pm 0.0107 $  & $\sigma_{\rm std} = 0.02698$\\
\hline
\hline
{\bf AGB stars}       & $1.01 < \theta < 1.56$ & $-0.11< \log g < 1.56$ \\
exponential fit       & N = 18 stars           & $\sigma_{\rm typ} = 0.00612$\\
$c_0$                 & $-0.8893 \pm 0.2198 $  & $\sigma_{\rm std} = 0.00892$\\
$\theta$              & $ 1.4950 \pm 0.3610 $  & $ r^2 = 0.985 $             \\
$\theta^2$            & $-0.4816 \pm 0.1461 $  &                             \\
\hline
\end{tabular}
\end{table}

\subsubsection{Residuals and error analysis}

To explore in more detail the reliability of the fitting functions, we computed
the unbiased residual standard deviation from the fits, $\sigma_{\rm std} =
0.0093$, and the typical error in the measured index, $\sigma_{\rm typ} =
0.0025$, for the stars employed in the computation of the global fitting
functions derived above. 

After the initial fit, we discovered that $\sigma_{\rm std}$ was larger than
what should be expected uniquely from index uncertainties (see also the values
of $\sigma_{\rm std}$ and $\sigma_{\rm typ}$ for different subgroups of stars
in Table~\ref{errors_param_ini}). Furthermore, a $F$~test of comparison of
variances showed that $\sigma_{\rm std}$ was statistically larger than
$\sigma_{\rm typ}$, leading to the introduction on an additional residual error
($\sigma_{\rm res}^2 = \sigma_{\rm std}^2 - \sigma_{\rm typ}^2$).  Such
residual error could be due to uncertainties in the input stellar atmospheric
parameters, not included in the error budget so far. In order to check this, we
have computed how errors in the input stellar parameters translate into
uncertainties in the predicted D$_{\rm CO}$. This depends both on the local
functional form of the fitting function and on the atmospheric parameters range
(e.g. hot stars have $T_\mathrm{eff}$ uncertainties larger than cooler stars).
For each star of the sample, we have derived the errors corresponding to the
uncertainties in effective temperature ($\sigma_{T_\mathrm{eff}}$), gravity
($\sigma_{\log g}$) and metallicity ($\sigma_{\rm [Fe/H]}$). The effect of the
uncertainties in the three stellar parameters were finally computed as
$\sigma_{\rm par}^2=\sigma_{T_\mathrm{eff}}^2+\sigma_{\log g}^2+\sigma_{\rm
[Fe/H]}^2$. Table~\ref{errors_param_ini} presents, summarized for the different
stellar groups, all the error estimations already discussed. Finally, in the
cases where the residual errors ($\sigma_{\rm res}$) were not explained by the
uncertainties in the stellar parameters ($\sigma_{\rm par}$), an extra residual
error was added to the original index error. At the end, no iterations were
needed for the globular cluster stars and AGB stars, one iteration was
required for the giants and two iterations for dwarfs (in any case, the
additional error for dwarf stars is lower than the necessary for giants). The
uncertainties of the final D$_{\rm CO}$ fitting functions are listed in
Table~\ref{errors_param}. Note that although in this table $\sigma_{\rm std}$
is still larger than $\sigma_{\rm typ}$ for the four initial subgroups of
stars analyzed in Table~\ref{errors_param_ini}, both standard deviations are
statistically comparable using the $F$~test of variances previously mentioned.

Finally, as the purpose of this paper is to predict reliable index values for
any combination of input atmospheric parameters, we have also computed the
random errors in such predictions. These uncertainties are given in
Table~\ref{table-absolute-errors} for some representative values of input
parameters. 

\begin{table*}
\caption{Uncertainties of the initial D$_{\rm CO}$ fitting functions for
different groups of stars and mean D$_{\rm CO}$ errors due to uncertainties in
the input atmospheric parameters. $N$: number of stars. $\sigma_{\rm std}$:
unbiased residual standard deviation for the fit. $\sigma_{\rm typ}$: typical
observational D$_{\rm CO}$ error for the subsample of stars.  $\sigma_{\rm
res}$: residual error. $\sigma_{T_\mathrm{eff}}$ and
$\sigma_{\rm [Fe/H]}$: mean D$_{\rm CO}$ errors due to uncertainties in the
input $T_\mathrm{eff}$ and [Fe/H] (no error due to uncertainties in the 
input $\log g$ is computed since the computed fitting functions do not require
gravity terms). $\sigma_{\rm par}$: total error
due to atmospheric parameters (quadratic sum of the previous errors).
$\sigma_{\rm std}$ is not explained by the $\sigma_{\rm typ}$ and
$\sigma_{\rm par}$ values for giants and dwarfs, and a residual error has been
considered for these group of stars}.
\label{errors_param_ini} 
\centering
\begin{tabular}{lrcccccc}
\hline\hline
                  & $N$ & $\sigma_{\rm std}$ & $\sigma_{\rm typ}$ &                   $\sigma_{\rm res}$ & $\sigma_{T_\mathrm{eff}}$ & $\sigma_{\rm [Fe/H]}$ & $\sigma_{\rm par}$ \\
\hline
Field dwarfs                           &  54 & 0.0059 & 0.0017 & 0.0056 & 0.0010 &  0.0009 & 0.0014 \\
Field giants                           & 147 & 0.0108 & 0.0023 & 0.0105 & 0.0062 &  0.0019 & 0.0065 \\
Globular cluster stars                 &  85 & 0.0276 & 0.0236 &        & 0.0139 &  0.0010 & 0.0140 \\
AGB stars                              &  19 & 0.0115 & 0.0061 &        & 0.0097 &  0.0000 & 0.0097 \\
                                       &     &        &        &        &        &         &        \\
All                                    & 305 & 0.0093 & 0.0025 &        & 0.0075 &  0.0014 & 0.0078 \\
\hline
\end{tabular}
\end{table*}

\begin{table*}
\caption{Uncertainties of the final D$_{\rm CO}$ fitting functions for
different groups of stars and same computed errors than in
Table~\ref{errors_param_ini}.}
\label{errors_param} 
\centering
\begin{tabular}{lrccccc}
\hline\hline
                  & $N$ & $\sigma_{\rm std}$ & $\sigma_{\rm typ}$ &                   $\sigma_{T_\mathrm{eff}}$ & $\sigma_{\rm [Fe/H]}$ & $\sigma_{\rm par}$ \\
\hline
Field dwarfs                           &  54 & 0.0086 & 0.0078 & 0.0010 &  0.0013 & 0.0017 \\
Field giants                           & 147 & 0.0123 & 0.0113 & 0.0062 &  0.0014 & 0.0064 \\
Globular cluster stars                 &  85 & 0.0252 & 0.0236 & 0.0135 &  0.0009 & 0.0135 \\
AGB stars                              &  19 & 0.0115 & 0.0061 & 0.0097 &  0.0000 & 0.0097 \\
                                       &     &        &        &        &         &        \\
All                                    & 305 & 0.0130 & 0.0114 & 0.0074 &  0.0012 & 0.0076 \\
\hline
\end{tabular}
\end{table*}

\begin{table}
\caption{Absolute errors in the fitting functions predictions for different
values of the atmospheric parameters. Input $\log g$ values varying with
effective temperature for dwarfs (V), giants (III) and supergiants (I) have 
been taken from \citet{Lang}.}
\label{table-absolute-errors}
\centering
\begin{tabular}{ccccc}
\hline\hline
                 &          &   & $\sigma[{\rm D_{CO}}]$ & \\
$T_\mathrm{eff}$ &  [Fe/H]  &   V    &  III   &   I         \smallskip\\
\hline
6000          &          & 0.002  & 0.001  & 0.001 \\
 & & & & \\                                          
5500          &  $+0.5$  & 0.010  & 0.002  & 0.002 \\
5500          &  $+0.0$  & 0.007  & 0.001  & 0.001 \\
5500          &  $-1.0$  & 0.006  & 0.001  & 0.001 \\
5500          &  $-2.0$  & 0.011  & 0.002  & 0.002 \\
 & & & & \\                                          
5000          &  $+0.5$  & 0.010  & 0.001  & 0.001 \\
5000          &  $+0.0$  & 0.007  & 0.001  & 0.001 \\
5000          &  $-1.0$  & 0.005  & 0.001  & 0.001 \\
5000          &  $-2.0$  & 0.011  & 0.001  & 0.001 \\
 & & & & \\                                          
4500          &  $+0.5$  & 0.011  & 0.004  & 0.004 \\
4500          &  $+0.0$  & 0.008  & 0.003  & 0.003 \\
4500          &  $-1.0$  & 0.007  & 0.003  & 0.003 \\
4500          &  $-2.0$  & 0.011  & 0.006  & 0.006 \\
 & & & & \\                                          
4000          &  $+0.5$  & 0.014  & 0.003  & 0.003 \\
4000          &  $+0.0$  & 0.012  & 0.002  & 0.002 \\
4000          &  $-1.0$  & 0.011  & 0.003  & 0.003 \\
4000          &  $-2.0$  & 0.014  & 0.005  & 0.005 \\
 & & & & \\                                          
3500          &  $+0.5$  & 0.018  & 0.003  & 0.007 \\
3500          &  $+0.0$  & 0.018  & 0.003  & 0.007 \\
3500          &  $-1.0$  & 0.018  & 0.003  & 0.007 \\
3500          &  $-2.0$  & 0.018  & 0.004  & 0.007 \\
 & & & & \\                                          
3000          &        & 0.005  & 0.014  & 0.014 \smallskip\\
\hline
\end{tabular}
\end{table}

\section{Summary}
\label{sec-summary}

The aim of this work was to obtain an accurate empirical calibration of the
behaviour of the CO feature at 2.3~$\mu$m for individual stars, with the
purpose of making it possible to obtain reliable predictions for the CO
strength for stellar populations in unresolved systems with a wide range of
ages and metallicities. The main results of this work can be summarized as
follows:

\begin{enumerate}

\item We present a new stellar library in the spectral region around the first
CO bandhead at 2.3~$\mu$m. It consists of 220 stars with stellar atmospheric
parameters in the range \mbox{$2485\leq T_\mathrm{eff} \leq 13404$~K},
\mbox{$-0.34\leq \log g \leq 5.30$~dex}, \mbox{$-2.63 \leq {\rm [Fe/H]} \leq
0.98$~dex}. 

\item We define a new line-strength index for the first CO bandhead at
2.3~$\mu$m, D$_{\rm CO}$, less sensitive to spectral resolution, wavelength
calibration, signal-to-noise ratio and flux calibration than previous
definitions.  

\item We compute empirical fitting functions for the D$_{\rm CO}$ to
parametrize the behaviour of the CO feature as a function of the stellar
atmospheric parameters. In this work we quantify, for the first time, the
metallicity dependence.

\end{enumerate}

We expect that the work presented in this paper will help researchers to start
exploiting in depth the so far poorly-explored and poorly-understood near-IR
spectral region centered at 2.3~$\mu$m, since, as we have shown, the strong CO
bandhead can be employed to extract useful physical information of composite
stellar populations.

\begin{acknowledgements} 
This work was supported by the Spanish research projects AYA2006--14318,
AYA2006--15698-C02-02, AYA2007--67752C03-03. EMQ acknowledges the Spanish
Ministry of Education and Science and the European Social Found for a
Formaci{\'o}n de Personal Investigador fellowship under the project
AYA2003--01840. AJC is a Juan de la Cierva Fellow of the Spanish Ministry of
Education and Science. This research was supported by a Marie Curie
Intra-European Fellowship within the 6th European Community Framework
Programme. Based on observations collected at the Centro Astron\'{o}mico
Hispano Alem\'{a}n (CAHA) at Calar Alto, operated jointly by the Max-Planck
Institut f\"{u}r Astronomie and the Instituto de Astrof\'{\i}sica de
Andaluc\'{\i}a (CSIC). Based on observations made with the Italian Telescopio
Nazionale Galileo (TNG) operated on the island of La Palma by the Fundaci\'{o}n
Galileo Galilei of the INAF (Istituto Nazionale di Astrofisica) at the Spanish
Observatorio del Roque de los Muchachos of the Instituto de Astrof\'{\i}sica de
Canarias. This research has made use of the SIMBAD database (operated at CDS,
Strasbourg, France), the NASA's Astrophysiscs Data System Bibliographic
Services, and the {\it Hipparcos} Input Catalogue.  The authors are grateful to
the allocation time committees for the generous concession of observing time.
We finally thank the anonymous referee for very useful comments and
suggestions.
\end{acknowledgements}


\bibliography{biblio,definiciones}

\begin{thebibliography}{82}
\expandafter\ifx\csname natexlab\endcsname\relax\def\natexlab#1{#1}\fi

\bibitem[{{Ali} {et~al.}(1995){Ali}, {Carr}, {Depoy}, {Frogel}, \&
  {Sellgren}}]{Ali}
{Ali}, B., {Carr}, J.~S., {Depoy}, D.~L., {Frogel}, J.~A., \& {Sellgren}, K.
  1995, \aj, 110, 2415

\bibitem[{{Alonso} {et~al.}(1999){Alonso}, {Arribas}, \&
  {Mart{\'{\i}}nez-Roger}}]{1999A&AS..140..261A}
{Alonso}, A., {Arribas}, S., \& {Mart{\'{\i}}nez-Roger}, C. 1999, \aaps, 140,
  261

\bibitem[{{Baldwin} {et~al.}(1973){Baldwin}, {Frogel}, \&
  {Persson}}]{1973ApJ...184..427B}
{Baldwin}, J.~R., {Frogel}, J.~A., \& {Persson}, S.~E. 1973, \apj, 184, 427

\bibitem[{{Bendo} \& {Joseph}(2004)}]{Bendo04}
{Bendo}, G.~J. \& {Joseph}, R.~D. 2004, \aj, 127, 3338

\bibitem[{{Bruzual} \& {Charlot}(2003)}]{2003MNRAS.344.1000B}
{Bruzual}, G. \& {Charlot}, S. 2003, \mnras, 344, 1000

\bibitem[{{Cardiel}(1999)}]{1999PhDT........12C}
{Cardiel}, N. 1999, Ph.D.~Thesis

\bibitem[{{Cardiel}(2007)}]{indexf}
{Cardiel}, N. 2007, in Highlights of Spanish astrophysics IV, ed.
  F.~{Figueras}, J.~{Girart}, M.~{Hernanz}, \& C.~{Jordi}, CD--ROM

\bibitem[{{Cardiel} {et~al.}(1998){Cardiel}, {Gorgas}, {Cenarro}, \&
  {Gonzalez}}]{1998A&AS..127..597C}
{Cardiel}, N., {Gorgas}, J., {Cenarro}, J., \& {Gonzalez}, J.~J. 1998, \aaps,
  127, 597

\bibitem[{{Cardiel} {et~al.}(2003){Cardiel}, {Gorgas}, {S{\' a}nchez-Bl{\'
  a}zquez}, {Cenarro}, {Pedraz}, {Bruzual}, \& {Klement}}]{Cardiel03}
{Cardiel}, N., {Gorgas}, J., {S{\' a}nchez-Bl{\' a}zquez}, P., {et~al.} 2003,
  \aap, 409, 511

\bibitem[{{Carretta} \& {Gratton}(1997)}]{CG97}
{Carretta}, E. \& {Gratton}, R.~G. 1997, \aaps, 121, 95

\bibitem[{{Cayrel de Strobel} {et~al.}(2001){Cayrel de Strobel}, {Soubiran}, \&
  {Ralite}}]{Cayrel01}
{Cayrel de Strobel}, G., {Soubiran}, C., \& {Ralite}, N. 2001, \aap, 373, 159

\bibitem[{{Cenarro} {et~al.}(2001{\natexlab{a}}){Cenarro}, {Cardiel}, {Gorgas},
  {Peletier}, {Vazdekis}, \& {Prada}}]{Cen2001a}
{Cenarro}, A.~J., {Cardiel}, N., {Gorgas}, J., {et~al.} 2001{\natexlab{a}},
  \mnras, 326, 959

\bibitem[{{Cenarro} {et~al.}(2001{\natexlab{b}}){Cenarro}, {Gorgas}, {Cardiel},
  {Pedraz}, {Peletier}, \& {Vazdekis}}]{Cen2001b}
{Cenarro}, A.~J., {Gorgas}, J., {Cardiel}, N., {et~al.} 2001{\natexlab{b}},
  \mnras, 326, 981

\bibitem[{{Cenarro} {et~al.}(2002){Cenarro}, {Gorgas}, {Cardiel}, {Vazdekis},
  \& {Peletier}}]{2002MNRAS.329..863C}
{Cenarro}, A.~J., {Gorgas}, J., {Cardiel}, N., {Vazdekis}, A., \& {Peletier},
  R.~F. 2002, \mnras, 329, 863

\bibitem[{{Cenarro} {et~al.}(2007){Cenarro}, {Peletier},
  {S{\'a}nchez-Bl{\'a}zquez}, {Selam}, {Toloba}, {Cardiel},
  {Falc{\'o}n-Barroso}, {Gorgas}, {Jim{\'e}nez-Vicente}, \&
  {Vazdekis}}]{MILES_PARAM}
{Cenarro}, A.~J., {Peletier}, R.~F., {S{\'a}nchez-Bl{\'a}zquez}, P., {et~al.}
  2007, \mnras, 374, 664

\bibitem[{{Crampin} \& {Hoyle}(1961)}]{1961MNRAS.122...27C}
{Crampin}, J. \& {Hoyle}, F. 1961, \mnras, 122, 27

\bibitem[{{Cushing} {et~al.}(2005){Cushing}, {Rayner}, \& {Vacca}}]{Cush}
{Cushing}, M.~C., {Rayner}, J.~T., \& {Vacca}, W.~D. 2005, \apj, 623, 1115

\bibitem[{{Davidge} {et~al.}(2008){Davidge}, {Beck}, \& {McGregor}}]{Davidge08}
{Davidge}, T.~J., {Beck}, T.~L., \& {McGregor}, P.~J. 2008, \apj, 677, 238

\bibitem[{{Davidge} \& {Jensen}(2007)}]{2007AJ....133..576D}
{Davidge}, T.~J. \& {Jensen}, J.~B. 2007, \aj, 133, 576

\bibitem[{{Doyon} {et~al.}(1994){Doyon}, {Joseph}, \&
  {Wright}}]{1994ApJ...421..101D}
{Doyon}, R., {Joseph}, R.~D., \& {Wright}, G.~S. 1994, \apj, 421, 101

\bibitem[{{F{\" o}rster Schreiber}(2000)}]{For}
{F{\" o}rster Schreiber}, N.~M. 2000, \aj, 120, 2089

\bibitem[{{Frogel} {et~al.}(1975){Frogel}, {Becklin}, {Neugebauer}, {Matthews},
  {Persson}, \& {Aaronson}}]{1975ApJ...195L..15F}
{Frogel}, J.~A., {Becklin}, E.~E., {Neugebauer}, G., {et~al.} 1975, \apjl, 195,
  L15

\bibitem[{{Frogel} {et~al.}(1995){Frogel}, {Kuchinski}, \&
  {Tiede}}]{1995AJ....109.1154F}
{Frogel}, J.~A., {Kuchinski}, L.~E., \& {Tiede}, G.~P. 1995, \aj, 109, 1154

\bibitem[{{Frogel} {et~al.}(1980){Frogel}, {Persson}, \&
  {Cohen}}]{1980ApJ...240..785F}
{Frogel}, J.~A., {Persson}, S.~E., \& {Cohen}, J.~G. 1980, \apj, 240, 785

\bibitem[{{Frogel} {et~al.}(1978){Frogel}, {Persson}, {Matthews}, \&
  {Aaronson}}]{1978ApJ...220...75F}
{Frogel}, J.~A., {Persson}, S.~E., {Matthews}, K., \& {Aaronson}, M. 1978,
  \apj, 220, 75

\bibitem[{{Frogel} {et~al.}(2001){Frogel}, {Stephens}, {Ram{\'{\i}}rez}, \&
  {DePoy}}]{Frogel2001}
{Frogel}, J.~A., {Stephens}, A., {Ram{\'{\i}}rez}, S., \& {DePoy}, D.~L. 2001,
  \aj, 122, 1896

\bibitem[{{Girardi} {et~al.}(2000){Girardi}, {Bressan}, {Bertelli}, \&
  {Chiosi}}]{2000A&AS..141..371G}
{Girardi}, L., {Bressan}, A., {Bertelli}, G., \& {Chiosi}, C. 2000, \aaps, 141,
  371

\bibitem[{{Goldader} {et~al.}(1997){Goldader}, {Joseph}, {Doyon}, \&
  {Sanders}}]{Goldader97}
{Goldader}, J.~D., {Joseph}, R.~D., {Doyon}, R., \& {Sanders}, D.~B. 1997,
  \apj, 474, 104

\bibitem[{{Gorgas} {et~al.}(1999){Gorgas}, {Cardiel}, {Pedraz}, \& {Gonz{\'
  a}lez}}]{1999A&AS..139...29G}
{Gorgas}, J., {Cardiel}, N., {Pedraz}, S., \& {Gonz{\' a}lez}, J.~J. 1999,
  \aaps, 139, 29

\bibitem[{{Gorgas} {et~al.}(1993){Gorgas}, {Faber}, {Burstein}, {Gonzalez},
  {Courteau}, \& {Prosser}}]{1993ApJS...86..153G}
{Gorgas}, J., {Faber}, S.~M., {Burstein}, D., {et~al.} 1993, \apjs, 86, 153

\bibitem[{{Hanson} {et~al.}(1996){Hanson}, {Conti}, \& {Rieke}}]{Han}
{Hanson}, M.~M., {Conti}, P.~S., \& {Rieke}, M.~J. 1996, \apjs, 107, 281

\bibitem[{{Hanson} {et~al.}(2005){Hanson}, {Kudritzki}, {Kenworthy}, {Puls}, \&
  {Tokunaga}}]{Han2}
{Hanson}, M.~M., {Kudritzki}, R.-P., {Kenworthy}, M.~A., {Puls}, J., \&
  {Tokunaga}, A.~T. 2005, \apjs, 161, 154

\bibitem[{{Harris}(1996)}]{Harris96}
{Harris}, W.~E. 1996, \aj, 112, 1487

\bibitem[{{Hill} {et~al.}(1999){Hill}, {Heisler}, {Sutherland}, \&
  {Hunstead}}]{1999AJ....117..111H}
{Hill}, T.~L., {Heisler}, C.~A., {Sutherland}, R., \& {Hunstead}, R.~W. 1999,
  \aj, 117, 111

\bibitem[{{Hopp} \& {Fern\'{a}ndez}(2002)}]{Hopp}
{Hopp}, U. \& {Fern\'{a}ndez}, M. 2002, Calar Alto Newsletter, 4

\bibitem[{{Ivanov} {et~al.}(2000){Ivanov}, {Rieke}, {Groppi}, {Alonso-Herrero},
  {Rieke}, \& {Engelbracht}}]{2000ApJ...545..190I}
{Ivanov}, V.~D., {Rieke}, G.~H., {Groppi}, C.~E., {et~al.} 2000, \apj, 545, 190

\bibitem[{{Ivanov} {et~al.}(2004){Ivanov}, {Rieke}, {Engelbracht},
  {Alonso-Herrero}, {Rieke}, \& {Luhman}}]{Ivanov}
{Ivanov}, V.~D., {Rieke}, M.~J., {Engelbracht}, C.~W., {et~al.} 2004, \apjs,
  151, 387

\bibitem[{{James} \& {Mobasher}(1999)}]{1999MNRAS.306..199J}
{James}, P.~A. \& {Mobasher}, B. 1999, \mnras, 306, 199

\bibitem[{{James} \& {Seigar}(1999)}]{1999A&A...350..791J}
{James}, P.~A. \& {Seigar}, M.~S. 1999, \aap, 350, 791

\bibitem[{{Johnson} \& {Mendez}(1970)}]{Joh}
{Johnson}, H.~J. \& {Mendez}, M.~E. 1970, \aj, 75, 785

\bibitem[{{Kleinmann} \& {Hall}(1986)}]{KH86}
{Kleinmann}, S.~G. \& {Hall}, D.~N.~B. 1986, \apjs, 62, 501

\bibitem[{{Kuchinski} \& {Frogel}(1995)}]{1995AJ....110.2844K}
{Kuchinski}, L.~E. \& {Frogel}, J.~A. 1995, \aj, 110, 2844

\bibitem[{{Kuchinski} {et~al.}(1995){Kuchinski}, {Frogel}, {Terndrup}, \&
  {Persson}}]{1995AJ....109.1131K}
{Kuchinski}, L.~E., {Frogel}, J.~A., {Terndrup}, D.~M., \& {Persson}, S.~E.
  1995, \aj, 109, 1131

\bibitem[{{Lan{\c c}on} \& {Rocca-Volmerange}(1992)}]{LR92}
{Lan{\c c}on}, A. \& {Rocca-Volmerange}, B. 1992, \aaps, 96, 593

\bibitem[{{Lan{\c c}on} \& {Wood}(2000)}]{Lan}
{Lan{\c c}on}, A. \& {Wood}, P.~R. 2000, \aaps, 146, 217

\bibitem[{{Lang}(1991)}]{Lang}
{Lang}, L. 1991, {\it Astrophysical Data: Planets and Stars}, Ed.
  Springer-Verlag, 1

\bibitem[{{Lejeune} {et~al.}(1997){Lejeune}, {Cuisinier}, \&
  {Buser}}]{1997A&AS..125..229L}
{Lejeune}, T., {Cuisinier}, F., \& {Buser}, R. 1997, \aaps, 125, 229

\bibitem[{{Lejeune} {et~al.}(1998){Lejeune}, {Cuisinier}, \&
  {Buser}}]{1998A&AS..130...65L}
{Lejeune}, T., {Cuisinier}, F., \& {Buser}, R. 1998, \aaps, 130, 65

\bibitem[{{Livingston} \& {Wallace}(1991)}]{SUN}
{Livingston}, W. \& {Wallace}, L. 1991, {An atlas of the solar spectrum in the
  infrared from 1850 to 9000 cm-1 (1.1 to 5.4 micrometer)} (NSO Technical
  Report, Tucson: National Solar Observatory, National Optical Astronomy
  Observatory, 1991)

\bibitem[{{Mannucci} {et~al.}(2001){Mannucci}, {Basile}, {Poggianti},
  {Cimatti}, {Daddi}, {Pozzetti}, \& {Vanzi}}]{2001MNRAS.326..745M}
{Mannucci}, F., {Basile}, F., {Poggianti}, B.~M., {et~al.} 2001, \mnras, 326,
  745

\bibitem[{{Maraston}(2005)}]{Maraston05}
{Maraston}, C. 2005, \mnras, 362, 799

\bibitem[{{Mayya}(1997)}]{1997ApJ...482L.149M}
{Mayya}, Y.~D. 1997, \apjl, 482, L149+

\bibitem[{{McWilliam} \& {Lambert}(1984)}]{1984PASP...96..882M}
{McWilliam}, A. \& {Lambert}, D.~L. 1984, \pasp, 96, 882

\bibitem[{{Mieske} \& {Kroupa}(2008)}]{2008ApJ...677..276M}
{Mieske}, S. \& {Kroupa}, P. 2008, \apj, 677, 276

\bibitem[{{Mobasher} \& {James}(1996)}]{1996MNRAS.280..895M}
{Mobasher}, B. \& {James}, P.~A. 1996, \mnras, 280, 895

\bibitem[{{Mobasher} \& {James}(2000)}]{2000MNRAS.316..507M}
{Mobasher}, B. \& {James}, P.~A. 2000, \mnras, 316, 507

\bibitem[{{Oliva} \& {Origlia}(1992)}]{lineas_OH2}
{Oliva}, E. \& {Origlia}, L. 1992, \aap, 254, 466

\bibitem[{{Origlia} {et~al.}(1993){Origlia}, {Moorwood}, \&
  {Oliva}}]{1993A&A...280..536O}
{Origlia}, L., {Moorwood}, A.~F.~M., \& {Oliva}, E. 1993, \aap, 280, 536

\bibitem[{{Origlia} \& {Oliva}(2000)}]{2000A&A...357...61O}
{Origlia}, L. \& {Oliva}, E. 2000, \aap, 357, 61

\bibitem[{{Puxley} {et~al.}(1997){Puxley}, {Doyon}, \& {Ward}}]{Puxley}
{Puxley}, P.~J., {Doyon}, R., \& {Ward}, M.~J. 1997, \apj, 476, 120

\bibitem[{{Ramirez} {et~al.}(1997){Ramirez}, {Depoy}, {Frogel}, {Sellgren}, \&
  {Blum}}]{Ram}
{Ramirez}, S.~V., {Depoy}, D.~L., {Frogel}, J.~A., {Sellgren}, K., \& {Blum},
  R.~D. 1997, \aj, 113, 1411

\bibitem[{{Ranada} {et~al.}(2007){Ranada}, {Singh}, {Gupta}, \& {Ashok}}]{BASI}
{Ranada}, A.~C., {Singh}, H.~P., {Gupta}, R., \& {Ashok}, N.~M. 2007, Bulletin
  of the Astronomical Society of India, 35, 87

\bibitem[{{Ridgway} {et~al.}(1994){Ridgway}, {Wynn-Williams}, \&
  {Becklin}}]{1994ApJ...428..609R}
{Ridgway}, S.~E., {Wynn-Williams}, C.~G., \& {Becklin}, E.~E. 1994, \apj, 428,
  609

\bibitem[{{Riffel} {et~al.}(2007){Riffel}, {Pastoriza},
  {Rodr{\'{\i}}guez-Ardila}, \& {Maraston}}]{Riffel07}
{Riffel}, R., {Pastoriza}, M.~G., {Rodr{\'{\i}}guez-Ardila}, A., \& {Maraston},
  C. 2007, \apjl, 659, L103

\bibitem[{{Rousselot} {et~al.}(2000){Rousselot}, {Lidman}, {Cuby}, {Moreels},
  \& {Monnet}}]{lineas_OH1}
{Rousselot}, P., {Lidman}, C., {Cuby}, J.-G., {Moreels}, G., \& {Monnet}, G.
  2000, \aap, 354, 1134

\bibitem[{{Rutledge} {et~al.}(1997){Rutledge}, {Hesser}, \& {Stetson}}]{RHS97}
{Rutledge}, G.~A., {Hesser}, J.~E., \& {Stetson}, P.~B. 1997, \pasp, 109, 907

\bibitem[{{S{\'a}nchez-Bl{\'a}zquez} {et~al.}(2006){S{\'a}nchez-Bl{\'a}zquez},
  {Peletier}, {Jim{\'e}nez-Vicente}, {Cardiel}, {Cenarro},
  {Falc{\'o}n-Barroso}, {Gorgas}, {Selam}, \& {Vazdekis}}]{MILES}
{S{\'a}nchez-Bl{\'a}zquez}, P., {Peletier}, R.~F., {Jim{\'e}nez-Vicente}, J.,
  {et~al.} 2006, \mnras, 371, 703

\bibitem[{{Shier} {et~al.}(1996){Shier}, {Rieke}, \& {Rieke}}]{Shier96}
{Shier}, L.~M., {Rieke}, M.~J., \& {Rieke}, G.~H. 1996, \apj, 470, 222

\bibitem[{{Silva} {et~al.}(2008){Silva}, {Kuntschner}, \&
  {Lyubenova}}]{Fornax_red}
{Silva}, D.~R., {Kuntschner}, H., \& {Lyubenova}, M. 2008, \apj, 674, 194

\bibitem[{{Stephens} \& {Frogel}(2004)}]{2004AJ....127..925S}
{Stephens}, A.~W. \& {Frogel}, J.~A. 2004, \aj, 127, 925

\bibitem[{{Terndrup} {et~al.}(1991){Terndrup}, {Frogel}, \&
  {Whitford}}]{1991ApJ...378..742T}
{Terndrup}, D.~M., {Frogel}, J.~A., \& {Whitford}, A.~E. 1991, \apj, 378, 742

\bibitem[{{Tinsley}(1972)}]{Tinsley72}
{Tinsley}, B.~M. 1972, \apj, 178, 319

\bibitem[{{Tinsley}(1978)}]{Tinsley78}
{Tinsley}, B.~M. 1978, \apj, 222, 14

\bibitem[{{Tinsley}(1980)}]{Tinsley80}
{Tinsley}, B.~M. 1980, Fundamentals of Cosmic Physics, 5, 287

\bibitem[{{Turon} {et~al.}(1992){Turon}, {Cr{\' e}z{\' e}}, {Egret}, \& {et
  al.}}]{Hipparcos}
{Turon}, C., {Cr{\' e}z{\' e}}, M., {Egret}, \& {et al.} 1992, The HIPPARCOS
  input catalogue. ESA Special Publication, 1136

\bibitem[{{Vanzi} \& {Rieke}(1997)}]{Vanzi97}
{Vanzi}, L. \& {Rieke}, G.~H. 1997, \apj, 479, 694

\bibitem[{{Vazdekis} {et~al.}(2003){Vazdekis}, {Cenarro}, {Gorgas}, {Cardiel},
  \& {Peletier}}]{2003MNRAS.340.1317V}
{Vazdekis}, A., {Cenarro}, A.~J., {Gorgas}, J., {Cardiel}, N., \& {Peletier},
  R.~F. 2003, \mnras, 340, 1317

\bibitem[{{Wallace} \& {Hinkle}(1996)}]{WH96}
{Wallace}, L. \& {Hinkle}, K. 1996, \apjs, 107, 312

\bibitem[{{Wallace} \& {Hinkle}(1997)}]{WH97}
{Wallace}, L. \& {Hinkle}, K. 1997, \apjs, 111, 445

\bibitem[{{Worthey}(1994)}]{1994ApJS...95..107W}
{Worthey}, G. 1994, \apjs, 95, 107

\bibitem[{{Worthey} {et~al.}(1994){Worthey}, {Faber}, {Gonzalez}, \&
  {Burstein}}]{1994ApJS...94..687W}
{Worthey}, G., {Faber}, S.~M., {Gonzalez}, J.~J., \& {Burstein}, D. 1994,
  \apjs, 94, 687

\bibitem[{{Zinn} \& {West}(1984)}]{ZW84}
{Zinn}, R. \& {West}, M.~J. 1984, \apjs, 55, 45

\end{thebibliography}
\bibliographystyle{aa}


\appendix
\section{Stellar library and cluster stars}

In Table~\ref{table_stellar_library} we list the stars of the stellar library
with their spectral type, K magnitude, effective temperature and the associated
uncertainty (T$_{\rm eff}$ and $\sigma[T_\mathrm{eff}]$), surface gravity and
its uncertainty ($\log g$ and $\sigma[\log g]$), metallicity and its
uncertainty ([Fe/H] and $\sigma$[Fe/H]), number of measurements (N), D$_{\rm
CO}$ index and their error $\sigma[{\rm D_{CO}]}$. The stars observed at the
TNG are labeled with a $\dag$ in the index measurement.

In Table~\ref{table_cluster_stars} we list the cluster stars from
\citet{Frogel2001} and \citet{2004AJ....127..925S} employed in the computation
of the empirical fitting functions. We present the derived atmospheric
parameters and their associated uncertainties, and the D$_{\rm CO}$ index and
error for each star.

In both tables, AGB stars are labeled with a $\star$ in the name of the
star.

\addtocounter{table}{1}
\longtab{1}{
\begin{longtable}{llcccccccclc}
\caption{Stellar library used in the fitting function procedure. Stars
marked with a $\dag$ symbol were observed at the TNG. AGB stars are marked with
a $\star$ symbol.}
\label{table_stellar_library}\\
\hline\hline
Name        & Spectral & $K$   & $T_\mathrm{eff}$ & $\sigma[T_\mathrm{eff}]$ & $\log g$ & $\sigma[\log g]$ & [Fe/H] & $\sigma$[Fe/H] & N & \multicolumn{1}{c}{D$_{\rm CO}$} & $\sigma[{\rm D_{CO}]}$ \\
            & Type     & (mag) &       (K)        &                          &  (dex)   &                  & (dex)  &                &   &              &                     \\           
\hline
\endfirsthead
\caption{continued.}\\
\hline\hline
Name        & Spectral & $K$   & $T_\mathrm{eff}$ & $\sigma[T_\mathrm{eff}]$ & $\log g$ & $\sigma[\log g]$ & [Fe/H] & $\sigma$[Fe/H] & N & \multicolumn{1}{c}{D$_{\rm CO}$} & $\sigma[{\rm D_{CO}]}$ \\
            & Type     & (mag) &       (K)        &                          &  (dex)   &                  & (dex)  &                &   &              &                     \\           
\hline
\endhead
\hline
\endfoot

\object{BD+012916}         &  KIIvw      & $+6.47$ &  4150 &  60.9 & 0.10 & 0.18 &  $-1.99$ & 0.09 & 2 & 1.092        &0.010 \\ 
\object{BD+233130}         &  G0         & $+6.95$ &  5039 &  75.0 & 2.42 & 0.40 &  $-2.55$ & 0.15 & 2 & 1.058        &0.012 \\
\object{BD+442051}         &  M2V        & $+4.77$ &  3696 &  60.9 & 5.00 & 0.18 &  $-1.50$ & 0.09 & 4 & 1.105$^\dag$ &0.010 \\
\object{G171-010 }         &  M6eV       & $+5.93$ &  2799 &  60.9 & 5.12 & 0.18 &     --   & 0.09 & 2 & 1.129        &0.009 \\
\object{HD001326B}         &  M6V        & $+5.95$ &  3344 &  60.9 & 5.30 & 0.18 &  $-1.40$ & 0.09 & 2 & 1.092        &0.009 \\
\object{HD004628 }         &  K2V        & $+3.68$ &  4960 &  75.0 & 4.60 & 0.40 &  $-0.29$ & 0.15 & 2 & 1.080        &0.009 \\
\object{HD010307 }         &  G2V        & $+3.57$ &  5838 &  60.9 & 4.28 & 0.18 &  $+0.03$ & 0.09 & 2 & 1.057        &0.009 \\
\object{HD013043 }         &  G2V        & $+5.38$ &  5695 &  60.9 & 3.68 & 0.18 &  $+0.10$ & 0.09 & 3 & 1.048        &0.009 \\
\object{HD013555 }         &  F5V        & $+4.12$ &  6378 &  60.9 & 4.01 & 0.18 &  $-0.35$ & 0.09 & 3 & 1.048        &0.009 \\
\object{HD014221 }         &  F4V        & $+5.25$ &  6342 &  60.9 & 3.91 & 0.18 &  $-0.35$ & 0.09 & 3 & 1.042        &0.009 \\
\object{HD014662 }         &  F7Ib       & $+4.18$ &  5933 & 117.6 & 1.30 & 0.21 &  $-0.03$ & 0.09 & 3 & 1.045        &0.011 \\
\object{HD015596 }         &  G5III-IV   & $+3.86$ &  4755 &  75.0 & 2.50 & 0.40 &  $-0.70$ & 0.15 & 3 & 1.097        &0.011 \\
\object{HD015798 }         &  F5V        & $+3.47$ &  6345 &  60.9 & 3.85 & 0.18 &  $-0.16$ & 0.09 & 3 & 1.057        &0.009 \\
\object{HD016901 }         &  G0Ib-II    & $+3.56$ &  5345 & 117.6 & 0.85 & 0.21 &  $+0.00$ & 0.09 & 3 & 1.053        &0.010 \\
\object{HD017361 }         &  K1.5III    & $+2.09$ &  4600 &  60.9 & 2.85 & 0.18 &  $-0.02$ & 0.09 & 3 & 1.120        &0.011 \\
\object{HD017382 }         &  K1V        & $+5.61$ &  5065 &  75.0 & 4.50 & 0.40 &  $-0.13$ & 0.15 & 3 & 1.066        &0.009 \\
\object{HD020619 }         &  G0         & $+5.46$ &  5652 &  60.9 & 4.48 & 0.18 &  $-0.26$ & 0.09 & 3 & 1.054        &0.010 \\
\object{HD020893 }         &  K3III      & $+2.19$ &  4340 &  60.9 & 2.04 & 0.18 &  $+0.08$ & 0.09 & 3 & 1.146        &0.011 \\
\object{HD021017 }         &  K4III      & $+2.88$ &  4410 &  60.9 & 2.36 & 0.18 &  $+0.00$ & 0.09 & 3 & 1.138        &0.011 \\
\object{HD021197 }         &  K5V        & $+5.12$ &  4657 & 117.6 & 4.59 & 0.21 &  $+0.33$ & 0.10 & 3 & 1.100        &0.010 \\
\object{HD021910 }         &  G8III-IV   & $+4.99$ &  4582 &  60.9 & 1.75 & 0.18 &  $-0.60$ & 0.09 & 3 & 1.092        &0.011 \\
\object{HD023841 }         &  K1III      & $+3.80$ &  4279 &  60.9 & 1.67 & 0.21 &  $-0.95$ & 0.09 & 2 & 1.130        &0.012 \\
\object{HD025329 }         &  K1Vsb      & $+6.20$ &  4787 &  75.0 & 4.58 & 0.40 &  $-1.72$ & 0.15 & 2 & 1.058        &0.009 \\
\object{HD026297 }         &  G5-6IVw    & $+6.12$ &  4316 &  75.0 & 1.06 & 0.40 &  $-1.67$ & 0.15 & 3 & 1.083        &0.011 \\
\object{HD026322 }         &  F2IV-V     & $+4.48$ &  7072 &  60.9 & 3.49 & 0.18 &  $+0.16$ & 0.09 & 3 & 1.043        &0.011 \\
\object{HD026846 }         &  K3III      & $+2.27$ &  4541 &  60.9 & 2.62 & 0.18 &  $+0.15$ & 0.09 & 2 & 1.137        &0.011 \\
\object{HD027371 }         &  K0III      & $+1.51$ &  4271 &  60.9 & 3.00 & 0.18 &  $+0.34$ & 0.09 & 2 & 1.105        &0.011 \\
\object{HD027819 }         &  A7V        & $+4.41$ &  8129 &  60.9 & 4.00 & 0.18 &  $-0.20$ & 0.09 & 3 & 1.047        &0.009 \\
\object{HD028305 }         &  G9.5III    & $+1.42$ &  4846 &  60.9 & 2.68 & 0.18 &  $+0.11$ & 0.09 & 3 & 1.106        &0.011 \\
\object{HD029139 }         &  K5III      & $-3.04$ &  3910 &  75.0 & 1.59 & 0.40 &  $-0.34$ & 0.15 & 3 & 1.188        &0.011 \\
\object{HD030959 }         &  M3Svar     & $-0.66$ &  3451 & 117.6 & 0.80 & 0.21 &  $-0.15$ & 0.10 & 3 & 1.220        &0.010 \\
\object{HD031295 }         &  A0V        & $+4.41$ &  8991 & 117.6 & 4.08 & 0.21 &  $-0.89$ & 0.10 & 3 & 1.046        &0.009 \\
\object{HD031767 }         &  K2II       & $+1.34$ &  4120 &  60.9 & 1.78 & 0.18 &  $+0.26$ & 0.09 & 3 & 1.164        &0.011 \\
\object{HD032147 }         &  K3V        & $+3.71$ &  4658 & 100.0 & 4.47 & 0.50 &  $+0.02$ & 0.30 & 3 & 1.100        &0.009 \\
\object{HD035155 }         &  S?I        & $+2.13$ &  3600 & 117.6 & 0.80 & 0.21 &  $-0.72$ & 0.10 & 2 & 1.237        &0.012 \\
\object{HD035369 }         &  G8III      & $+2.06$ &  4863 &  75.0 & 2.50 & 0.40 &  $-0.26$ & 0.15 & 3 & 1.097        &0.011 \\
\object{HD035601}$^\star$  &  M1.5Ia     & $+1.66$ &  3550 &  60.9 & 0.00 & 0.18 &  $+0.00$ & 0.09 & 4 & 1.301$^\dag$ &0.010 \\
\object{HD036003 }         &  K5V        & $+4.88$ &  4465 &  60.9 & 4.61 & 0.18 &  $+0.09$ & 0.10 & 3 & 1.093        &0.009 \\
\object{HD036395 }         &  M1V        & $+4.00$ &  3590 &  60.9 & 4.90 & 0.21 &  $-0.45$ & 0.09 & 3 & 1.116        &0.009 \\
\object{HD037160 }         &  G8III-IV   & $+1.80$ &  4668 &  75.0 & 2.46 & 0.40 &  $-0.50$ & 0.15 & 3 & 1.105        &0.011 \\
\object{HD037536}$^\star$  &  M2Iabs     & $+0.97$ &  3789 & 117.6 & 0.70 & 0.21 &  $-0.15$ & 0.10 & 3 & 1.294        &0.010 \\
\object{HD037828 }         &  K0         & $+4.06$ &  4296 &  75.0 & 1.14 & 0.40 &  $-1.38$ & 0.15 & 3 & 1.110        &0.011 \\
\object{HD037984 }         &  K1III      & $+2.21$ &  4404 &  60.9 & 2.45 & 0.18 &  $-0.26$ & 0.09 & 3 & 1.129        &0.011 \\
\object{HD038656 }         &  G8III      & $+2.24$ &  4928 &  60.9 & 2.52 & 0.18 &  $-0.22$ & 0.09 & 3 & 1.085        &0.011 \\
\object{HD039364 }         &  G8III/IV   & $+1.40$ &  4550 &  60.9 & 2.10 & 0.18 &  $-0.94$ & 0.09 & 3 & 1.125        &0.012 \\
\object{HD039801}$^\star$  &  M2Iab      & $-3.56$ &  3547 &  60.9 & 0.00 & 0.21 &  $+0.03$ & 0.10 & 3 & 1.292        &0.010 \\
\object{HD040657 }         &  K1.5III    & $+1.64$ &  4370 &  60.9 & 2.42 & 0.18 &  $-0.58$ & 0.09 & 3 & 1.136        &0.011 \\
\object{HD041597 }         &  G8III      & $+2.89$ &  4700 &  75.0 & 2.38 & 0.40 &  $-0.54$ & 0.15 & 3 & 1.123        &0.011 \\
\object{HD041636 }         &  G9III      & $+3.97$ &  4709 &  60.9 & 2.50 & 0.18 &  $-0.20$ & 0.09 & 3 & 1.116        &0.011 \\
\object{HD042474}$^\star$  &  M2Iabpe... & $+1.85$ &  3789 & 117.6 & 0.70 & 0.21 &  $-0.36$ & 0.10 & 2 & 1.277        &0.010 \\
\object{HD042543}$^\star$  &  M1Ia-ab    & $+0.80$ &  3789 & 117.6 & 0.00 & 0.21 &  $-0.42$ & 0.10 & 3 & 1.298        &0.010 \\
\object{HD044007 }         &  G5IVw      & $+6.97$ &  4969 &  75.0 & 2.26 & 0.40 &  $-1.47$ & 0.15 & 3 & 1.071        &0.012 \\
\object{HD044889 }         &  K0I        & $+3.60$ &  3775 & 117.6 & 0.40 & 0.21 &  $-0.20$ & 0.10 & 4 & 1.186        &0.014 \\
\object{HD045829}$^\star$  &  K0Iab      & $+3.35$ &  4500 & 117.6 & 0.20 & 0.21 &  $-0.01$ & 0.09 & 2 & 1.194        &0.010 \\
\object{HD047914 }         &  K5III      & $+1.70$ &  3962 &  60.9 & 1.50 & 0.18 &  $+0.05$ & 0.09 & 3 & 1.172        &0.011 \\
\object{HD048329 }         &  G8Ib       & $+0.12$ &  4150 &  60.9 & 0.80 & 0.21 &  $+0.20$ & 0.09 & 2 & 1.170        &0.012 \\
\object{HD048433 }         &  K1III      & $+1.93$ &  4460 & 100.0 & 1.88 & 0.50 &  $-0.25$ & 0.30 & 3 & 1.120        &0.011 \\
\object{HD048565 }         &  F8         & $+5.80$ &  5929 &  60.9 & 3.59 & 0.18 &  $-0.70$ & 0.09 & 3 & 1.048        &0.009 \\
\object{HD049161 }         &  K4III      & $+1.58$ &  4176 &  60.9 & 1.69 & 0.18 &  $+0.08$ & 0.09 & 3 & 1.153        &0.011 \\
\object{HD049331}$^\star$  &  M1II       & $+0.56$ &  3600 & 117.6 & 0.70 & 0.21 &  $-0.03$ & 0.10 & 2 & 1.284        &0.010 \\
\object{HD052005 }         &  K4Iab      & $+2.10$ &  4117 &  60.9 & 0.60 & 0.18 &  $-0.20$ & 0.09 & 2 & 1.197        &0.011 \\
\object{HD052973 }         &  G0Ibvar    & $+2.13$ &  5659 & 117.6 & 1.37 & 0.21 &  $+0.34$ & 0.10 & 3 & 1.045        &0.010 \\
\object{HD054810 }         &  K0III      & $+2.44$ &  4697 &  60.9 & 2.35 & 0.18 &  $-0.30$ & 0.09 & 2 & 1.111        &0.011 \\
\object{HD057264 }         &  G8III      & $+2.75$ &  4620 &  60.9 & 2.72 & 0.18 &  $-0.33$ & 0.09 & 3 & 1.116        &0.011 \\
\object{HD058207 }         &  K0III      & $+1.56$ &  4788 &  60.9 & 2.55 & 0.18 &  $-0.12$ & 0.09 & 3 & 1.113        &0.011 \\
\object{HD058521}$^\star$  &  M5Ib-IIvar & $-0.68$ &  3238 &  60.9 & 0.00 & 0.18 &  $-0.19$ & 0.09 & 4 & 1.315$^\dag$ &0.010 \\
\object{HD060179 }         &  A1V        & $+1.64$ & 10286 & 117.6 & 4.00 & 0.21 &  $+0.98$ & 0.10 & 2 & 1.049        &0.009 \\
\object{HD060522 }         &  M0III-IIIb & $+0.23$ &  3899 &  60.9 & 1.20 & 0.18 &  $+0.12$ & 0.09 & 2 & 1.177        &0.015 \\
\object{HD061064 }         &  F6III      & $+4.21$ &  6449 &  60.9 & 3.21 & 0.21 &  $+0.42$ & 0.09 & 3 & 1.048        &0.011 \\
\object{HD061603 }         &  K5III      & $+2.17$ &  3870 &  60.9 & 1.50 & 0.18 &  $+0.24$ & 0.09 & 3 & 1.204        &0.012 \\
\object{HD061606 }         &  K2V        & $+4.88$ &  4833 & 117.6 & 4.55 & 0.21 &  $+0.07$ & 0.09 & 3 & 1.077        &0.009 \\
\object{HD061772 }         &  K3III      & $+1.33$ &  3995 &  60.9 & 1.47 & 0.18 &  $+0.08$ & 0.09 & 2 & 1.203        &0.011 \\
\object{HD062345 }         &  G8IIIa     & $+1.52$ &  5017 &  60.9 & 2.63 & 0.18 &  $-0.08$ & 0.09 & 3 & 1.086        &0.011 \\
\object{HD062721 }         &  K5III      & $+1.23$ &  3954 &  60.9 & 1.52 & 0.18 &  $-0.22$ & 0.09 & 2 & 1.188        &0.011 \\
\object{HD063352 }         &  K0III      & $+2.87$ &  4226 &  60.9 & 2.20 & 0.18 &  $-0.31$ & 0.09 & 3 & 1.153        &0.011 \\
\object{HD063791 }         &  G0         & $+5.42$ &  4629 &  75.0 & 1.76 & 0.40 &  $-1.63$ & 0.15 & 2 & 1.074        &0.011 \\
\object{HD064332 }         &  S?I        & $+2.30$ &  3500 & 117.6 & 0.50 & 0.21 &  $-0.34$ & 0.10 & 2 & 1.242        &0.010 \\
\object{HD065714 }         &  G8III      & $+3.91$ &  4840 &  60.9 & 1.50 & 0.18 &  $+0.27$ & 0.09 & 2 & 1.097        &0.012 \\
\object{HD066141 }         &  K2III      & $+1.44$ &  4258 &  60.9 & 1.90 & 0.18 &  $-0.30$ & 0.09 & 3 & 1.145        &0.011 \\
\object{HD068284 }         &  F8V        & $+6.26$ &  5860 &  60.9 & 3.98 & 0.18 &  $-0.57$ & 0.09 & 3 & 1.050        &0.009 \\
\object{HD069267 }         &  K4III      & $+0.19$ &  4043 &  60.9 & 1.51 & 0.18 &  $-0.12$ & 0.09 & 4 & 1.170$^\dag$ &0.011 \\
\object{HD070272 }         &  K5III      & $+0.37$ &  3900 &  60.9 & 1.05 & 0.18 &  $+0.04$ & 0.09 & 3 & 1.180        &0.011 \\
\object{HD072184 }         &  K2III      & $+3.50$ &  4624 &  60.9 & 2.61 & 0.18 &  $+0.12$ & 0.09 & 2 & 1.123        &0.013 \\
\object{HD072324 }         &  G9III      & $+3.97$ &  4887 &  60.9 & 2.13 & 0.18 &  $+0.16$ & 0.09 & 3 & 1.099        &0.011 \\
\object{HD072905 }         &  G1.5Vb     & $+4.17$ &  5864 &  60.9 & 4.48 & 0.18 &  $-0.04$ & 0.09 & 3 & 1.061        &0.010 \\
\object{HD073394 }         &  G5IIIw     & $+4.95$ &  4500 &  60.9 & 1.10 & 0.18 &  $-1.38$ & 0.09 & 4 & 1.088$^\dag$ &0.011 \\
\object{HD073593 }         &  G8IV       & $+2.96$ &  4717 &  60.9 & 2.25 & 0.18 &  $-0.12$ & 0.09 & 4 & 1.120$^\dag$ &0.011 \\
\object{HD074000 }         &  F6VI       & $+8.39$ &  6166 &  60.9 & 4.19 & 0.18 &  $-2.02$ & 0.09 & 2 & 1.061        &0.010 \\
\object{HD074395 }         &  G2Iab      & $+2.85$ &  5250 & 117.6 & 1.30 & 0.21 &  $-0.05$ & 0.09 & 4 & 1.076$^\dag$ &0.010 \\
\object{HD074442 }         &  K0III      & $+1.56$ &  4657 &  60.9 & 2.51 & 0.18 &  $-0.06$ & 0.09 & 3 & 1.120        &0.011 \\
\object{HD075732 }         &  G8V        & $+4.01$ &  5079 &  75.0 & 4.48 & 0.40 &  $+0.16$ & 0.15 & 3 & 1.076        &0.009 \\
\object{HD076813 }         &  G9III      & $+3.15$ &  6072 & 117.6 & 4.20 & 0.21 &  $-0.82$ & 0.10 & 4 & 1.082$^\dag$ &0.010 \\
\object{HD076932 }         &  F7-8IV-V   & $+4.36$ &  5866 & 100.0 & 3.96 & 0.50 &  $-0.93$ & 0.30 & 3 & 1.057        &0.009 \\
\object{HD078712 }         &  M6SI       & $-1.87$ &  3202 &  60.9 & 0.00 & 0.18 &  $-0.11$ & 0.09 & 4 & 1.216$^\dag$ &0.010 \\
\object{HD078732 }         &  G8II       & $+3.20$ &  4900 & 117.6 & 2.00 & 0.21 &  $+0.24$ & 0.10 & 8 & 1.108        &0.011 \\
\object{HD079211 }         &  M0V        & $+4.14$ &  3710 &  60.9 & 4.71 & 0.18 &  $-0.40$ & 0.10 & 3 & 1.093        &0.010 \\
\object{HD079452 }         &  G6III      & $+3.86$ &  4829 &  60.9 & 2.35 & 0.18 &  $-0.84$ & 0.09 & 4 & 1.086$^\dag$ &0.012 \\
\object{HD081192 }         &  G7III      & $+4.11$ &  4705 &  75.0 & 2.50 & 0.40 &  $-0.62$ & 0.15 & 4 & 1.101$^\dag$ &0.011 \\
\object{HD082074 }         &  G6IV       & $+4.15$ &  5055 & 117.6 & 3.30 & 0.21 &  $-0.48$ & 0.10 & 2 & 1.088        &0.011 \\
\object{HD082885 }         &  G8IV-V     & $+3.69$ &  5488 &  60.9 & 4.61 & 0.18 &  $+0.00$ & 0.09 & 2 & 1.061        &0.009 \\
\object{HD083425 }         &  K3III      & $+1.58$ &  4120 &  60.9 & 2.00 & 0.18 &  $-0.35$ & 0.09 & 4 & 1.170$^\dag$ &0.011 \\
\object{HD083618 }         &  K3III      & $+0.87$ &  4231 &  60.9 & 1.74 & 0.18 &  $-0.08$ & 0.09 & 4 & 1.161$^\dag$ &0.011 \\
\object{HD083632 }         &  K2III      & $+4.72$ &  4214 &  60.9 & 1.00 & 0.21 &  $-1.39$ & 0.09 & 4 & 1.160$^\dag$ &0.011 \\
\object{HD085235 }         &  A3IV       & $+4.37$ & 11200 & 117.6 & 3.55 & 0.21 &  $-0.40$ & 0.10 & 3 & 1.049        &0.009 \\
\object{HD085503 }         &  K0III      & $+1.36$ &  4472 &  75.0 & 2.33 & 0.40 &  $+0.23$ & 0.15 & 3 & 1.136        &0.011 \\
\object{HD085773 }         &  G:w?I      & $+7.95$ &  4463 &  60.9 & 0.98 & 0.18 &  $-2.17$ & 0.09 & 1 & 1.056        &0.014 \\
\object{HD087737 }         &  A0Ib       & $+3.29$ &  9625 &  60.9 & 1.98 & 0.21 &  $-0.04$ & 0.10 & 4 & 1.053$^\dag$ &0.010 \\
\object{HD087822 }         &  F4V        & $+5.13$ &  6590 &  60.9 & 4.15 & 0.18 &  $+0.14$ & 0.09 & 4 & 1.049$^\dag$ &0.009 \\
\object{HD089484 }         &  K1IIIb     & $-0.81$ &  4470 &  60.9 & 2.35 & 0.18 &  $-0.49$ & 0.09 & 3 & 1.119        &0.011 \\
\object{HD089822B}         &  A0sp?III   & $+3.39$ &  5538 & 117.6 & 2.44 & 0.21 &  $+0.51$ & 0.10 & 2 & 1.050        &0.010 \\
\object{HD092523 }         &  K3III      & $+1.55$ &  4090 &  60.9 & 1.96 & 0.18 &  $-0.38$ & 0.09 & 2 & 1.163        &0.011 \\
\object{HD093487 }         &  F8         & $+6.76$ &  5250 &  60.9 & 1.80 & 0.18 &  $-1.05$ & 0.09 & 2 & 1.068        &0.012 \\
\object{HD095578 }         &  M0III      & $+0.80$ &  3700 &  60.9 & 1.40 & 0.18 &  $-0.23$ & 0.09 & 2 & 1.206        &0.011 \\
\object{HD095735 }         &  M2V        & $+3.25$ &  3551 &  60.9 & 4.90 & 0.21 &  $-0.20$ & 0.09 & 8 & 1.108$^\dag$ &0.009 \\
\object{HD096360 }         &  M?I        & $+2.76$ &  3550 & 117.6 & 0.50 & 0.21 &  $-0.58$ & 0.10 & 4 & 1.244$^\dag$ &0.012 \\
\object{HD097907 }         &  K3III      & $+2.43$ &  4351 &  60.9 & 2.07 & 0.18 &  $-0.10$ & 0.09 & 2 & 1.162        &0.011 \\
\object{HD099648 }         &  G8II-III   & $+2.83$ &  4850 & 117.6 & 1.90 & 0.21 &  $+0.36$ & 0.10 & 3 & 1.097        &0.011 \\
\object{HD099998 }         &  K4III      & $+1.24$ &  3863 &  60.9 & 1.79 & 0.18 &  $-0.16$ & 0.09 & 4 & 1.186$^\dag$ &0.011 \\
\object{HD101501 }         &  G8Vvar     & $+3.58$ &  5401 &  60.9 & 4.60 & 0.18 &  $-0.13$ & 0.09 & 3 & 1.054        &0.009 \\
\object{HD102224 }         &  K0III      & $+0.98$ &  4383 &  75.0 & 2.02 & 0.40 &  $-0.46$ & 0.15 & 2 & 1.148        &0.011 \\
\object{HD102328 }         &  K3III      & $+2.63$ &  4390 &  60.9 & 2.09 & 0.18 &  $+0.35$ & 0.09 & 2 & 1.155        &0.011 \\
\object{HD103095 }         &  G8Vp       & $+4.37$ &  5025 &  60.9 & 4.56 & 0.18 &  $-1.36$ & 0.09 & 4 & 1.043$^\dag$ &0.002 \\
\object{HD103877 }         &  AmV        & $+5.88$ &  7341 & 117.6 & 4.00 & 0.21 &  $+0.40$ & 0.09 & 1 & 1.038        &0.009 \\
\object{HD104307 }         &  K2III      & $+3.68$ &  4451 & 117.6 & 2.00 & 0.21 &  $-0.01$ & 0.10 & 2 & 1.159        &0.012 \\
\object{HD105262 }         &  B9         & $+6.75$ &  8542 & 117.6 & 1.50 & 0.21 &  $-1.37$ & 0.10 & 2 & 1.056        &0.010 \\
\object{HD107213 }         &  F8Vs       & $+5.13$ &  6298 &  60.9 & 4.01 & 0.18 &  $+0.36$ & 0.09 & 4 & 1.051$^\dag$ &0.009 \\
\object{HD110014 }         &  K2III      & $+2.01$ &  4399 &  60.9 & 1.47 & 0.18 &  $+0.10$ & 0.09 & 3 & 1.151        &0.011 \\
\object{HD111631 }         &  M0.5V      & $+4.88$ &  3785 &  60.9 & 4.75 & 0.21 &  $+0.10$ & 0.09 & 4 & 1.109$^\dag$ &0.011 \\
\object{HD113092 }         &  K2III      & $+2.11$ &  4283 &  60.9 & 1.95 & 0.18 &  $-0.37$ & 0.09 & 3 & 1.148        &0.011 \\
\object{HD113285 }         &  M8III      & $-1.06$ &  2485 & 117.6 & 0.00 & 0.18 &     --   &  --  & 5 & 1.250        &0.012 \\
\object{HD114038 }         &  K1III      & $+2.72$ &  4530 &  60.9 & 2.71 & 0.18 &  $-0.04$ & 0.09 & 4 & 1.132$^\dag$ &0.011 \\
\object{HD114095 }         &  G5         & $+5.88$ &  4650 &  60.9 & 2.40 & 0.18 &  $-0.70$ & 0.09 & 2 & 1.108        &0.011 \\
\object{HD114330 }         &  AV1        & $+4.30$ &  9509 & 117.6 & 3.80 & 0.21 &  $-0.01$ & 0.10 & 2 & 1.057        &0.011 \\
\object{HD114946 }         &  G6V        & $+3.11$ &  5171 &  60.9 & 3.64 & 0.18 &  $+0.13$ & 0.09 & 3 & 1.093        &0.010 \\
\object{HD114961}$^\star$  &  M7III      & $+1.50$ &  3012 & 112.8 & 0.00 & 0.18 &  $-0.81$ & 0.09 & 8 & 1.295$^\dag$ &0.010 \\
\object{HD117176 }         &  G5V        & $+3.50$ &  5525 &  60.9 & 3.81 & 0.18 &  $-0.10$ & 0.09 & 2 & 1.067        &0.009 \\
\object{HD117876 }         &  G8III      & $+3.87$ &  4782 &  75.0 & 2.25 & 0.40 &  $-0.50$ & 0.15 & 3 & 1.106        &0.012 \\
\object{HD119228 }         &  M2III      & $+0.33$ &  3600 &  60.9 & 1.60 & 0.18 &  $+0.30$ & 0.09 & 8 & 1.203$^\dag$ &0.011 \\
\object{HD119667 }         &  K5         & $+4.18$ &  3700 & 117.6 & 1.00 & 0.21 &  $-0.35$ & 0.10 & 4 & 1.219$^\dag$ &0.011 \\
\object{HD120933 }         &  K5III      & $-0.00$ &  3820 &  60.9 & 1.52 & 0.18 &  $+0.50$ & 0.09 & 8 & 1.226$^\dag$ &0.012 \\
\object{HD121130 }         &  M3III      & $-0.24$ &  3672 & 117.6 & 1.25 & 0.21 &  $-0.24$ & 0.10 & 2 & 1.216        &0.012 \\
\object{HD121299 }         &  K2III      & $+2.85$ &  4710 &  60.9 & 2.64 & 0.18 &  $-0.03$ & 0.09 & 4 & 1.125$^\dag$ &0.011 \\
\object{HD122563 }         &  F8IV       & $+3.73$ &  4566 &  75.0 & 1.12 & 0.40 &  $-2.63$ & 0.15 & 4 & 1.053$^\dag$ &0.011 \\
\object{HD122956 }         &  G6IV-Vw    & $+5.89$ &  4635 &  75.0 & 1.49 & 0.40 &  $-1.75$ & 0.15 & 2 & 1.084        &0.010 \\
\object{HD123299 }         &  A0III      & $+3.63$ &  9894 &  60.9 & 2.90 & 0.21 &  $+0.12$ & 0.10 & 5 & 1.048        &0.011 \\
\object{HD123657 }         &  M4III      & $-0.23$ &  3450 &  60.9 & 0.85 & 0.21 &  $+0.00$ & 0.09 & 4 & 1.231$^\dag$ &0.013 \\
\object{HD124186 }         &  K4III      & $+3.59$ &  4347 &  60.9 & 2.10 & 0.18 &  $+0.24$ & 0.09 & 3 & 1.144        &0.011 \\
\object{HD124850 }         &  F7IV       & $+2.80$ &  6116 &  60.9 & 3.87 & 0.18 &  $-0.11$ & 0.09 & 3 & 1.051        &0.009 \\
\object{HD124897 }         &  K2IIIp     & $-2.91$ &  4361 &  75.0 & 1.93 & 0.40 &  $-0.53$ & 0.15 & 2 & 1.137        &0.013 \\
\object{HD126327 }         &  M7.5III    & $+1.74$ &  2819 &  60.9 & 0.00 & 0.18 &  $-0.58$ & 0.09 & 4 & 1.282$^\dag$ &0.013 \\
\object{HD126681 }         &  G3V        & $+7.63$ &  5536 &  60.9 & 4.65 & 0.18 &  $-1.25$ & 0.09 & 2 & 1.068        &0.010 \\
\object{HD126778 }         &  K0III      & $+5.84$ &  4847 &  60.9 & 2.34 & 0.21 &  $-0.62$ & 0.09 & 2 & 1.120        &0.011 \\
\object{HD127243 }         &  G3IV       & $+3.15$ &  4978 &  75.0 & 3.20 & 0.40 &  $-0.59$ & 0.15 & 3 & 1.091        &0.012 \\
\object{HD130694 }         &  K4III      & $+1.10$ &  4040 &  60.9 & 1.85 & 0.18 &  $-0.34$ & 0.09 & 3 & 1.173        &0.011 \\
\object{HD130705 }         &  K4II-III   & $+3.95$ &  4336 &  60.9 & 2.10 & 0.18 &  $+0.41$ & 0.09 & 4 & 1.151$^\dag$ &0.011 \\
\object{HD131430 }         &  K2/K3III   & $+2.20$ &  4190 &  60.9 & 2.18 & 0.18 &  $+0.04$ & 0.09 & 4 & 1.159$^\dag$ &0.011 \\
\object{HD131918 }         &  K4III      & $+2.09$ &  3970 &  60.9 & 1.49 & 0.18 &  $+0.22$ & 0.09 & 3 & 1.173        &0.012 \\
\object{HD132345 }         &  K3III-IVp  & $+3.26$ &  4374 &  60.9 & 1.60 & 0.18 &  $+0.23$ & 0.09 & 3 & 1.155        &0.013 \\
\object{HD134063 }         &  G5III      & $+5.47$ &  4885 &  60.9 & 2.34 & 0.21 &  $-0.69$ & 0.09 & 4 & 1.109$^\dag$ &0.011 \\
\object{HD135722 }         &  G8III      & $+1.22$ &  4847 &  75.0 & 2.56 & 0.40 &  $-0.44$ & 0.15 & 2 & 1.086        &0.011 \\
\object{HD136726 }         &  K4III      & $+1.93$ &  4120 &  60.9 & 2.03 & 0.18 &  $+0.07$ & 0.09 & 4 & 1.190$^\dag$ &0.014 \\
\object{HD137471 }         &  M1III      & $+1.03$ &  3422 &  60.9 & 1.10 & 0.18 &  $+0.07$ & 0.10 & 3 & 1.194        &0.010 \\
\object{HD137704 }         &  K4III      & $+2.11$ &  4095 &  60.9 & 1.97 & 0.18 &  $-0.27$ & 0.09 & 4 & 1.166$^\dag$ &0.011 \\
\object{HD137759 }         &  K2III      & $+0.77$ &  4498 &  60.9 & 2.38 & 0.18 &  $+0.05$ & 0.09 & 3 & 1.125        &0.012 \\
\object{HD137909 }         &  F0p        & $+3.45$ &  8541 & 117.6 & 4.25 & 0.21 &  $+0.83$ & 0.10 & 3 & 1.047        &0.009 \\
\object{HD138481 }         &  K5III      & $+1.21$ &  3890 &  60.9 & 1.64 & 0.18 &  $+0.20$ & 0.09 & 4 & 1.166$^\dag$ &0.018 \\
\object{HD139641 }         &  G7.5IIIb   & $+3.10$ &  5030 &  60.9 & 3.22 & 0.18 &  $-0.55$ & 0.09 & 2 & 1.081        &0.011 \\
\object{HD140160 }         &  A0p...V    & $+5.20$ &  9164 & 117.6 & 3.30 & 0.21 &  $-0.25$ & 0.10 & 2 & 1.050        &0.011 \\
\object{HD141527 }         &  G0Iab:pe   & $+4.56$ &  6816 &  60.9 & 0.48 & 0.18 &  $-0.50$ & 0.09 & 4 & 1.042        &0.014 \\
\object{HD141714 }         &  G3.5III    & $+2.66$ &  5230 &  60.9 & 3.02 & 0.18 &  $-0.29$ & 0.09 & 5 & 1.075        &0.012 \\
\object{HD145675 }         &  K0V        & $+4.71$ &  5264 &  75.0 & 4.66 & 0.40 &  $+0.34$ & 0.15 & 4 & 1.080$^\dag$ &0.009 \\
\object{HD146051 }         &  M0.5III    & $-1.17$ &  3793 & 117.6 & 1.40 & 0.21 &  $+0.32$ & 0.10 & 2 & 1.189        &0.011 \\
\object{HD147923 }         &  M...       & $+3.46$ &  3600 & 117.6 & 0.80 & 0.21 &  $-0.19$ & 0.10 & 4 & 1.221$^\dag$ &0.011 \\
\object{HD148783 }         &  M6III      & $+0.29$ &  3279 & 112.8 & 0.20 & 0.21 &  $-0.06$ & 0.09 & 4 & 1.249$^\dag$ &0.012 \\
\object{HD148897 }         &  G8pII      & $+1.96$ &  4284 & 117.6 & 1.15 & 0.21 &  $-0.75$ & 0.09 & 2 & 1.116        &0.011 \\
\object{HD149009 }         &  K5III      & $+2.04$ &  3910 &  60.9 & 1.60 & 0.18 &  $+0.30$ & 0.09 & 3 & 1.209        &0.012 \\
\object{HD149661 }         &  K0V        & $+3.91$ &  5168 &  60.9 & 4.63 & 0.18 &  $+0.04$ & 0.09 & 4 & 1.075$^\dag$ &0.009 \\
\object{HD150012 }         &  F5III-IV   & $+5.26$ &  6505 &  60.9 & 3.90 & 0.18 &  $+0.16$ & 0.09 & 3 & 1.058        &0.009 \\
\object{HD150680 }         &  F9IV       & $+1.28$ &  5672 &  60.9 & 3.74 & 0.18 &  $+0.01$ & 0.09 & 2 & 1.055        &0.009 \\
\object{HD151203 }         &  M3IIIab    & $+0.89$ &  3640 &  60.9 & 0.70 & 0.18 &  $-0.10$ & 0.10 & 3 & 1.218        &0.011 \\
\object{HD154733 }         &  K3III      & $+2.48$ &  4279 &  60.9 & 2.10 & 0.18 &  $+0.00$ & 0.09 & 4 & 1.165$^\dag$ &0.011 \\
\object{HD155763 }         &  B6III      & $+3.60$ & 13397 & 117.6 & 4.24 & 0.18 &  $-0.95$ & 0.10 & 2 & 1.049        &0.010 \\
\object{HD156014 }         &  M5Ib-II    & $-1.99$ &  3161 & 112.8 & 0.00 & 0.21 &  $+0.00$ & 0.09 & 2 & 1.267        &0.012 \\
\object{HD156026 }         &  K5V        & $+3.86$ &  4541 &  60.9 & 4.54 & 0.18 &  $-0.37$ & 0.09 & 2 & 1.106        &0.009 \\
\object{HD156283 }         &  K3IIvar    & $-0.02$ &  4460 &  60.9 & 2.33 & 0.18 &  $+0.18$ & 0.09 & 3 & 1.176        &0.011 \\
\object{HD157910 }         &  G5III      & $+4.39$ &  5137 &  60.9 & 1.83 & 0.18 &  $-0.32$ & 0.09 & 3 & 1.077        &0.011 \\
\object{HD160933 }         &  F9V        & $+4.82$ &  5684 &  75.0 & 3.90 & 0.40 &  $-0.32$ & 0.15 & 2 & 1.061        &0.009 \\
\object{HD161096 }         &  K2III      & $+0.43$ &  4543 &  60.9 & 2.16 & 0.18 &  $+0.08$ & 0.09 & 2 & 1.138        &0.016 \\
\object{HD161797 }         &  G5IV       & $+1.51$ &  5411 &  75.0 & 3.87 & 0.40 &  $+0.16$ & 0.15 & 3 & 1.067        &0.010 \\
\object{HD161817 }         &  A2VI(HB)   & $+6.29$ &  7759 &  60.9 & 2.95 & 0.18 &  $-0.95$ & 0.09 & 2 & 1.046        &0.010 \\
\object{HD163990 }         &  M6Svar     & $+0.19$ &  3365 & 117.6 & 0.70 & 0.21 &  $+0.01$ & 0.10 & 3 & 1.247        &0.011 \\
\object{HD163993 }         &  G8III      & $+1.73$ &  5028 &  60.9 & 2.70 & 0.18 &  $+0.03$ & 0.09 & 3 & 1.087        &0.012 \\
\object{HD164058 }         &  K5III      & $-1.16$ &  3930 &  60.9 & 1.26 & 0.18 &  $-0.05$ & 0.09 & 4 & 1.189$^\dag$ &0.012 \\
\object{HD164136 }         &  F2II       & $+2.77$ &  6799 & 117.6 & 2.63 & 0.21 &  $-0.30$ & 0.10 & 3 & 1.044        &0.011 \\
\object{HD164349 }         &  K0.5IIb    & $+1.93$ &  4446 &  60.9 & 1.50 & 0.18 &  $+0.39$ & 0.09 & 4 & 1.157$^\dag$ &0.011 \\
\object{HD166208 }         &  G8III...   & $+2.93$ &  4919 &  75.0 & 2.52 & 0.40 &  $+0.08$ & 0.15 & 3 & 1.075        &0.011 \\
\object{HD167768 }         &  G3III      & $+3.89$ &  5235 &  60.9 & 1.61 & 0.21 &  $-0.68$ & 0.09 & 4 & 1.085$^\dag$ &0.011 \\
\object{HD168322 }         &  G8.5IIIb   & $+3.93$ &  4793 &  60.9 & 2.00 & 0.18 &  $-0.40$ & 0.09 & 3 & 1.111        &0.012 \\
\object{HD168720 }         &  M1III      & $+0.74$ &  3810 &  60.9 & 1.10 & 0.18 &  $+0.00$ & 0.10 & 4 & 1.211$^\dag$ &0.012 \\
\object{HD168723 }         &  K0III-IV   & $+1.05$ &  4859 &  75.0 & 3.13 & 0.40 &  $-0.19$ & 0.15 & 2 & 1.099        &0.011 \\
\object{HD173819 }         &  K0Ibpvar   & $+2.15$ &  4421 & 117.6 & 0.00 & 0.21 &  $-0.88$ & 0.10 & 3 & 1.121        &0.012 \\
\object{HD174638 }         &  B7Ve....   & $+3.19$ & 12136 &  60.9 & 2.50 & 0.18 &  $+0.43$ & 0.09 & 3 & 1.041        &0.011 \\
\object{HD175865 }         &  M5III      & $-1.83$ &  3520 &  60.9 & 0.50 & 0.18 &  $+0.14$ & 0.10 & 4 & 1.250$^\dag$ &0.012 \\
\object{HD181096 }         &  F6IV:      & $+6.47$ &  6276 &  75.0 & 4.09 & 0.40 &  $-0.26$ & 0.15 & 2 & 1.047        &0.009 \\
\object{HD182835 }         &  F2Ib       & $+4.01$ &  7350 & 721.4 & 2.15 & 0.32 &  $+0.09$ & 0.29 & 2 & 1.050        &0.010 \\
\object{HD184499 }         &  G0V        & $+5.07$ &  5738 & 100.0 & 4.02 & 0.50 &  $-0.66$ & 0.30 & 4 & 1.044$^\dag$ &0.009 \\
\object{HD184786 }         &  M4.5III    & $+0.74$ &  3467 & 117.6 & 0.60 & 0.21 &  $-0.04$ & 0.10 & 4 & 1.234$^\dag$ &0.012 \\
\object{HD185144 }         &  K0V        & $+2.90$ &  5260 &  75.0 & 4.55 & 0.40 &  $-0.24$ & 0.15 & 4 & 1.061$^\dag$ &0.009 \\
\object{HD187216 }         &  R...       & $+6.02$ &  3500 & 117.6 & 0.40 & 0.21 &  $-2.48$ & 0.10 & 4 & 1.138$^\dag$ &0.010 \\
\object{HD187921 }         &  K0var      & $+3.80$ &  6000 & 117.6 & 1.00 & 0.21 &  $+0.28$ & 0.10 & 3 & 1.046        &0.012 \\
\object{HD188119 }         &  G8III      & $+1.73$ &  4915 &  75.0 & 2.61 & 0.40 &  $-0.32$ & 0.15 & 3 & 1.098        &0.012 \\
\object{HD191277 }         &  K3III      & $+2.72$ &  4459 &  60.9 & 2.71 & 0.18 &  $+0.30$ & 0.09 & 4 & 1.131$^\dag$ &0.011 \\
\object{HD195593 }         &  F5Iab      & $+3.65$ &  6700 & 721.4 & 1.95 & 0.18 &  $+0.12$ & 0.09 & 3 & 1.046        &0.012 \\
\object{HD199799 }         &  M1I        & $+1.33$ &  3400 & 117.6 & 0.30 & 0.21 &  $-0.24$ & 0.10 & 4 & 1.241$^\dag$ &0.012 \\
\object{HD202447 }         &  G0III+...  & $+2.34$ &  6087 &  60.9 & 3.24 & 0.18 &  $+0.09$ & 0.09 & 5 & 1.085        &0.010 \\
\object{HD209369 }         &  F5V        & $+3.96$ &  6217 &  60.9 & 3.85 & 0.18 &  $-0.26$ & 0.09 & 2 & 1.051        &0.009 \\
\object{HD216228 }         &  K0III      & $+1.27$ &  4768 &  75.0 & 2.49 & 0.40 &  $+0.01$ & 0.15 & 2 & 1.118        &0.011 \\
\object{HD217382 }         &  K4III      & $+1.49$ &  4035 &  60.9 & 1.24 & 0.18 &  $-0.25$ & 0.09 & 2 & 1.182        &0.011 \\
\object{HD223047}$^\star$  &  G5Ib       & $+2.47$ &  4990 & 117.6 & 1.50 & 0.21 &  $+0.18$ & 0.10 & 2 & 1.140        &0.010 \\
\object{HD232078 }         &  K4-5III    & $+4.19$ &  4008 &  60.9 & 0.30 & 0.18 &  $-1.73$ & 0.09 & 4 & 1.139$^\dag$ &0.011 \\
\end{longtable}
}

\addtocounter{table}{1}
\longtab{2}{
\begin{longtable}{lcccccccc}
\caption{Cluster stars used in the fitting function procedure. AGB stars
are maked with a $\star$ symbol.}
\label{table_cluster_stars}\\
\hline\hline
\hline
Name     & $T_\mathrm{eff}$ & $\sigma[T_\mathrm{eff}]$ & $\log g$ & $\sigma[\log g]$ & [Fe/H] & $\sigma$[Fe/H] & D$_{\rm CO}$ & $\sigma[{\rm D_{CO}]}$ \\
         &       (K)        &                          &   (dex)  &                  &  (dex) &                &              &          \\           
\hline
\endfirsthead
\caption{continued.}\\
\hline\hline
Name     & $T_\mathrm{eff}$ & $\sigma[T_\mathrm{eff}]$ & $\log g$ & $\sigma[\log g]$ & [Fe/H] & $\sigma$[Fe/H] & D$_{\rm CO}$ & $\sigma[{\rm D_{CO}]}$ \\
         &       (K)        &                          &   (dex)  &                  &  (dex) &                &              &          \\           
\hline
\endhead
\hline
\endfoot
Liller1-6        & 3612 & 127.0 & $+0.05$ & 0.11 & $-0.61$ & 0.09 & 1.250 & 0.047 \\ 
Liller1-7        & 3671 &  96.0 & $+0.07$ & 0.13 & $-0.61$ & 0.09 & 1.178 & 0.043 \\ 
Liller1-157$^\star$     & 4011 & 121.0 & $-0.11$ & 0.11 & $-0.61$ & 0.09 & 1.248 & 0.047 \\ 
Liller1-158      & 3671 &  96.0 & $-0.02$ & 0.11 & $-0.61$ & 0.09 & 1.245 & 0.047 \\ 
Liller1-162      & 3627 & 119.0 & $+0.29$ & 0.11 & $-0.61$ & 0.09 & 1.214 & 0.045 \\ 
Liller1-166      & 3973 & 118.0 & $+0.56$ & 0.13 & $-0.61$ & 0.09 & 1.150 & 0.041 \\ 
Liller1-299      & 3150 & 134.0 & $-0.07$ & 0.11 & $-0.61$ & 0.09 & 1.255 & 0.047 \\ 
M69-1            & 3830 & 106.0 & $+0.04$ & 0.10 & $-0.78$ & 0.03 & 1.215 & 0.023 \\ 
M69-II-37        & 3716 &  97.0 & $+0.09$ & 0.12 & $-0.78$ & 0.03 & 1.178 & 0.022 \\ 
M69-I-40         & 3917 & 113.0 & $+0.25$ & 0.12 & $-0.78$ & 0.03 & 1.166 & 0.022 \\ 
M69-2            & 3864 & 109.0 & $+0.30$ & 0.11 & $-0.78$ & 0.03 & 1.215 & 0.023 \\ 
M69-3            & 3864 & 109.0 & $+0.30$ & 0.11 & $-0.78$ & 0.03 & 1.212 & 0.023 \\ 
M69-4            & 3899 & 112.0 & $+0.41$ & 0.12 & $-0.78$ & 0.03 & 1.219 & 0.024 \\ 
M71-29           & 3641 & 108.0 & $+0.09$ & 0.12 & $-0.84$ & 0.06 & 1.203 & 0.017 \\ 
M71-30           & 3992 & 120.0 & $+0.68$ & 0.13 & $-0.84$ & 0.06 & 1.185 & 0.016 \\ 
M71-B            & 3764 & 100.0 & $+0.25$ & 0.12 & $-0.84$ & 0.06 & 1.188 & 0.016 \\ 
M71-46           & 4011 & 121.0 & $+0.80$ & 0.11 & $-0.84$ & 0.06 & 1.195 & 0.016 \\ 
M71-A4           & 4153 & 134.0 & $+0.79$ & 0.13 & $-0.84$ & 0.06 & 1.178 & 0.016 \\ 
M71-1=H          & 3796 & 103.0 & $+0.86$ & 0.12 & $-0.84$ & 0.06 & 1.175 & 0.016 \\ 
M71-2=I          & 4565 & 173.0 & $+1.08$ & 0.11 & $-0.84$ & 0.06 & 1.149 & 0.015 \\ 
M71-3=113        & 3954 & 116.0 & $+0.83$ & 0.13 & $-0.84$ & 0.06 & 1.192 & 0.016 \\ 
M71-4=45         & 4011 & 121.0 & $+0.87$ & 0.13 & $-0.84$ & 0.06 & 1.171 & 0.016 \\ 
M71-5=64         & 4153 & 134.0 & $+1.42$ & 0.10 & $-0.84$ & 0.06 & 1.159 & 0.016 \\ 
M71-6=66         & 4649 & 182.0 & $+1.39$ & 0.10 & $-0.84$ & 0.06 & 1.137 & 0.015 \\ 
M71-8=21         & 4458 & 162.0 & $+1.45$ & 0.13 & $-0.84$ & 0.06 & 1.138 & 0.015 \\ 
NGC0104-A02      & 3533 & 128.0 & $+0.08$ & 0.13 & $-0.78$ & 0.02 & 1.214 & 0.035 \\ 
NGC0104-W12      & 3780 & 102.0 & $+0.05$ & 0.10 & $-0.78$ & 0.02 & 1.225 & 0.036 \\ 
NGC0104-A19      & 3554 & 128.0 & $+0.10$ & 0.12 & $-0.78$ & 0.02 & 1.181 & 0.034 \\ 
NGC0104-V07      & 3764 & 100.0 & $+0.21$ & 0.13 & $-0.78$ & 0.02 & 1.188 & 0.034 \\ 
NGC0104-V06      & 3864 & 109.0 & $+0.43$ & 0.12 & $-0.78$ & 0.02 & 1.172 & 0.033 \\ 
NGC0104-L168     & 3882 & 110.0 & $+0.43$ & 0.12 & $-0.78$ & 0.02 & 1.215 & 0.035 \\ 
NGC0104-5529     & 3973 & 118.0 & $+0.64$ & 0.13 & $-0.78$ & 0.02 & 1.205 & 0.035 \\ 
NGC0104-2426     & 4070 & 126.0 & $+0.93$ & 0.12 & $-0.78$ & 0.02 & 1.170 & 0.033 \\ 
NGC0104-1505     & 4070 & 126.0 & $+0.96$ & 0.12 & $-0.78$ & 0.02 & 1.184 & 0.034 \\ 
NGC0104-4418     & 4091 & 128.0 & $+0.95$ & 0.11 & $-0.78$ & 0.02 & 1.188 & 0.034 \\ 
NGC0104-1510     & 4153 & 134.0 & $+1.01$ & 0.14 & $-0.78$ & 0.02 & 1.168 & 0.033 \\ 
NGC0104-2416     & 4219 & 140.0 & $+1.33$ & 0.15 & $-0.78$ & 0.02 & 1.177 & 0.034 \\ 
NGC0104-6408     & 4383 & 155.0 & $+1.44$ & 0.14 & $-0.78$ & 0.02 & 1.163 & 0.033 \\ 
NGC0288-A96      & 4070 & 126.0 & $+0.50$ & 0.10 & $-1.14$ & 0.03 & 1.144 & 0.016 \\ 
NGC0288-A78      & 4132 & 132.0 & $+0.67$ & 0.11 & $-1.14$ & 0.03 & 1.148 & 0.016 \\ 
NGC0288-C20      & 4153 & 134.0 & $+0.71$ & 0.11 & $-1.14$ & 0.03 & 1.118 & 0.015 \\ 
NGC0288-A77      & 4219 & 140.0 & $+0.83$ & 0.10 & $-1.14$ & 0.03 & 1.135 & 0.016 \\ 
NGC0288-A245     & 4433 & 160.0 & $+1.10$ & 0.12 & $-1.14$ & 0.03 & 1.124 & 0.015 \\ 
NGC0362-III11    & 4011 & 121.0 & $+0.37$ & 0.10 & $-1.09$ & 0.03 & 1.162 & 0.035 \\ 
NGC0362-IV100    & 4091 & 128.0 & $+0.50$ & 0.10 & $-1.09$ & 0.03 & 1.145 & 0.034 \\ 
NGC0362-III63    & 4031 & 123.0 & $+0.52$ & 0.10 & $-1.09$ & 0.03 & 1.169 & 0.036 \\ 
NGC0362-III44    & 4091 & 128.0 & $+0.66$ & 0.10 & $-1.09$ & 0.03 & 1.119 & 0.033 \\ 
NGC0362-III70    & 4310 & 148.0 & $+0.75$ & 0.10 & $-1.09$ & 0.03 & 1.088 & 0.031 \\ 
NGC5927-100$^\star$     & 3847 & 107.0 & $+0.01$ & 0.11 & $-0.64$ & 0.01 & 1.246 & 0.046 \\ 
NGC5927-799      & 3847 & 107.0 & $+0.38$ & 0.12 & $-0.64$ & 0.01 & 1.243 & 0.045 \\ 
NGC5927-627      & 3864 & 109.0 & $+0.51$ & 0.11 & $-0.64$ & 0.01 & 1.203 & 0.043 \\ 
NGC5927-532      & 3992 & 120.0 & $+0.80$ & 0.12 & $-0.64$ & 0.01 & 1.167 & 0.041 \\ 
NGC5927-622      & 4175 & 136.0 & $+1.00$ & 0.15 & $-0.64$ & 0.01 & 1.152 & 0.040 \\ 
NGC5927-536      & 4310 & 148.0 & $+1.27$ & 0.15 & $-0.64$ & 0.01 & 1.191 & 0.042 \\ 
NGC6388-1        & 3954 & 116.0 & $-0.34$ & 0.13 & $-0.74$ & 0.18 & 1.238 & 0.038 \\ 
NGC6388-3        & 3899 & 112.0 & $-0.05$ & 0.11 & $-0.74$ & 0.18 & 1.230 & 0.038 \\ 
NGC6388-4        & 3701 &  96.0 & $+0.09$ & 0.13 & $-0.74$ & 0.18 & 1.204 & 0.037 \\ 
NGC6440-KF-1     & 3847 & 107.0 & $+0.41$ & 0.13 & $-0.62$ & 0.10 & 1.179 & 0.018 \\ 
NGC6440-KF-2     & 3813 & 105.0 & $+0.46$ & 0.11 & $-0.62$ & 0.10 & 1.210 & 0.018 \\ 
NGC6440-KF-3     & 3716 &  97.0 & $+0.47$ & 0.12 & $-0.62$ & 0.10 & 1.196 & 0.018 \\ 
NGC6440-KF-4     & 3686 &  94.0 & $+0.50$ & 0.12 & $-0.62$ & 0.10 & 1.193 & 0.018 \\ 
NGC6440-KF-5     & 3864 & 109.0 & $+0.53$ & 0.11 & $-0.62$ & 0.10 & 1.239 & 0.019 \\ 
NGC6440-KF-6     & 3747 &  99.0 & $+0.57$ & 0.12 & $-0.62$ & 0.10 & 1.188 & 0.018 \\ 
NGC6440-KF-8     & 3747 &  99.0 & $+0.67$ & 0.11 & $-0.62$ & 0.10 & 1.196 & 0.018 \\ 
NGC6440-KF-7     & 3796 & 103.0 & $+0.67$ & 0.12 & $-0.62$ & 0.10 & 1.201 & 0.018 \\ 
NGC6528-7        & 3864 & 109.0 & $+0.30$ & 0.11 & $-0.61$ & 0.08 & 1.220 & 0.029 \\ 
NGC6528-11       & 3732 &  98.0 & $+0.05$ & 0.11 & $-0.61$ & 0.08 & 1.242 & 0.030 \\ 
NGC6528-22$^\star$      & 3813 & 105.0 & $+0.37$ & 0.12 & $-0.61$ & 0.08 & 1.275 & 0.031 \\ 
NGC6528-6$^\star$      & 3917 & 113.0 & $+0.41$ & 0.12 & $-0.61$ & 0.08 & 1.248 & 0.030 \\ 
NGC6553-20       & 3780 & 102.0 & $-0.16$ & 0.12 & $-0.60$ & 0.04 & 1.209 & 0.034 \\ 
NGC6553-19       & 3551 & 138.0 & $-0.02$ & 0.10 & $-0.60$ & 0.04 & 1.227 & 0.035 \\ 
NGC6553-25       & 3747 &  99.0 & $+0.14$ & 0.14 & $-0.60$ & 0.04 & 1.229 & 0.036 \\ 
NGC6553-16$^\star$      & 3917 & 113.0 & $+0.27$ & 0.11 & $-0.60$ & 0.04 & 1.288 & 0.038 \\ 
NGC6553-26$^\star$      & 3847 & 107.0 & $+0.32$ & 0.13 & $-0.60$ & 0.04 & 1.281 & 0.038 \\ 
NGC6553-14$^\star$      & 3813 & 105.0 & $+0.38$ & 0.12 & $-0.60$ & 0.04 & 1.265 & 0.037 \\ 
NGC6553-2        & 3656 & 105.0 & $+0.42$ & 0.12 & $-0.60$ & 0.04 & 1.192 & 0.034 \\ 
NGC6624-KF-1     & 3917 & 113.0 & $+0.40$ & 0.11 & $-0.70$ & 0.03 & 1.230 & 0.034 \\ 
NGC6624-KF-2     & 3917 & 113.0 & $+0.53$ & 0.13 & $-0.70$ & 0.03 & 1.160 & 0.030 \\ 
NGC6624-KF-3     & 3917 & 113.0 & $+0.54$ & 0.13 & $-0.70$ & 0.03 & 1.162 & 0.030 \\ 
NGC6624-KF-4     & 3899 & 112.0 & $+0.82$ & 0.14 & $-0.70$ & 0.03 & 1.218 & 0.033 \\ 
NGC6624-KF-5     & 4070 & 126.0 & $+0.96$ & 0.15 & $-0.70$ & 0.03 & 1.197 & 0.032 \\ 
NGC6712-LM5$\star$     & 4111 & 130.0 & $+0.55$ & 0.10 & $-0.94$ & 0.03 & 1.223 & 0.024 \\ 
NGC6712-LCO1     & 4132 & 132.0 & $+0.58$ & 0.11 & $-0.94$ & 0.03 & 1.211 & 0.023 \\ 
NGC6712-LCO3     & 4219 & 140.0 & $+0.61$ & 0.10 & $-0.94$ & 0.03 & 1.188 & 0.023 \\ 
NGC6712-LM8      & 4111 & 130.0 & $+0.71$ & 0.11 & $-0.94$ & 0.03 & 1.157 & 0.022 \\ 
NGC6712-LM10     & 4264 & 144.0 & $+0.72$ & 0.10 & $-0.94$ & 0.03 & 1.194 & 0.023 \\ 
NGC6712-B66      & 4196 & 138.0 & $+0.85$ & 0.12 & $-0.94$ & 0.03 & 1.195 & 0.023 \\ 
Terzan2-1        & 4241 & 142.0 & $+0.40$ & 0.10 & $-0.65$ & 0.14 & 1.158 & 0.034 \\ 
Terzan2-2        & 3973 & 118.0 & $+0.41$ & 0.11 & $-0.65$ & 0.14 & 1.205 & 0.037 \\ 
Terzan2-3$^\star$      & 3899 & 112.0 & $+0.43$ & 0.12 & $-0.65$ & 0.14 & 1.260 & 0.040 \\ 
Terzan2-4        & 4175 & 136.0 & $+0.44$ & 0.11 & $-0.65$ & 0.14 & 1.215 & 0.037 \\ 
Terzan2-5        & 3936 & 115.0 & $+0.56$ & 0.12 & $-0.65$ & 0.14 & 1.200 & 0.037 \\ 
Terzan2-7        & 4111 & 130.0 & $+0.61$ & 0.10 & $-0.65$ & 0.14 & 1.199 & 0.037 \\ 
Terzan2-8        & 4132 & 132.0 & $+0.62$ & 0.10 & $-0.65$ & 0.14 & 1.151 & 0.034 \\ 
\end{longtable}
}

\end{document}